\documentclass[aps,eqsecnum,a4paper]{revtex4}
\usepackage{makeidx}
\usepackage{amsmath}
\usepackage{amsfonts}
\usepackage{amssymb}
\usepackage{mathrsfs}
\usepackage[mathscr]{euscript}
\usepackage[dvips]{graphicx}
\usepackage{fancyhdr}
\usepackage[hypertex=true]{hyperref} %[backref=page][hyperindex=false]
\def\overstrike#1#2{{\setbox0\hbox{$#2$}\hbox to \wd0{\hss
    $#1$\hss}\kern-\wd0\box0}}
\def\opm{~\overstrike{\bigcirc}{\pm}~}
\def\omp{~\overstrike{\bigcirc}{\mp}~}

\numberwithin{equation}{subsection}

\begin{document}

\title{Theory of directional pulse propagation: detailed calculations}
\author{P. Kinsler}
\affiliation{
  Department of Physics$^*$, Imperial College,
  Prince Consort Road,
  London SW7 2BW, 
  United Kingdom.
}

\begin{abstract}

I construct combined electric and magnetic field variables
 which independently represent energy flows in
 the forward and backward directions respectively, 
 and use these to re-formulate Maxwell's equations.
The emphasis is on detailed calculations, 
 with a more general overview being published in 
 Phys. Rev. A72 and arXiv.
These directional variables enable us to not only judge the effect
 and significance of backward-travelling field components,
 but also to discard them when appropriate.
They thereby have the potential to simplify numerical simulations, 
 leading to potential speed gains of up to 100\% 
 over standard FDTD or PSSD simulations.
These field variables are also used to derive both
 envelope equations useful for narrow-band pulse propagation,
 and a second order wave equation.
Alternative definitions are also presented, 
 along with their associated wave equations.

\end{abstract}

\lhead{\includegraphics[height=5mm,angle=0]{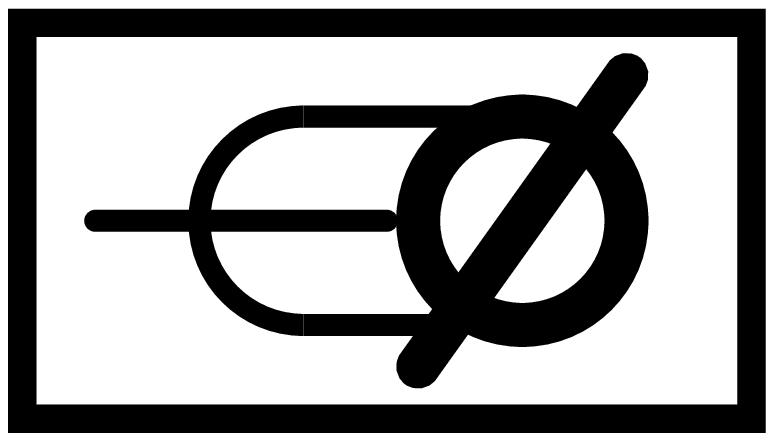}~~DETAILED CALCULATION}
\chead{~}
\rhead{
\href{mailto:Dr.Paul.Kinsler@physics.org}{Dr.Paul.Kinsler@physics.org}\\
\href{http://www.kinsler.org/physics/}{http://www.kinsler.org/physics/}
}
\lfoot{\thesection . \thesubsection}%; ~~~~ (\yymmdddate\today:\currenttime) }
%\rfoot{{\large {\em Not for redistribution}}}

\date{\today}
\maketitle
\thispagestyle{fancy}

\newcommand{\xxref}[1]{{\bf See:#1:}}
\newcommand{\xxlabel}[1]{[{\bf Label:{#1}}]}

This report should be read along with the paper 
 Phys.Rev.A72, 063807 (2005)
 {\em ``Theory of directional pulse propagation''},
 by  
 P. Kinsler, S.B.P. Radnor, G.H.C. New,
 for proper context.

 This document is primarily intended as a complete (as possible) record
of the calculational steps that were necessarily abbreviated (or omitted) from that published work; 
 it also contains a great many other relevant calculations along
 with some speculation.
It is an edited version of a longer document from which on-going work has been excised; and, as a
"work in progress", despite my best efforts, may contain occasional mistakes. 
Please contact me if
you have any comments, corrections or queries.

~

\noindent
WWW: QOLS Group 
\href{http://www.qols.ph.ic.ac.uk/}{http://www.qols.ph.ic.ac.uk/} \\
WWW: Physics Dept. 
\href{http://www.ph.ic.ac.uk/}{http://www.ph.ic.ac.uk/} \\
WWW: Imperial College
\href{http://www.ic.ac.uk/}{http://www.ic.ac.uk/} \\
Email: Paul Kinsler 
\href{mailto:Dr.Paul.Kinsler@physics.org}{Dr.Paul.Kinsler@physics.org}\\
Email: G.H.C. New
\href{mailto:g.new@ic.ac.uk}{g.new@ic.ac.uk}

% ----------------------------------------------------------------------
\tableofcontents

%\newpage
\chead{Directional fields: ...}
% ----------------------------------------------------------------------

~\\

\section{Introduction}

We define field variables with directional properties, 
 and use them to show how to optimise both our understanding of 
 pulse propagation and numerical simulations.
Simple plane-polarized versions of these were originally introduced by 
 Fleck \cite{Fleck-1970prb}, 
 using $\epsilon^{1/2} E_x \pm \mu^{1/2} H_y$.
However, 
 these were little used and the bulk of the paper primarily addressed 
 generating ultra-short pulses by Q-switched lasers, 
 discussing numerical techniques and simulation results. 
Here we derive wave equations and present results using generalised 
 versions of these field variables.  
In the course of our investigations, 
 we use these variables to illuminate several interesting features 
 of $z$-propagated optical pulse simulations
 which are usually overlooked in a ``forward-only'' approximation.

The main novelty of these new field variables is that they correspond to 
 forward and backward directed energy fluxes; 
 and are constructed from combinations of the electric and magnetic fields, 
 along with the properties of the propagation medium.

I also derive wave equations describing the propagation of 
 these field components, 
 in fully vectorised and plane-polarized versions.  
In addition to these exact re-expressions of Maxwell's equations, 
 I also present second order  and
 envelope-based wave equations for $G^{\pm}$, 
 in both stationary and moving frames.
A further advantage of this approach is that it is as easy
 to include magneto-optic effects as  
 electro-optics ones (i.e. dispersion and nonlinearity).  
For example, 
 I will later show that the plane polarized first order wave equations 
 describing the $z$-propagation of $G^{\pm}$ in the spectral ($\omega$) 
 domain are --
~
\begin{eqnarray}
  \partial_z G^{\pm}
&=&
 \mp 
  \imath \omega 
  \alpha_r \beta_r 
  ~
  G^\pm
 ~~
 \mp 
  \frac{\imath \omega \beta_r}
       {2}
  \alpha_c
  *
  \left[ G^+ + G^- \right]
 ~~
 -
  \frac{\imath \omega \alpha_r}
       {2}
  \beta_c
  *
  \left[ G ^+ - G^- \right]
,
\label{eqn-intro-egpropag}
\end{eqnarray}
where $\alpha_r=\sqrt{\epsilon_L}, \beta_r=\sqrt{\mu_L}$
 contain the linear response of the medium, 
 and $\alpha_c$ contains nonlinearity that affects the electric field, 
 and $\beta_c$ is its magnetic counterpart.
Using these forward and backward  directed $G^{\pm}$ variables has
 three main advantages:
\\
\textbullet ~
%(1) 
The forward directed
field variable $G^{+}$ is the appropriate physical choice
to use when studying the forward propagating part of the field --
as is the case for most pulse propagation investigations.
\\
\textbullet ~
%(2) 
Because we can choose to simulate only  the forward directed field 
variable $G^{+}$, we have half the calculations to do, and 
hence gain a speed advantage.  By using a 
$z$-propagated PSSD algorithm\cite{Tyrrell-KN-2005jmo},
we also get a fast and 
flexible treatment of dispersive and nonlinear effects; 
hence our simulations significantly outperform 
standard FDTD approaches.
\\
\textbullet ~
%(3) 
Most 
other approaches (even the recent \cite{Kolesik-M-2004pre}) assume 
the backward propagating parts of the field are negligible.  Since 
our $G^{-}$ describes the backward directed part,
we have a clear and physically appropriate basis on which to calculate
it, and hence can (when appropriate) rigorously justify the common 
forward-only
approximation.

We can do even better by changing to 
a  co-moving frame.  This simplifies the $G^+$ propagation,
by removing the frequency-like oscillations, but has the price of 
making the $G^-$ oscillations faster.  However, in the limit where 
we decide to approximate by ignoring the 
backward propagating $G^-$ part of the field, this price
becomes irrelevant.

% --------  -------- 

\subsection{Why am I doing this? An anecdotal history}

Fleck's definitions were brought to my attention by Geoff New 
 shortly after I started working with at Imperial in 2001.
Because they gave rise to additional $d/dt$ terms in the polarization,
 when compared to standard forms, 
 he wondered whether there was a link with the 
 Brabec and Krausz derivation\cite{Brabec-K-1997prl}
 of similar-looking corrections to an envelope propagation equation.
Although it turns out that this was not the case, 
 I continued working on the idea.
At first I focussed on generating a second order wave equation of $G^\pm$, 
 in analogy to the one usually presented for the $E$ field.
Although I generalized my GFEA derivation 
 \cite{Kinsler-N-2003pra,Kinsler-FCPP} to include Fleck-like terms, 
 there seemed no practical use for the technique.

In 2002 a PhD student in our group (JCA Tyrrell) started working on 
 Maxwell's equations solvers, 
 and applying them to pulse propagation in nonlinear media.
He developed the PSSD method \cite{Tyrrell-KN-2005jmo}
 for solving Maxwell's equations,
 where the field is held as a function of time
 and propagated forward in space.
This is common in envelope-based nonlinear optics simulations, 
 but had not been applied to full  Maxwell's solvers before.
This got me thinking about representing Maxwell's equations using 
 $G^\pm$, 
 and put the project on track to being of practical use, 
 although there were many outstanding questions on interpretation of 
 the fields and the wave equations
While trying to clarify these issues, 
 I became aware of work by Kolesik and co-workers 
 \cite{Kolesik-MM-2002prl,Kolesik-M-2004pre}), 
 who projected out Maxwell's equations into a directional form.
Their treatment motivated me on to vectorize the definitions, 
 and get a PhD student (SBP Radnor) to start on simulations.
As we worked on the paper eventually published as \cite{Kinsler-RN-2005pra}, 
 I further generalised the definitions to include the longitudinal 
 parts of the fields, 
 fully developed the variant forms of the variables, 
 and derived envelope versions of the wave equations.

As it stands, 
 these kinds of directional wave equations,
 either using my $G^\pm$ variables, 
 or the Kolesik et.al. approach
 make traditional envelope theories based on 
 the second order wave equation for $E$ utterly redundant.
This is true even for 
 the generalised ones such as \cite{Brabec-K-1997prl,Kinsler-N-2003pra}
 which incorporate wideband corrections.
Directional wave equations simple provide propagation equations
 without introducing the many approximations inherent in theories based on  
 second order forms, 
 as are detailed exaustively in \cite{Kinsler-N-2003pra,Kinsler-FCPP}.
This is treated in more detail in my report 
 {\em ``Field and envelope methods in nonlinear optics''}.

I am occasionally asked why I picked $G^\pm$ instead of (e.g.) $F^\pm$
 to name these directional fields --
 after all, 
 $F$ follows alphabetically from the electric field (letter) $E$, 
 and it is the first letter of Fleck's surname,
 so it might seem the more obvious choice.
Unfortunately I can't remember why; 
 but there was a reason why I passed over $F$, 
 I think something to do with some other $F$ I was using at the time.
I should have picked $K^\pm$, anyway.

% --------  -------- 

\subsection{An alternative approach}\label{S-intro-KolesikMoloney}

% --------  

\subsubsection{Kolesik, Moloney, and Mlejnek; PRL {\bf 89}, 283902 (2002);  and projection operators}

Kolesik et.al. \cite{Kolesik-MM-2002prl} also use not-dissimilar
 combinations of field variables.  
They introduce projection operators for the forward and backward parts 
 of the field in their eqn (KMM~1), which 
 I rewrite slightly using $\vec{u} = \vec{k} / \left| \vec{k} \right|$:
~
\begin{eqnarray}
  \mathscr{P}^{\pm} 
  \vec{D}(\vec{k})
&=&
  \frac{1}{2}
  \left[
    \vec{D}(\vec{k})
   \mp
      \textrm{sgn}(kz) \frac{k}{\omega(k)}
     \vec{u} 
      \times \vec{H} (\vec{k})
  \right]
\\
  \mathscr{P}^{\pm} 
  \vec{H}(\vec{k})
&=&
  \frac{1}{2}
  \left[
    \vec{H}(\vec{k}))
   \pm
      \textrm{sgn}(kz) \frac{\omega(k)}{k}
     \vec{u} 
      \times \vec{D} (\vec{k})
  \right]
.
\end{eqnarray}
~
They then use these, with Maxwell's equations to generate their UPPE 
in eqn (KMM~7):
~
\begin{eqnarray}
  \partial_t \vec{D}_f (\vec{k})
&=&
 -
  \imath \omega(k) \vec{D}_f (\vec{k})
 +
  \frac{\imath}{2}
  \omega(k)
  \left[
    \vec{P}_{NL}(\vec{D},\vec{k})
   -
    \vec{u} 
      \left(
        \vec{u} \cdot \vec{P}_{NL}(\vec{D},\vec{k})
      \right)
  \right]
.
\label{eqn-Kolesik-7}
\end{eqnarray}

There are three main difference compared with my generalised 
 Fleck-style $G^\pm$ approach.  
They:

\begin{description}

\item[1.] describe things in terms of projection operators,

\item[2.] project out {\em both} a directional $D$ and a directional $H$,

\item[3.] construct their (effective) directional variables in a different
          (but related) way.

\end{description}

Another difference is that, when implimenting a solution, 
 they keep the {\em time derivatives} on the LHS, 
 and manipulate the $\nabla \times$ terms on the RHS; 
 so their UPPE gives a time derivative of the field, 
 and so their fields are propagated through time. 
This contrasts with my approach which keeps the $\nabla \times $ on the LHS,
 and manipulates the time derivatives -- 
 hence my fields are propagated through space.  
Their choice is (most probably) why they choose to work with 
 the displacement field $D$ rather than $E$.
In the time domain, 
 the use of $D$ avoids the need for the $D = \epsilon \ast E$
 convolution of the field with the permittivity, 
 at least on the time-derivative side of the Maxwell's equations.  
They (presumably) remove the space derivatives resulting from the 
 $\nabla \times$ terms when going into the spatial frequency ($k$) domain, 
 leading to the appearance of $K$ terms on the RHS.

However it is not so easy to avoid complication 
 when solving Maxwell's equations.  
Their use of $D$ simplifies the time derivative side, 
 but there is a cost to pay on the space derivative ($\nabla \times$) side, 
 where they need to reconstruct $E$ from $D$ (see eqn (DMN~2)),
 and similarly evaluate the nonlinearity  (see eqn (DMN~3)).  
Even then, 
 just prior to eqn (DMN~7), Kolesik et.al. explain how they go into the 
 frequency domain anyway!  
As an aside, 
 they might easily have used $B$ and not $H$ in their equations, 
 and so included the possibility of magnetic media in their model.

In contrast, I leave the space derivative ($\nabla \times$) sides alone,
 and work with the time derivative side. 
This has the advantage that when I go into the frequency domain, 
 the convolution becomes a simple product,
 and I do not have to resort to reconstructions of $E$ and $P_{NL}$.
It is interesting to compare the ``$\partial_t$'' UPPE with my similar-looking
``$\partial_z$'' eqn (\ref{eqn-stationaryGpropagation}) that follows later,
 where they neglect to introduce the advantages of a co-moving frame 
 (cf my eqn(\ref{eqn-movingGpropagation})).  
See also \ref{Ss-altfleck-vsKolesik}, 
 where I attempt a conversion between my formulation and theirs.

\noindent
{\bf SUMMARY: } 
Maxwell-like theory with combined $G$-like variables, but no co-moving frame.  
Temporal propagation only.

% --------  

\subsubsection{Kolesik and Moloney; PRE {\bf 70}, 036604 (2004). }

This work \cite{Kolesik-M-2004pre} is an extension of the method 
 proposed in \cite{Kolesik-MM-2002prl}, 
 and contains a variety of applications.
It discusses both $t$ and $z$ propagated approaches, 
 and includes decomposition into transverse modes.
They develop an envelope version of their wave equation, 
 but do not discuss using multiple envelopes.
In general they only solves for $E$, 
 not both $E$ and $H$ separately, 
 there is also no examination (or calculation) of 
 the significance of backward propagating fields.
They separate the polarization $P$ from displacement $D$; 
 and includes current terms.
See also \ref{Ss-altfleck-vsKolesik}, 
 where I attempt a conversion between my formulation and theirs.

% --------  -------- 

\subsection{Moving frames}

In Fidel et.al. 1997 \cite{Fidel-HKZ-1997jcm} 
 {\em "Hybrid Ray FDTD moving window approach to pulse propagation"},
 they look at a variety of moving window and moving frame  approaches
 to solving for pulse propagation with Maxwell's equations using FDTD.  
They do introduce a co-moving frame, and derive the appropriate 
 discretized Maxwell's equations; 
 but they do not use combined $E$ and $B$
 variables as done by Fleck, 
 and as done in the following sections here.

\noindent
{\bf SUMMARY: } 
Maxwell-like theory with 
co-moving frame, but no combined $G$-like variables.

% --------  -------- 

\subsection{Beltrami Fields, Photon Wave Functions: $\vec{E}+\imath \vec{B}$}

The $\vec{G}^\pm$ fields I constuct here 
 bear a superficial resemblance to Beltrami variables 
 \cite{Weiglhofer-L-1994pre,Lakhtakia-1994ijimw,Hillion-1995jpa,Moses-1971siamjam},
 which are defined along the lines of
 $\vec{Q} = \sqrt{\epsilon} \vec{E} + \imath \sqrt{\mu} \vec{H}$.
However, 
 they differ in two important respects.  
First, 
 for a given Beltrami $\vec{Q}$ you immediately know what 
 the component $\vec{E}$ and $\vec{H}$ fields are, 
 whereas you require both $\vec{G}^\pm$ to do the same.
Secondly, 
 $\vec{Q}$ does not assume any preferred direction, 
 whereas the construction of $\vec{G}^\pm$ requires
 a direction $\vec{u}$ to be chosen.
It's interesting to wonder whether 
 the $E\pm \imath H$ approaches might generalise
 to something like $E\pm e^{\imath \phi} H$).
However, 
 since $E \pm \imath H$ is unique for a given $E$ and $H$; 
 and $E \pm H$ is not unique without both components, 
 so it's not obvious how it might work.

Note that 
some discussions of Beltrami fields refer to ``force free'' systems, 
 where $\nabla \times B = \alpha B$,
 which are not (as far as I can see) applicable to an optics regime.

% --------  -------- 

%\bibliography{fleck}

% ----------------------------------------------------------------------
%\newpage

\section{The material properties: reference and correction parameters}
\label{S-medium}

Maxwell's equations describe propagation of the EM field through a 
 medium described by permittivity $\epsilon$ and permeability $\mu$.
In addition, 
 it is common to add in a ``polarization'' term $\vec{P}$, 
 which is used to relate the displacement field $\vec{D}$ to $\vec{E}$ in 
 the usual way: 
 $\vec{D} = \epsilon \ast \vec{E} + \vec{P}$.  
In my treatment,
 I hide this polarization inside the definition of $\epsilon$
 (writing instead just $\vec{D} = \epsilon \ast \vec{E}$ for the same system); 
 similarly any magnetic polarization would be hidden inside $\mu$ 
 and not written out separately.  
Note that I do make a split of the total permittivity
 (and also the permeability) into two pieces, 
 a ``reference'' part and a ``correction'' part, 
 and these will not necessarily be the same as the traditional 
 division.

\noindent
Note 1: I use the ``$\ast$'' notation for convolutions 
 $a(t) \ast b(t) = \int a(t-t') b(t') dt'$.

\noindent
Note 2: Quantities in the  frequency domain are indicated by tildes.

% ---------  --------- ---------
\subsection{Reference and correction contributions}\label{Ss-referection}

We first denote the linear response of the medium to be $\epsilon_L$, 
 and its nonlinear response to be $\epsilon_{NL}$.
The nonlinear term may include some time response, 
 such as e.g. $\epsilon_{NL}(t) = \int \epsilon_3(\tau) E(t-\tau)^2 d\tau$.
The effect of the total permittivity can be expressed
 in either the time domain or the frequency domain (indicated by tildes).
It is
~
\begin{eqnarray}
 \epsilon(t) * E(t)
&=&
  \epsilon_L(t) * E(t)
 +
  \epsilon_{NL}(t) E(t)
\label{eqn-defs-epsilon-t}
\\
\textrm{or} 
~~~~
~~~~
  \tilde{\epsilon}(\omega)
  *
  \tilde{E}(\omega)
&=&
  \tilde{\epsilon}_L(\omega) 
  \tilde{E}(\omega)
 + 
  \tilde{\epsilon}_{NL}(\omega)
  *
  \tilde{E}(\omega)
.
\label{eqn-defs-epsilon-w}
\end{eqnarray}
Similar expressions can be written down for the case of magnetic dispersion
 or magnetic nonlinearities,
~
\begin{eqnarray}
 \mu(t) * E(t)
&=&
  \mu_L(t) * E(t)
 +
  \mu_{NL}(t) E(t)
\label{eqn-defs-mu-t}
\\
\textrm{or} 
~~~~
~~~~
  \tilde{\mu}(\omega)
  *
  \tilde{E}(\omega)
&=&
  \tilde{\mu}_L(\omega) 
  \tilde{E}(\omega)
 + 
  \tilde{\mu}_{NL}(\omega)
  *
  \tilde{E}(\omega)
.
\label{eqn-defs-mu-w}
\end{eqnarray}

The definitions of ${G}^{\pm}$ 
 (and their generalized vector counterparts $\vec{G}^{\pm}$, introduced below)
 depend on the properties of
 the propagation medium through the
 permittivity $\epsilon$ and permeability $\mu$.  
In principle it might seem attractive to define $\vec{G}^{\pm}$ using
 the exact values of $\epsilon, \mu$ (including the nonlinearity), 
 but we will usually want to be able to 
 uniquely reconstruct the fields $E$ and $H$ 
 from our new fields.
We do this by including as much as possible of the 
 dispersive properties in the reference parameters
 (and hence the definitions of $\vec{G}^{\pm}$), 
 leaving nonlinear properties (and potentially some residual dispersion) 
 for correction terms.
For example, 
 if the true dispersion was multi-valued or zero in places, 
 it is possible to define the reference to include an
 approximation to the true dispersion without those inconvenient properties,
 and use the correction term to compensate.

The frequency domain is the starting point for
 calculating the necessary parameters.
This is because the formulae are simplest in the $\omega$ domain, 
 where any linear response is described using products rather 
 than convolutions.
This will mean that we calculate the time-domain parameters by 
 (back) Fourier transforming the frequency domain ones.
The frequency domain definitions are
~
\begin{eqnarray}
  \tilde{\epsilon} * \tilde{E}
&=&
  \tilde{\epsilon}_r(\omega) \tilde{E}(\omega) 
 +
  \tilde{\epsilon}_c(\omega) * \tilde{E}
\\
&=&
   \tilde{\alpha}_r^2(\omega) \tilde{E}(\omega)
 + \tilde{\alpha}_r(\omega) ~ \tilde{\alpha}_c(\omega) * \tilde{E}(\omega),
\label{eqn-defs-alphaX}
\\
  \tilde{\mu} * \tilde{H}
&=&
  \tilde{\mu}_r(\omega) \tilde{H}(\omega)
 +
  \tilde{\mu}_c(\omega) * \tilde{H}(\omega)
\\
&=&
    \tilde{\beta}_r^2(\omega) \tilde{H}(\omega)
  + \tilde{\beta}_r(\omega) ~ \tilde{\beta}_c(\omega) * \tilde{H}(\omega),
\label{eqn-defs-betaX}
\end{eqnarray}
where the correction parameters $\tilde{\epsilon}_c$ and $\tilde{\mu}_c$
 represent the discrepancy between the true values and the reference.
The smaller these correction terms are, 
 the better the match.
By using these frequency dependent parameters in the generalized
  definitions of ${G}^{\pm}$, 
  we will be able to propagate pulses using only the ${G}^{+}$ variable,
  a gain in both mathematical simplicity and computational speed.
Note that since the definitions of ${G}^{\pm}$ depend 
 (roughly speaking) on the square roots of 
 $\tilde{\epsilon}$ and $\tilde{\mu}$, 
 we have introduced the $\tilde{\alpha}$ and $\tilde{\beta}$ parameters, 
 which will feature prominently 
 (along with their time domain counterparts $\alpha, \beta$),
 in the generalized definitions of 
 $\vec{G}^{\pm} $ that follow.

%In the time domain, we have parameters 
% $\alpha_r(t) = \mathscr{F}^{-1} [ \tilde{\alpha}_r (\omega) ]$
% and 
% $\beta_r(t)  = \mathscr{F}^{-1} [ \tilde{\beta}_r  (\omega) ]$, 
% where 
% $\mathscr{F}$ denotes the Fourier transform.

To recap,
 I split the medium properties into two parts:

{\bf First,} 
 there are the ``reference'' parts $\epsilon_r$ and $\mu_r$. 
These are the contributions from the material properties 
 which will be incorporated into the definition(s) of $G^{\pm}$. 
These will only contain {\em linear} response terms, 
 to guarantee we can reconstruct $E, H$ from $G^\pm$

{\bf Second,} 
 there are the ``correction'' parts $\epsilon_c(t)$ and $\mu_c(t)$, 
 which will contain the difference between the chosen reference and 
 the true behaviour of the material.
These will only contain any {\em nonlinear} response terms, 
 but may also contain an residual linear response also.

We might also choose to split the correction terms up into more pieces -- 
 e.g. a dispersive part, and a nonlinear part:
 e.g., 
~
\begin{eqnarray}
  \tilde{\alpha}_c * \tilde{E}
&=&
  \tilde{\alpha}_c^{D} \tilde{E}
 +
  \tilde{\alpha}_c^{NL} * \tilde{E}
.
\end{eqnarray}

% ---------  --------- ---------
\subsection{Nonlinearity}\label{Ss-medium-nl}

Since it is usually impractical to include nonlinearities in the 
 reference parameters,  
 these will normally appear in the correction terms $\epsilon_c$, $\mu_c$.
As an example, 
 consider a $n$-th order (electric) nonlinearity,
 in which case 
 $  \epsilon_c(t) =  \chi^{(n)}(t)
 \ast
  E(t)^{n-1}
$, and 
~
\begin{eqnarray}
  \tilde{\alpha}_r (\omega) \tilde{\alpha}_c (\omega) 
%  *
%  \tilde{E}(\omega)
&=&
  \mathscr{F} 
    \left[
      \chi^{(n)}(t) 
     \ast
       E(t)^{n-1}
    \right]
%  *
%  \tilde{E}(\omega)
\\
  \tilde{\alpha}_c (\omega) 
%  *
%  \tilde{E}(\omega)
&=&
  \left[ 
    \tilde{\alpha}_r (\omega)
 \right]^{-1}
 .
  \mathscr{F} 
    \left[
      \chi^{(n)}(t) 
     \ast
       E(t)^{n-1}
    \right]
%  *
%  \tilde{E}(\omega)
,
\label{eqn-nonlinear-alphac-w}
\\
  \alpha_c (t) 
% E(t)
&=&
 \mathscr{F}^{-1}
 \left\{
   \left[ 
     \tilde{\alpha}_r (\omega)
   \right]^{-1}
  .
  \tilde{\chi}^{(n)}(\omega) 
  .
      \mathscr{F}
        \left[    
          E(t)^{n-1}
        \right]
 \right\}
%  E(t)
,
\label{eqn-nonlinear-alphac-t}
\\
~
~
\textrm{where (see later) } 
~~~~
~~~~
  E(t) 
&=&
  \frac{1}{2}
  \left\{
    \mathscr{F}^{-1} \left[ \tilde{\alpha}_r^{-1} \right]
  \right\}
 \ast
  \left[ {G}^{+} + {G}^{-} \right]
\end{eqnarray}
where $\mathscr{F}[...]$ is the Fourier transform (FT) from time to frequency,
 and $E(t)$ can be found from eqn. (\ref{eqn-S-defs-Evector}).
If the reference parameters $\tilde{\alpha}_r$ contain
 dispersion (which will be the typical case), 
 we can see from eqn. (\ref{eqn-nonlinear-alphac-w}) that this will make
  $\tilde{\alpha}_c (\omega)$ dispersive even if $\chi^{(n)}$ is 
 instantaneous.
In the case of an instantaneous nonlinearity, 
 this adds more computational work (an extra two FTs), 
 although for non-instantaneous ones we needed the FTs anyway.
If the nonlinearity is instantaneous {\em and} the 
 reference parameters are non-dispersive,
 we have simply 
  $  \alpha_c^{NL}(t) 
  =  \alpha_r^{-n} . \chi^{(n)} . 
     2^{-n+1} \left[ {G}^{+} + {G}^{-} \right]^{n-1}$.

In practise, therefore, 
 calculation of an instantaneous nonlinear term will involve 
 calculating $\chi^{(n)} E(t)^{n}$, 
 fourier tranforming (denoted by $\mathscr{F}$) the result, 
 then dividing it by the (possibly frequency dependent) reference 
 parameter $\alpha_r$.  
If the reference parameter $\alpha_r$ is a constant, 
 then the fourier transform is redundant and the nonlinear
 step can be handled while remaining entirely in the time domain.
Note that there is no way we can avoid convolutions if the 
 reference $\alpha_r$ contains any frequency dependence -- 
 even if our nonlinearity is instantaneous.  
This is because to calculate $\alpha_c$ I have to remove 
 the reference $\alpha_r$ from $\tilde{\epsilon}_c$; 
 since $\alpha_c(\omega) = \tilde{\epsilon}_c(\omega) / \alpha_r(\omega)$.

%\end{section}
% ----------------------------------------------------------------------

%\newpage

\section{Definitions of $\vec{G}^{\pm}$}\label{S-definitions}

Here I go significantly further than Fleck\cite{Fleck-1970prb} 
 in two respects: 
 I {\em vectorise the definitions}, 
 and {\em  allow for dispersive effects}.
An important addition is the inclusion of a longitudinal field component,
 which is required in order to retain a full vector description
 of the EM field.
Originally I tried to generate a motivation for the directional $G^\pm$ fields
 described above by taking Maxwells equations and moving to a co-moving frame. 
However, 
 it seems now that while they certainly seem to describe parts of the
 field propagating in opposite directions, 
 they do not seem tied to any particular choice of frame.

After a brief discussion of what Fleck did, I  
 define a vector form $\vec{G}^{\pm}$ of the Fleck-style fields $G^\pm$.  
This requires the definition of a ``direction of propagation''
 denoted by the constant unit vector $\vec{u}$.  
I can then achieve the swapping of the transverse components of $\vec{H}$  
 (since $G_x \sim E_x \pm H_y$) by using 
 the cross product of $\vec{u}$ with the magnetic field variable $\vec{H}$.

Secondly I use ``reference'' material parameters 
 which I use to construct my $G^\pm$ fields.  
This allows me to distinguish between the parameters used in my construction 
 of $G^\pm$ from the actual material properties, 
 which may be too complicated to use in that context.
Generally, 
 I include all of the linear response of the medium in these
 reference parameters, 
 which makes it easiest to define $\vec{G}^{\pm}$ in the frequency domain.
This is because the linear time-response convolution in the time-domain
 becomes a simple product in the frequency domain.  
A good (accurate) choice of reference medium for a particular problem
 will ensure that the contribution due to the backward directed 
 part of the field (${G}^{-}$) is small.
In general we would want the reference part to include as much about the 
 material as is practicable, 
 to ensure the smallest possible correction term(s) for 
 propagation.

Note that whether or not the $\epsilon, \mu$ contain dispersion, 
 we here restrict ourself to the case where they are {\em scalar}. 
Also,
 I let the {\em argument} of the vector field quantities tell us 
 whether we are in the time or frequency domain, 
 to avoid too much notational clutter.
As a final note, 
 in the following calculations the assumption will be made that 
 the media are bulk, 
 without (implicit) interfaces hidden in $\epsilon$ or $\mu$.  
This means that $\partial_z \epsilon = \partial_z \mu = 0$ is assumed.

% ---------  --------- ---------
\subsection{Fleck's approach}\label{Ss-definitions-fleck}

Fleck defines a relative permittivity $\epsilon$, 
 and magnetic permeability $\mu$ in a dispersionless host medium -- 
 it is clear that these are relative parameters because 
 he later defines $\eta=\left( \epsilon \mu \right) ^{1/2}$
 as the refractive index of the of host medium  (just after eqn (F1.4b)).  
He allows for a Conductivity $\sigma$; 
 and a Polarization $P$ which characterises
 the active atoms and which may be amplifying or absorbing.  
The electric field is $E_x$ and the magnetic field $B_y = \mu H_y$.  
As usual, 
 the speed of light $c$ is just related to 
 the permittivity and permeability of the vacuum,
 i.e. $\left( \epsilon_0 \mu_0 \right) ^{-1/2}$.
Fleck simplifies plane polarised Maxwell field equations, 
 treating them only along the $z$ direction, 
 resulting in his eqn.(F1.1a,b)
~
\begin{eqnarray}
\partial_t 
  \frac{\epsilon}{c}
  E_x
&=&
-\partial_z H_y
-\frac{4\pi}{c}
 \partial_t P
-\frac{4\pi}{c}
 \sigma
 E_x,
\label{Fleck1.1a}
\\
\partial_t 
  \frac{\mu}{c}
  H_y
&=&
-\partial_z E_x
\label{Fleck1.1b}
.
\end{eqnarray}

Then Fleck defines new combined EM field variables $E^{\pm}$ in eqn.(F1.3),
 but I write them $G^{\pm}$ to avoid confusion with the $E$ field variables.

\begin{eqnarray}
E^{\pm}
\equiv
G^{\pm}
&=&
  \epsilon^{1/2} E_x 
 \pm 
  \mu^{1/2} H_y
\label{Fleck1.3}
.
\end{eqnarray}

Because of the relative phases of $E$ and $H$ in these definitions, if the the
two fields $G^+$ and $G^-$ are constants multiplied by a standard
exponentially oscillating carrier wave, they are associated with an energy
flux (i.e. Poynting vector) in opposite directions along the $z$ axis; this will presumably also hold nearly true when replacing the constant amplitude with 
``slowly varying'' pulse envelopes. See also section \ref{S-eflux}.

Fleck uses his definitions to obtain equations of motion for his 
 field variables $G^+$ and $G^-$ ($E^{\pm}$), 
 recorded in his eqn.(F1.4).  
They are
~
\begin{eqnarray}
\frac{\eta}{c}\partial_t G^{+}
+
\partial_z G^{+}
&=&
 - \frac{4\pi \mu^{1/2}}{c}
   \partial_t P
 - \frac{2 \pi \sigma}{c}
   \left(
     \frac{\mu}{\epsilon}
   \right)^{1/2}
   \left( 
       G^{+} + G^{-}
   \right)
\label{Fleck1.4a}
,\\
\frac{\eta}{c}\partial_t G^{-}
-
\partial_z G^{-}
&=&
 - \frac{4\pi \mu^{1/2}}{c}
   \partial_t P
 - \frac{2 \pi \sigma}{c}
   \left(
     \frac{\mu}{\epsilon}
   \right)^{1/2}
   \left( 
       G^{+} + G^{-}
   \right)
.
\label{Fleck1.4b}
\end{eqnarray}

I now scale eqns.(\ref{Fleck1.4a},\ref{Fleck1.4b}) (Fleck (F1.4a, F1.4b)),
 setting 
 $G'^\pm = \left(\eta / c \right) G^\pm$, 
 $P' = \left(4 \pi \mu^{1/2} / c \right) P$, and 
 $\sigma' = \left(2 \pi \sigma /
 \epsilon \right)^{1/2}$.  
In the following, I will drop the prime ($'$) marks and use $G$ for $G'$ (etc)
 in order to reduce the visual clutter and make equations easier to read.  
{\em For this subsection only}, 
 from this point onwards each occurence of $G^\pm$, $P$, 
 or $\sigma$ should be understood
 to refer to the prime variables $G'^\pm$, $P'$ \& $\sigma'$ -- 
 {\em unless explicitly stated otherwise}.
 The scaled equations are
~
\begin{eqnarray}
\partial_t G^{+}
+
\partial_z G^{+}
&=&
 - \partial_t P
 - \sigma 
   \left( 
       G^{+} + G^{-}
   \right)
\label{scaled-a}
,\\
\partial_t G^{-}
-
\partial_z G^{-}
&=&
 - \partial_t P
 - \sigma 
   \left( 
       G^{+} + G^{-}
   \right)
\label{scaled-b}
.
\end{eqnarray}

% --------- --------- --------- 
\subsection{Alternative forms}\label{Ss-definitions-vector-comments}

The ``standard'' definition for directional fields
 handles the electric field $E$ most easily; 
 although rather non-intuitively
 we will find that it makes it hard to include longitudinal $E$ components.
Fortunately, 
 other definitions are possible, 
 and these follow the standard ``Primary $\vec{E}$'' definition 
 presented next in subsection \ref{Ss-definitions-vector}, 
 in subsections
 \ref{Ss-definitions-vectorH}, 
 \ref{Ss-definitions-vectorD}, and
 \ref{Ss-definitions-vectorB}. 

Use of an alternative definition can protentially shift the non-transverse
 effects to a field variable where it causes less trouble. 
Note that circularly polarized forms can be generated for all types
 the same way as is described for primary-$\vec{E}$ type $\vec{G}^\pm$ 
 variables in  subsubsection \ref{Sss-definitions-vector-circularpolarization}.

These alternative expressions are represented only in the frequency domain; 
 time domain forms are presented for primary-$\vec{E}$ only; 
 for the other forms it is a simple matter to construct their
 time domain representation.
In the case of dispersionless reference media 
 (i.e. constant $\epsilon_r(\omega), \mu(\omega)$), 
 the time and frequency forms look very similar.
The set of alternative definitions was (partly) inspired by PDD suggesting 
 (private communication, 24 February 2003) trying
 alternative combinations --  
 such as swapping $D$ for $E$ and/or $B$ for $H$.

% ---------  --------- ---------
\subsection{Primary $\vec{E}$: the standard vector form}\label{Ss-definitions-vector}

This vector definition of directional fields most closely mirrors the simple
 form suggested by Fleck.  
For completeness, 
 I will write them down in both time and frequency domains.
Note that the use of $\vec{u} \times \vec{H}$ means that the $\vec{G}^\pm$
 will not contain any information about $\vec{u} \cdot \vec{H}$, 
 the longitudinal part of $\vec{H}$ --
 thus we insist $\vec{G}^\pm$ be magnetically transverse 
 and define a $\vec{G}^{\circ}$.
The vector fields $\vec{G}^{\pm}$ are (time domain)
~
\begin{eqnarray}
  \vec{G}^{\pm} (t)
&=&
  \alpha_r(t) \ast \vec{E}(t) + \vec{u} \times \beta_r(t) \ast \vec{H}(t)
,
\label{eqn-S-defs-Gvector}
\\
  {G}^{\circ} (t)
&=&
  \left[ \vec{u} \cdot \beta_r(t) \ast \vec{H}(t) \right]
\label{eqn-S-defs-Gvector0}
\\
\textrm{so} ~~~~ ~~~~
  \alpha_r(t)
  \ast
  \vec{E} (t)
&=&
  \frac{1}{2} 
  \left[ 
    \vec{G}^+(t) + \vec{G}^-(t) 
  \right]
\label{eqn-S-defs-Evector}
\\
\textrm{and} ~~~~ ~~~~
  \vec{u} \times 
  \beta_r(t)
  \ast
  \vec{H} (t)
&=& 
  \frac{1}{2} 
  \left[ 
    \vec{G}^+(t) - \vec{G}^- (t)
  \right]
\label{eqn-S-defs-Hvector}
,
\end{eqnarray}
or,
in the frequency domain, where we can avoid the convolutions
~
\begin{eqnarray}
  \vec{G}^{\pm} (\omega)
&=&
  \tilde{\alpha}_r (\omega)
  ~ 
  \vec{E}(\omega) 
 \pm
  \vec{u} \times 
  \sqrt{\mu_r(\omega)} 
  ~
  \vec{H}(\omega)
,
\label{eqn-S-defs-Gvector-w}
\\
  {G}^{\circ} (\omega)
&=&
  \left[
    \vec{u} \cdot
    \tilde{\beta}_r (\omega)
    ~
    \vec{H}(\omega)
  \right]
,
\label{eqn-S-defs-Gvector0-w}
\\
\textrm{so} ~~~~ ~~~~
  \tilde{\alpha}_r (\omega)
  \vec{E} (\omega)
&=&
  \frac{1}{2}
  \left[ 
    \vec{G}^+(\omega) + \vec{G}^-(\omega) 
  \right]
\label{eqn-S-defs-Evector-w}
\\
\textrm{and} ~~~~ ~~~~
  \vec{u} \times 
    \tilde{\beta}_r (\omega)
    \vec{H} (\omega)
&=& 
  \frac{1}{2}
  \left[  
    \vec{G}^+(\omega) - \vec{G}^- (\omega)
  \right]
\\
\textrm{or} ~~~~ ~~~~
    \tilde{\beta}_r (\omega)
    \vec{H} (\omega)
&=&
  \frac{1}{2}
  \vec{u}
  \times
  \left[  
    \vec{G}^+(\omega) - \vec{G}^- (\omega)
  \right]
 +
  \vec{u}
  {G}^{\circ}
,
\label{eqn-S-defs-Hvector-w}
\end{eqnarray}
since $\vec{u} \times \vec{u} \times \vec{A} 
= [ \vec{u} \cdot \vec{A} ] - \vec{A}$.
In general usage I will remain in the frequency domain, because there I
 avoid complications due to convolutions (or worse, deconvolutions);
 in addition the notation is simpler.

Definition: THF -- transverse $H$ (magnetic intensity) field 
 (i.e. $\vec{u} \cdot \vec{H} = 0$).

A derivation of the wave equations using this form is in 
section \ref{S-1storderE}.

% --------- 
\subsubsection{Divergence}\label{Sss-definitions-vector-divergence}

In the normal description of EM, 
 the divergence $\nabla \cdot \vec{D} = \rho$.
However the picture here is based on $\vec{E}$,  
 so we need a way to construct $\vec{D}$ from $\vec{E}$.
We follow the following iterative procedure, 
 which should work for the usual case of weak nonlinearities
~
\begin{eqnarray}
  \vec{D}(\omega)
&=&
  \tilde{\alpha}_r^2 \vec{E}(\omega)
 +
  \tilde{\alpha}_r \tilde{\alpha}_c *  \vec{E}(\omega)
\\
\Longrightarrow
~~~~ ~~~~
  \tilde{\alpha}_r^2 \vec{E}(\omega)
&=&
  \vec{D}(\omega)
 -
  \tilde{\alpha}_r \tilde{\alpha}_c *  \vec{E}(\omega)
\\
  \vec{E}(\omega)
&=&
  \tilde{\alpha}_r^{-2} \vec{D}(\omega)
 -
  \tilde{\alpha}_r^{-1} \tilde{\alpha}_c *  \vec{E}(\omega)
\\
&=&
  \tilde{\alpha}_r^{-2} \vec{D}(\omega)
 -
  \tilde{\alpha}_r^{-1} \tilde{\alpha}_c *  
  \left[
    \tilde{\alpha}_r^{-2} \vec{D}(\omega)
   -
    \tilde{\alpha}_r^{-1} \tilde{\alpha}_c *  \vec{E}(\omega)
  \right]
\\
&=&
  \tilde{\alpha}_r^{-2} \vec{D}(\omega)
 -
  \tilde{\alpha}_r^{-3} \tilde{\alpha}_c *  \vec{D}(\omega)
 +
   \tilde{\alpha}_r^{-2} \tilde{\alpha}_c * \tilde{\alpha}_c *  \vec{E}(\omega)
\\
&=&
  \tilde{\alpha}_r^{-2} \vec{D}(\omega)
 -
  \tilde{\alpha}_r^{-3} \tilde{\alpha}_c *  \vec{D}(\omega)
 +
   \tilde{\alpha}_r^{-2} \tilde{\alpha}_c * \tilde{\alpha}_c *
  \left[
    \tilde{\alpha}_r^{-2} \vec{D}(\omega)
   -
    \tilde{\alpha}_r^{-1} \tilde{\alpha}_c *  \vec{E}(\omega)
  \right]
\\
&=&
  \tilde{\alpha}_r^{-2} \vec{D}(\omega)
 -
  \tilde{\alpha}_r^{-3} \tilde{\alpha}_c *  \vec{D}(\omega)
 +
   \tilde{\alpha}_r^{-4} \tilde{\alpha}_c * \tilde{\alpha}_c * \vec{D}(\omega)
   -
    \tilde{\alpha}_r^{-3} 
    \tilde{\alpha}_c * \tilde{\alpha}_c * \tilde{\alpha}_c *
    \vec{E}(\omega)
\\
&=&
  \sum_{i=0}^{\infty}
    \tilde{\alpha}_r^{-2-i}
    \left[ - \tilde{\alpha}_c * \right]^i
    \vec{D}(\omega)
\end{eqnarray}

Thus,
~
\begin{eqnarray}
  \nabla \cdot \vec{E}(\omega)
&=&
  \nabla \cdot 
  \sum_{i=0}^{\infty}
    \tilde{\alpha}_r^{-2-i}
    \left[ - \tilde{\alpha}_c * \right]^i
    \vec{D}(\omega)
\\
&=&
  \sum_{i=0}^{\infty}
    \tilde{\alpha}_r^{-2-i}
    \left[ - \tilde{\alpha}_c * \right]^i
    \nabla \cdot 
    \vec{D}(\omega)
\\
&=&
  \tilde{\alpha}_r^{-2}
  \sum_{i=0}^{\infty}
    \tilde{\alpha}_r^{-i}
    \left[ - \tilde{\alpha}_c * \right]^i
    \rho(\omega)
\\
\textrm{with the shorthand notation}
~~~~ ~~~~
&=&
  \tilde{\alpha}_{\Sigma}^{-2}
    \rho(\omega)
\end{eqnarray}

It is worth noting that in the alternative constructions 
 following this ``Primary $\vec{E}$'' one, 
 it is far simpler to calculate the divergence.
In any case,
 we can now write the divergence of $\vec{G}^\pm $ as follows
~
\begin{eqnarray}
  \nabla \cdot \vec{G}^\pm (\omega)
&=&
  \nabla \cdot \tilde{\alpha}_r \vec{E} (\omega)
 ~~
 \pm
  \nabla \cdot \vec{u} \times \tilde{\beta}_r \vec{H} (\omega) 
\\
&=&
  \tilde{\alpha}_{\Sigma}^{-2}
  \tilde{\alpha}_r
  \nabla \cdot \vec{D}
 ~~
 \pm
  \tilde{\beta}_r
  \vec{H} \cdot 
  \left(
    \nabla \times \vec{u} 
  \right)
 ~~
 \mp
  \tilde{\beta}_r 
  \vec{u} \cdot 
  \left(
    \nabla \times \vec{H} 
  \right)
\\
&=&
  \tilde{\alpha}_{\Sigma}^{-2}
  \tilde{\alpha}_r
  \rho
 ~~
 \pm
  0
 ~~
 \mp
  \tilde{\beta}_r 
  \vec{u} \cdot 
  \left(
    -
    \imath \omega
    \tilde{\alpha}^2
    \vec{E}
    +
    \vec{J}
  \right)
\\
&=&
  \tilde{\alpha}_{\Sigma}^{-2}
  \tilde{\alpha}_r
  \rho
 ~~
 \pm
  \imath \omega
  \tilde{\alpha}^2
  \tilde{\beta}_r 
  \vec{u} \cdot 
    \vec{E}
 ~~
 \mp
  \tilde{\beta}_r 
  \vec{u} \cdot 
    \vec{J}
.
\\
&=&
  \tilde{\alpha}_{\Sigma}^{-2}
  \alpha_r
  \rho
 ~~
 \pm
  \frac{\imath \omega}{2}
  \tilde{\alpha}^2
  \tilde{\alpha}_r^{-1} 
  \tilde{\beta}_r 
  \vec{u} \cdot 
  \left[
    \vec{G}^+  + \vec{G}^-
  \right]
 ~~
 \mp
  \tilde{\beta}_r 
  \vec{u} \cdot 
    \vec{J}
.
\end{eqnarray}

Thus I never use the (no monopoles condition) $\nabla \cdot \vec{B} = 0$.
Clearly, if $\vec{G}^{\pm}$ are pure transverse 
 (not just THF, which is built into the definition, 
 but no longitudinal $E$ fields either), 
 then
~
\begin{eqnarray}
  \nabla \cdot \vec{G}^\pm(\omega)
&=&
  \tilde{\alpha}_{\Sigma}^{-2}
  \tilde{\alpha}_r
  \rho
\end{eqnarray}

NB:
~
\begin{eqnarray}
  \nabla \cdot \vec{G}^+(\omega)
 -
  \nabla \cdot \vec{G}^-(\omega)
&=&
  \tilde{\alpha}_{\Sigma}^{-2}
  \tilde{\alpha}_r
  \rho
 ~~
 +
  \frac{\imath \omega}{2}
  \tilde{\alpha}^2
  \tilde{\alpha}_r^{-1} 
  \tilde{\beta}_r 
  \vec{u} \cdot 
  \left[
    \vec{G}^+(\omega)  + \vec{G}^-(\omega)
  \right]
 -
  \tilde{\beta}_r 
  \vec{u} \cdot 
    \vec{J}(\omega)
\\
&&
~~~~ ~~~~
 -
  \tilde{\alpha}_{\Sigma}^{-2}
  \tilde{\alpha}_r
  \rho
 ~~
 +
  \frac{\imath \omega}{2}
  \tilde{\alpha}^2
  \tilde{\alpha}_r^{-1} 
  \tilde{\beta}_r 
  \vec{u} \cdot 
  \left[
    \vec{G}^+  + \vec{G}^-
  \right]
 -
  \tilde{\beta}_r 
  \vec{u} \cdot 
    \vec{J}
\\
&=&
 + 
  \imath \omega
  \tilde{\alpha}_{\Sigma}^2
  \tilde{\alpha_r}^{-1} 
  \tilde{\beta}_r 
  \vec{u} \cdot 
  \left[
    \vec{G}^+(\omega)   +   \vec{G}^-(\omega) 
  \right]
 -
  2
  \tilde{\beta}_r 
  \vec{u} \cdot 
    \vec{J}(\omega) 
.
\end{eqnarray}

% --------- 
\subsubsection{Circularly polarized}\label{Sss-definitions-vector-circularpolarization}

Handling circularly polarized light is easy using $\vec{G}^\pm$ fields.
Just as for $E$ and $H$ fields,
 you define
~
\begin{eqnarray}
  \vec{E}
&=&
  \vec{x}
  E_x
  \cos (\omega t - \phi)
 +
  \vec{y}
  E_y
  \cos (\omega t )
,
\\
  \vec{H}
&=&
 -
  \vec{x}
  H_x
  \cos (\omega t)
 +
  \vec{y}
  H_y
  \cos (\omega t - \phi)
\end{eqnarray}

we get
~ 
\begin{eqnarray}
  \vec{E} \pm \vec{z} \times \vec{H}
&=&
  \vec{x}
  \left [ E_x \pm H_y \right]
  \cos (\omega t)
 +
  \vec{y}
  \left [ E_y \pm H_x \right]
  \cos (\omega t - \phi)
\\
  \vec{G}^\pm
&=&
  \vec{x}
  G_x^\pm
  \cos (\omega t - \phi)
 +
  \vec{y}
  G_y^\pm
  \cos (\omega t)
\end{eqnarray}

Similar definitions will hold for choices other than the standard 
 ``primary $\vec{E}$''
 $\vec{G}^\pm$ fields. 

% --------- 
\subsubsection{Comparison with $E$ and $H$}\label{Sss-definitions-vector-cf}

Assume
~
\begin{eqnarray}
  G^+
&=&
  A e^{\imath \left( kz - \omega t \right)},
~~~~ ~~~~
  G^- = 0
\\
  E
&=&
  \frac{A}{2 \alpha_r}
  e^{\imath \left( kz - \omega t \right)},
~~~~ ~~~~
  H
= 
  \frac{A}{2 \beta_r}
  e^{\imath \left( kz - \omega t \right)},
\\
\textrm{so}
~~~~ ~~~~
  E/H
&=&
  \frac{A}{2 \alpha_r}
  .
  \frac{2 \beta_r}{A}
~~~~ ~~~~
=
  \frac{\beta_r}{\alpha_r}
\end{eqnarray}

Compare with the results we get from an $E$ carrier-based approach
~
\begin{eqnarray}
  E
&=&
  B e^{\imath \left( kz - \omega t \right)},
\\
\textrm{to get $H$ we use}
~~~~ ~~~~
  \mu \partial_t H
&=&
  \partial_z E
\\
\textrm{so}
~~~~ ~~~~
  \mu \partial_t H
&=&
  \imath k
 .
  B e^{\imath \left( kz - \omega t \right)},
\\
  H
&=&
  \frac{1}{\mu}
  \frac{ \imath k }
       { \left(-\imath \omega \right)}
 .
  B e^{\imath \left( kz - \omega t \right)},
\\
&=&
 -
  \frac{1}{\beta^2}
  \alpha \beta
 .
  B e^{\imath \left( kz - \omega t \right)},
~~~~ ~~~~
=
 -
  \frac{\alpha}{\beta}
 .
  B e^{\imath \left( kz - \omega t \right)},  
\\
\textrm{so}
~~~~ ~~~~
  E/H
&=&
  B
  .
  \frac{\beta}{\alpha B}
~~~~ ~~~~
=
  \frac{\beta}{\alpha}
\end{eqnarray}

% --------- --------- --------- 
\subsection{Primary $\vec{H}$}\label{Ss-definitions-vectorH}

If we wanted to swap the roles of $\vec{H}$ and $\vec{E}$, 
we might instead define the field variables  as
~
\begin{eqnarray}
  \vec{G}'^{\pm} (\omega)
&=&
  \tilde{\beta}_r (\omega)
  ~ \vec{H}  (\omega)
 \pm
   \vec{u} \times 
  ~ \tilde{\alpha}_r (\omega)
  ~ \vec{E}  (\omega)
,
\label{eqn-S-defs-Gpvector}
\\
  {G}'^{\circ} (\omega)
&=&
   \vec{u} \cdot
  ~ \tilde{\alpha}_r (\omega)
  ~ \vec{E}  (\omega)
,
\label{eqn-S-defs-Gpvector0}
\\
\textrm{so} ~~~~ ~~~~
  \vec{u} \times 
  \tilde{\alpha}_r (\omega) \vec{E} (\omega)
&=&
  \frac{1}{2} 
  \left[ \vec{G}'^+(\omega)   -   \vec{G}'^-(\omega) \right]
\\
\textrm{or} ~~~~ ~~~~
  \tilde{\alpha}_r (\omega) \vec{E} (\omega)
&=&
  \frac{1}{2} 
  \vec{u} 
  \times
  \left[ \vec{G}'^+(\omega)   -   \vec{G}'^-(\omega) \right]
 +
  \vec{u}
  {G}'^{\circ}
\label{eqn-S-defs-EvectorGp}
\\
\textrm{and} ~~~~ ~~~~
  \tilde{\beta}_r (\omega)
  \vec{H} (\omega)
&=& 
  \frac{1}{2} 
  \left[ \vec{G}'^+(\omega)   +   \vec{G}'^-(\omega) \right]
\label{eqn-S-defs-HvectorGp}
.
\end{eqnarray}

We would do this so that the magnetic terms not included in the 
reference medium (i.e. in the definitions of $\vec{G}'^\pm$)
would lose the non-transverse correction terms (see later), although
terms like those would then appear on the electric terms
(and I would define a transverse electric fields (TEF) approximation
instead).  

A derivation of the wave equations using this form is in 
section \ref{S-1storderH}.

%All I have done here is swap the roles of $E$ and $H$, so the 
%first order wave equation for $G'$ could be easily found merely
%by swapping $G \leftrightarrow G'$, $\alpha \leftrightarrow \beta$, and
%so on (but watch for role of current $J$, if it's included).

% --------- 
\subsubsection{Divergence}\label{Sss-definitions-vectorH-divergence}

Here we need to calculate $\vec{B}$ from $\vec{H}$, 
 giving us something like $\vec{B} = \tilde{\beta}_{\Sigma}^{-2} \vec{H}$;
 just as for the Primary $\vec{E}$ form
 where we needed to calculate $\vec{D}$ from $\vec{E}$.
However, 
 rather than repeat an almost identical calculation, 
 I simply note that $\nabla \cdot \vec{B} = 0$ (no magnetic monopoles), 
 which will cause the term to vanish anyway.

The divergence calculation is
~
\begin{eqnarray}
  \nabla \cdot \vec{G}'^\pm (\omega)
&=&
  \nabla \cdot \tilde{\beta}_r \vec{H} (\omega)
 \pm
  \nabla \cdot 
  \left( 
    \vec{u} \times \tilde{\alpha}_r \vec{E} (\omega)
 \right)
\\
&=&
  \tilde{\beta}_{\Sigma}^{-2}
  \tilde{\beta}_r
  \nabla \cdot \vec{B}
 \pm
  \tilde{\alpha}_r
  \nabla \cdot 
  \left( 
    \vec{u} \times  \vec{E}
 \right)
\\
&=&
  0
 \mp
  \tilde{\alpha}_r
  \vec{u}
   \cdot
  \left( 
    \nabla \times  \vec{E}
  \right)
 \pm
  \tilde{\alpha}_r
  \vec{E}
  \left( 
    \nabla \times  \vec{u}
  \right)
\\
&=&
 \mp
  \tilde{\alpha}_r
  \vec{u}
   \cdot
  \left( 
    \imath \omega
    \tilde{\beta}^2 \vec{H}
  \right)
\\
&=&
 \mp
  \tilde{\alpha}_r
  \tilde{\beta}_r^{-1}
  \tilde{\beta}^2
 ~
    \imath \omega
  \left( 
     \vec{u}
     \cdot
     \tilde{\beta_r} \vec{H}
  \right)
\\
&=&
 \mp
  \frac{\imath \omega}{2}
  \tilde{\alpha}_r
  \tilde{\beta}_r^{-1}
  \tilde{\beta}^2
  ~
     \vec{u}
     \cdot
  \left[ 
   \vec{G}^{+} (\omega)   +   \vec{G}^{-} (\omega)
  \right]
.
\end{eqnarray}

So no longitudinal magnetic intensity field $H$ means zero divergence.
Clearly, if $\vec{G}'^{\pm}$ are pure transverse 
 (not just TEF, which is built into the definition, 
 but no longitudinal $H$ fields either), 
 then the divergence is zero.

% --------- --------- --------- 
\subsection{Primary $\vec{D}$}\label{Ss-definitions-vectorD}

If we were more interested in $\vec{D}$ than $\vec{E}$ 
(or $\vec{B}$ than $\vec{H}$), we might instead define
variables  as
~
\begin{eqnarray}
  \vec{F}^{\pm} (\omega)
&=&
  \tilde{\alpha}_r^{-1} (\omega) 
  ~ \vec{D} (\omega)
 \pm 
  \vec{u} \times 
  ~ \tilde{\beta}_r^{-1} (\omega)
  ~ \vec{B} (\omega)
,
\label{eqn-S-defs-Fvector}
\\
  {F}^{\circ} (\omega)
&=&
  \vec{u} \cdot
  ~ \tilde{\beta}_r^{-1} (\omega)
  ~ \vec{B} (\omega)
,
\label{eqn-S-defs-Fvector0}
\\
\textrm{so} ~~~~ ~~~~
  \vec{D} (\omega)
&=&
  \frac{\tilde{\alpha}_r (\omega) }{2} 
  \left[ \vec{F}^+(\omega)   +   \vec{F}^-(\omega) \right]
\label{eqn-S-defs-EvectorF}
\\
\textrm{and} ~~~~ ~~~~
  \vec{u} \times 
  ~ \vec{B} (\omega)
&=& 
  \frac{\tilde{\beta}_r (\omega)}{2} 
  \left[ \vec{F}^+(\omega)   -   \vec{F}^-(\omega) \right]
\\
\textrm{and} ~~~~ ~~~~
  ~ \vec{B} (\omega)
&=& 
  \frac{\tilde{\beta}_r (\omega)}{2} 
  \vec{u} \times 
  \left[ \vec{F}^+(\omega)   -   \vec{F}^-(\omega) \right]
 +
  \tilde{\beta}_r (\omega)
  \vec{u}
  {F}^{\circ} (\omega)
.
\label{eqn-S-defs-BvectorF}
\end{eqnarray}

We would do this so that the displacement field (and magnetic fields) 
could be easily reconstructed from $ \vec{F}^\pm$; and, in analogy
to the THF and TEF mentioned above, we would use a transverse magnetic $B$
field approximation (TBF).

A derivation of the wave equations using this form is in 
section \ref{S-1storderD}.

%Note that I might instead define 
%(with $\tilde{c}_r = 1/ \tilde{\alpha}_r \tilde{\beta})_r$),
%~
%\begin{eqnarray}
%  \tilde{c}_r(\omega) 
%  \vec{F}^{\pm} (\omega)
%&=&
%  \tilde{\alpha}_r^{-1} (\omega)
%  ~ \vec{D} (\omega)
% \pm 
%  \vec{u} \times 
%  ~ \tilde{\beta}_r^{-1} (\omega)
%  ~ \vec{B} (\omega)
%,
%\label{eqn-S-defs-Fvector-alt}
%\\
%  \tilde{c}_r(\omega) 
%  {F}^{\circ} (\omega)
%&=&
%  \vec{u} \cdot
%  ~ \tilde{\beta}_r^{-1} (\omega)
%  ~ \vec{B} (\omega)
%,
%\end{eqnarray}
%if you preferred matching up $\alpha$'s with $D$ and $\beta$'s with $B$.

% --------- 
\subsubsection{Divergence}\label{Sss-definitions-vectorD-divergence}

\begin{eqnarray}
  \nabla \cdot \vec{F}^\pm (\omega)
&=&
  \nabla \cdot \tilde{\alpha}_r^{-1} \vec{D} (\omega)
 ~~
 \pm
  \nabla \cdot \vec{u} \times \tilde{\beta}_r^{-1} \vec{B} (\omega)
\\
&=&
  \tilde{\alpha}_r^{-1}
  \nabla \cdot \vec{D}
 ~~
 \pm
  \tilde{\beta}_r^{-1}
  \vec{B} \cdot 
  \left(
    \nabla \times \vec{u} 
  \right)
 ~~
 \mp
  \tilde{\beta}_r^{-1}
  \vec{u} \cdot 
  \left(
    \nabla \times \vec{B} 
  \right)
\\
&=&
  \tilde{\alpha}_r^{-1}
  \rho
 ~~
 \pm
  0
 ~~
 \mp
  \tilde{\beta}_r^{-1}
  \tilde{\beta}^2
  \vec{u} \cdot 
  \left(
    \nabla \times \vec{H} 
  \right)
\\
&=&
  \tilde{\alpha}_r^{-1}
  \rho
 ~~
 \mp
  \tilde{\beta}_r^{-1}
  \tilde{\beta}^2
  \vec{u} \cdot 
  \left(
    -
    \imath \omega 
    \vec{D}
    +
    \vec{J}
  \right)
\\
&=&
  \tilde{\alpha}_r^{-1}
  \rho
 ~~
 \pm
  \frac{\imath \omega}{2} 
  \tilde{\alpha}_r
  \tilde{\beta}_r^{-1}
  \tilde{\beta}^2
  \vec{u} \cdot 
  \left[
    \vec{F}^+ (\omega)   +   \vec{F}^- (\omega)
  \right]
  ~~
 \mp
  \tilde{\beta}_r^{-1}
  \tilde{\beta}^2
    \vec{J} (\omega)
.
\end{eqnarray}

Thus I never use the (no monopoles condition) $\nabla \cdot \vec{B} = 0$.
Clearly, if $\vec{F}^{\pm}$ are pure transverse 
 (not just TBF, which is built into the definition, 
 but no longitudinal $D$ fields either), 
 then the divergence is depends only on the charge density
~
\begin{eqnarray}
  \nabla \cdot \vec{F}^\pm
&=&
  \tilde{\alpha}_r^{-1}
  \rho
\end{eqnarray}

NB:
~
\begin{eqnarray}
  \nabla \cdot \vec{F}^+
 -
  \nabla \cdot \vec{F}^-
&=&
  \imath \omega
  \tilde{\alpha}_r
  \tilde{\beta}_r^{-1}
  \tilde{\beta}^2
  \vec{u} \cdot 
  \left[
    \vec{F}^+   +   \vec{F}^-
  \right]
  ~~
  -
  2
  \tilde{\beta}_r^{-1}
  \tilde{\beta}^2
    \vec{J}
\end{eqnarray}

% --------- --------- --------- 
\subsection{Primary $\vec{B}$}\label{Ss-definitions-vectorB}

If we wanted to swap the roles of $\vec{B}$ and $\vec{D}$, 
we might instead define
variables  as
~
\begin{eqnarray}
  \vec{F}'^{\pm} (\omega)
&=&
  \tilde{\beta}_r^{-1} (\omega)
  ~ \vec{B} (\omega)
 \pm 
  \vec{u} \times   
  ~\tilde{\alpha}_r^{-1} (\omega)
  ~\vec{D} (\omega)
,
\label{eqn-S-defs-Fpvector}
\\
  {F}'^{\circ} (\omega)
&=&
  \vec{u} \cdot
  ~\tilde{\alpha}_r^{-1} (\omega)
  ~\vec{D} (\omega)
,
\label{eqn-S-defs-Fpvector0}
\\
\textrm{so} ~~~~ ~~~~
  \vec{u} \times 
  ~ \vec{D} (\omega)
&=&
  \frac{\tilde{\alpha}_r (\omega)}{2} 
  \left[ \vec{F}'^+(\omega)   -   \vec{F}^-(\omega) \right]
\\
\textrm{or} ~~~~ ~~~~
  \vec{D} (\omega)
&=&
  \frac{\tilde{\alpha}_r (\omega)}{2} 
  \vec{u} \times 
  \left[ \vec{F}'^+(\omega)   -   \vec{F}^-(\omega) \right]
 +
  \tilde{\alpha}_r (\omega)
  \vec{u}
  {F}'^{\circ} (\omega)
\label{eqn-S-defs-DvectorFp}
\\
\textrm{and} ~~~~ ~~~~
  \vec{B} (\omega)
&=& 
  \frac{\tilde{\beta}_r (\omega)}{2} 
  \left[ \vec{F}'^+(\omega)   +   \vec{F}'^-(\omega) \right]
\label{eqn-S-defs-BvectorFp}
.
\end{eqnarray}

In analogy
to the THF and TEF mentioned above, we would use a transverse displacement
field approximation (TDF).

A derivation of the wave equations using this form is (not yet) in 
section \ref{S-1storderB}.

% --------- 
\subsubsection{Divergence}\label{Sss-definitions-vectorB-divergence}

\begin{eqnarray}
  \nabla \cdot \vec{F}'^\pm (\omega)
&=&
  \nabla \cdot \tilde{\beta}_r^{-1} \vec{B} (\omega)
 \pm
  \nabla \cdot 
  \left( 
    \vec{u} \times \tilde{\alpha}_r^{-1} \vec{D} (\omega)
 \right)
\\
&=&
  0
 \mp
  \tilde{\alpha}^2
  \tilde{\alpha}_r^{-1}
  \vec{u}
   \cdot
  \left( 
    \nabla \times  \vec{E}
  \right)
 \pm
  \tilde{\alpha}^2
  \tilde{\alpha}_r^{-1}
  \vec{E}
  \left( 
    \nabla \times  \vec{u}
  \right)
\\
&=&
 \mp
  \tilde{\alpha}^2
  \tilde{\alpha}_r^{-1}
  \vec{u}
   \cdot
  \left( 
    \imath \omega
    \vec{B}
  \right)
\\
&=&
 \mp
    \imath \omega
  \tilde{\alpha}^2
  \tilde{\alpha}_r^{-1}
  \tilde{\beta}_r
  \left( 
    \vec{u}
    \cdot
    \tilde{\beta}_r^{-1}
    \vec{B}
  \right)
\\
&=&
 \mp
  \frac{\imath \omega}{2}
  \tilde{\alpha}^2
  \tilde{\alpha}_r^{-1}
  \tilde{\beta}_r
  ~
    \vec{u}
    \cdot
  \left[
    \vec{F}^{+} (\omega)   +   \vec{F}^{-} (\omega)
  \right]
.
\end{eqnarray}

Clearly, if $\vec{F}'^{\pm}$ are pure transverse 
 (not just TDF, which is built into the definition, 
 but no longitudinal $B$ fields either), 
 then the divergence is zero.

% --------  

\subsection{Comparison to Kolesik et.al}\label{Ss-altfleck-vsKolesik}

(needs updating with alteration of $F$, $F'$ definitions at 20050727/8)

Now let us compare the equations Kolesik equations for $D$ with those 
 for the $\vec{G}^\pm$ fields.  
They have
~
\begin{eqnarray}
  \vec{D}(\vec{k}))^{\pm} 
&=&
  \frac{1}{2}
  \left[
    \vec{D}(\vec{k}))
   \mp
      \textrm{sgn}(kz) \frac{k}{\omega(k)}
     \vec{u} 
      \times \vec{H} (\vec{k})
  \right]
\\
\Longrightarrow
~~~~ ~~~~
  \frac{2}{\sqrt{\mu}}
  \left(
    \sqrt{\mu}
    \vec{D}^{\pm}
  \right)
&=&
  \vec{D}
 \mp
  \frac{1}{c}
  \vec{u} 
  \times
  \vec{H}
\\
  2
  \left(
    \sqrt{\mu}
    \vec{D}^{\pm}
  \right)
&=&
  \sqrt{\mu}
  \vec{D}
 \mp
  \mu \sqrt{\epsilon}
  \vec{u} 
  \times
  \vec{H}
\\
&=&
  \sqrt{\mu}
  \vec{D}
 \mp
  \sqrt{\epsilon}
  \vec{u} 
  \times
  \left( \mu \vec{H} \right)
\\
&=&
  \sqrt{\mu}
  \vec{D}
 \mp
  \sqrt{\epsilon}
  \vec{u} 
  \times
  \vec{B}
\end{eqnarray}

Barring my dropping of the $\textrm{sgn}(kz)$ factor,
 their $D^\pm$ can clearly be identified with my ``primary $D$''
 formulation of directional fields (see \ref{Ss-definitions-vectorD}).
Of course, there are differences, 
 notably that their dispersion is only applied to the magnetic field
(either $H$ or $B$); 
 my ``primary $E$'' formulation applied $\epsilon$ dispersion to $E$ only, 
 and $\mu$ dispersion to $H$ only.

They also present an equation for a directional $H$ which
 corresponds to the ``primary $B$'' form ($F'$) as opposed
to the $D$ ($F$) one.

% ---------  --------- ---------
\subsection{Moving Frame}\label{Ss-movingframe}

It is often useful to transform 
 the following wave equation derivations into a moving frame.
The frame will be defined by the speed $c_f = 1/ \alpha_f \beta_f$ 
 in analogy to the reference and correction parameters above.
A sensible choice of $c_f$ is probably helpful, 
 e.g. $c_f = c_r$ or possibly $c_f = v_g$, the pulse group velocity.

The (non relativistic) $z$-direction frame translation itself is 
~
\begin{eqnarray}
  t'
&=& 
  t - z / c_f
\\
  z' &=& z
\end{eqnarray}

so that 
~
\begin{eqnarray}
  \partial_{t'}
&=& 
  \partial_t
\\
  \partial_{z'}
&=&
  -c_f^{-1} \partial_t + \partial_z
\label{eqn-z-frametranslation}
\end{eqnarray}

When doing vector calculations, I will need a vector form of the frame 
translation.  I use 
~
\begin{eqnarray}
  \nabla \times \vec{Q}
&=&
  \nabla' \times \vec{Q}
 - 
  \alpha_f \beta_f \vec{u} \times \partial_t Q
\label{eqn-vector-frametranslation}
\end{eqnarray}

Assuming $\vec{u}$ is along the $z$-direction --
~
\begin{eqnarray}
  \nabla \times \vec{Q}
&=&
  \hat{x}  
  \left( \partial_y Q_z - \partial_z Q_y \right)
 +
  \hat{y}  
  \left( \partial_z Q_x - \partial_x Q_z \right)
 +
  \hat{z}  
  \left( \partial_x Q_y - \partial_y Q_x \right)
\\
&=&
  \hat{x} 
  \left( \partial_y Q_z - \partial_{z'} Q_y - \alpha_f \beta_f  \partial_t Q_y \right)
 +
  \hat{y} 
  \left( \partial_{z'} Q_x - \partial_x Q_z + \alpha_f \beta_f  \partial_t Q_x \right)
 +
  \hat{z} 
  \left( \partial_x Q_y - \partial_y Q_x \right)
\\
&=&
  \hat{x} 
  \left( \partial_y Q_z - \partial_{z'} Q_y \right)
 +
  \hat{y} 
  \left( \partial_{z'} Q_x - \partial_x Q_z \right)
 +
  \hat{z} 
  \left( \partial_x Q_y - \partial_y Q_x \right)
 - 
  \hat{x} \alpha_f \beta_f  \partial_t Q_y 
 + 
  \hat{y} \alpha_f \beta_f  \partial_t Q_x
\\
&=&
  \nabla' \times \vec{Q}
 - 
  \hat{x} \alpha_f \beta_f  \partial_t Q_y 
 + 
  \hat{y} \alpha_f \beta_f  \partial_t Q_x
\end{eqnarray}

I will often find it useful to define the ratio of the reference and
 frame speeds: 
~
\begin{eqnarray}
\xi &=& \alpha_f \beta_f / \alpha_r \beta_r
\end{eqnarray}

This moving frame is very useful in a space-propagated model since 
 the pulse is held as a function of time, 
 and so it will stay (nearly) centered while propagating forward.  
Although it has no sensible limit as the frame speed tends to zero, 
 we can recover that case by setting $\alpha_f \beta_f=0$ (or $\xi=0$) 
 and rewriting the primed variables as unprimed ones.

% ---------  --------- ---------
\subsubsection{Frequency domain}\label{Ss-movingframe-w}

In the frequency domain, the space derivative transforms like 
~
\begin{eqnarray}
  \partial_{z'}
&=& 
 ~ 
  \imath \omega \tilde{\alpha}_f \tilde{\beta}_f
 + 
  \partial_z
\end{eqnarray}

which seems to leave open the intriguing possibility of a {\em frequency}
 dependent (``dispersive'') moving frame by allowing 
 $\tilde{\alpha}_f$ and $\tilde{\beta}_f$ a frequency dependence.
I could put {\em all} the linear evolution in the reference,
 and match it to the frame 
 (so that $\alpha_f=\alpha_r$ and $\beta_f=\beta_r$ and $\xi=1$).
This means that the only alterations to the pulse profile would be those 
 due to nonlinear effects.

We might hope to use a dispersive frame in the following way.  
Imagine a pulse being propagated in a strongly dispersive medium with 
 some nonlinearity;
 and that the pulse might be expected to broaden by 
 a large factor of (e.g.) twenty during propagation.  
This would mean we would need a much wider time window in our simulation 
 than suggested by the initial pulse width.
In a dispersive frame matched to a dispersive reference medium, 
 our simulated pulse would not broaden {\em at all} due to dispersion, 
 so we then hope retain a narrow time window throughout the simulation,
 hence saving computer time.  
The drawback is that our nonlinear term absorbs the reference dispersion, 
 and this may cause significant pulse broadening of itself, 
 perhaps of similar extent to the dispersion would.  
Assuming it did give an advantage, 
 at the end of the simulation we would have to take our 
 dispersive-frame pulse and convert back to a non-dispersive frame 
 to see the ``lab'' pulse profile.
Fortunately the computational cost of this conversion is minimal.

% --------- --------- --------- 
\subsection{Pseudo-Beltrami forms}\label{Ss-altfleck-pseudobeltrami}

%% --------- --------- ---------
%\subsection{Miscellaneous: Weiglehofer and Lakhtakia ``Beltrami'' style definitions}
%\label{Ss-firstorder-weilah}

Weiglhofer and Lakhtakia \cite{Weiglhofer-L-1994pre} present a 
 number of quantities related to the EM field using their 
 Beltrami style definitions.  
If I rearrange the factors of $\mu$'s and $\epsilon$, 
 I might rewrite these as
~
\begin{eqnarray}
  \textrm{free field:} 
  ~~~~ ~~~~
  \epsilon^{1/2}
  Q_\pm 
&=&
  \epsilon^{1/2} E 
 \pm
  \imath \mu^{1/2} H
\\
  \textrm{coupled field:} 
  ~~~~ ~~~~
  \epsilon^{1/2}
  F_\pm
&=&
  \epsilon^{-1/2} D 
 \pm
  \imath
  \mu^{-1/2} B 
\\
  \textrm{currents:} 
  ~~~~ ~~~~
  \mu^{-1/2}
  W_\pm
&=&
 -
  \mu^{-1/2}
  J_m 
 \pm 
  \imath 
  \epsilon^{-1/2} 
  J_e
\\
  \textrm{charges:} 
  ~~~~ ~~~~
  \mu^{-1/2}
  w_\pm
&=&
 -
  \mu^{-1/2}
  \rho_m 
 \pm 
  \imath 
  \epsilon^{-1/2} 
  \rho_e
\end{eqnarray}

Analogously, and since we use $\vec{u} \times$ the curl-$H$ equation 
(which has $D$, $J_e$ on the RHS) --
~
\begin{eqnarray}
  \textrm{free field:} 
  ~~~~ ~~~~
  \vec{G}_\pm 
&=&
  \epsilon^{1/2} 
  \vec{E}
 \pm
  \mu^{1/2} 
    \vec{u} 
   \times
    \vec{H}
\\
  \textrm{coupled field:} 
  ~~~~ ~~~~
  \vec{F}_\pm
&=&
  \epsilon^{-1/2} 
    \vec{u} 
   \times
    \vec{D}
 \pm
  \mu^{-1/2} 
  \vec{B}
\\
  \textrm{currents:} 
  ~~~~ ~~~~
  \vec{K}_\pm
&=&
  \mu^{1/2} 
    \vec{u} 
   \times
    \vec{J}_e
 \pm 
  \epsilon^{1/2} 
    \vec{J}_m
\\
  \textrm{charges:} 
  ~~~~ ~~~~
  L_\pm
&=&
  \epsilon^{-1/2} 
    \rho_e
 \pm 
  \mu^{-1/2} 
  \rho_m 
 ~~~~ ~~~~
 ??
\end{eqnarray}

It might also be interesting to  construct Beltrami-like 
 combinations of the directional $G^\pm$ fields, 
 i.e.
~
\begin{eqnarray}
  \vec{Q}'^{\pm} 
&=&
  \vec{G}^\pm + \imath \vec{G}^\mp
.
\end{eqnarray}

% --------- --------- ---------
\subsection{Miscellaneous: Force on a moving charge}
\label{Ss-firstorder-movingcharge}

If I define a velocity $\vec{v} = v \vec{u}$, there is a curious
similarity between the force on a moving charge and the construction
of my $\vec{G}^+$ field variable.  We have --
~
\begin{eqnarray}
  \frac{1}{q}
  \vec{F}
&=&
  \vec{E} 
 +
  \vec{v}
  \times
 \vec{B}
\\
  \frac{\epsilon^{1/2}}{q}
  \vec{F}
&=&
  \epsilon^{1/2}
  \vec{E} 
 +
  v
  \epsilon^{1/2}
  \vec{u}
  \times
    \mu
    \vec{H}
\\
&=&
  \epsilon^{1/2}
  \vec{E} 
 +
  v
  \epsilon^{1/2}
  \mu^{1/2}
  \vec{u}
  \times
    \mu^{1/2}
    \vec{H}
\\
&=&
  \epsilon^{1/2}
  \vec{E} 
 +
  \frac{v}{c}
  \vec{u}
  \times
    \mu^{1/2}
    \vec{H}
\end{eqnarray}

So if $v=c$,
~
\begin{eqnarray}
  \frac{\epsilon^{1/2}}{q}
  \vec{F}
&=&
  \epsilon^{1/2}
  \vec{E} 
 +
  \vec{u}
  \times
    \mu^{1/2}
    \vec{H}
\\
&=&
  \vec{G}^+ 
\end{eqnarray}

This seems to be a remarkable coincidence!

\newpage
% ----------------------------------------------------------------------

\section{First order evolution equations: Primary $\vec{E}$}\label{S-1storderE}

Here I derive coupled wave equations for the $\vec{G}^\pm$ fields
 defined in \ref{Ss-definitions-vector}.
I also discuss how things look in various frames of reference, 
 and in subsection \ref{Ss-1storderE-gstationaryvector} generate 
 first-order differential equations for 
 forward and backward propagating fields.  
In the following I do not introduce any envelope-carrier decomposition 
 for the field variables; 
 although the introduction of a moving frame plays a related role.  
Note that the Vector Stationary Frame  
 subsection (\ref{Ss-1storderE-gstationaryvector}) also contains 
 the plane-polarized special case, 
 as well as a transformation to the moving frame as well; 
 and should be considered to contain the authoritative results.  
Note there are still some sign discrepancies between 
 the vector calculations and the plane polarized ones.

%It is possible that in some sense there is a subtle 
% correspondance between the two (``frame'' or ``carrier'') approaches.

% --------- --------- ---------

\subsection{Plane-polarized stationary frame (Example only)}\label{Ss-1storderE-gstationary}

Note that is better to follow the (authoritative) vector stationary
frame derviation (see \ref{Ss-1storderE-gstationaryvector}) instead.
However, his is a simpler derivation that gives a nice 
overview of the approach.  It is superceeded by the full vectorized
(and moving-frame) derivation(s) below.
I begin with the source-free Maxwell's equations --
~
\begin{eqnarray}
\frac{\partial H_y (t)}{\partial z}
&=&
  -\frac{\partial }{\partial t} \epsilon \ast E_x (t)
,
\\
\frac{\partial E_x (t)}{\partial z}
&=&
  -\frac{\partial }{\partial t}  \mu \ast H_y (t)
\end{eqnarray}

Because the convolutions $\epsilon \ast E_x$ and $\mu \ast  H_y$ 
 complicate the analysis, 
 I will fourier transform into the frequency domain where their linear parts 
 become simple products. 
Using the definitions in subsection \ref{S-medium}, 
 and carelessly dropping the tildes off the $\alpha$ and $\beta$ parameters, 
 I can proceed with the calculation --
~
\begin{eqnarray}
  \frac{\partial }{\partial z}
  \tilde{H}_y
&=&
  \imath \omega
    \tilde{\epsilon} \tilde{E_x}
~~~~ ~~~~
=
  \imath \omega
    \left( \alpha_r^2 \tilde{E_x} + \alpha_r \alpha_c * \tilde{E_x} \right) 
,
\\
  \frac{\partial }{\partial z}
  \tilde{E_x}
&=&
  \imath \omega
    \tilde{\mu}  \tilde{H}_y
~~~~ ~~~~
=
  \imath \omega
    \left( \beta_r^2 \tilde{H}_y + \beta_r \beta_c * \tilde{H}_y\right)^2  
\\
\Longrightarrow ~~~~ ~~~~
  \frac{\partial }{\partial z}
  \beta_r \tilde{H}_y
&=&
  \imath \omega \alpha_r \beta_r 
  \left[
    \alpha_r \tilde{E_x}
  \right]
 +
  \imath \omega
  \beta_r \alpha_c * 
  \left[
    \alpha_r \tilde{E_x}
  \right]
\label{eqn-defs-noframes-dbH}
,
\\
  \frac{\partial }{\partial z}
  \alpha_r  \tilde{E_x}
&=&
  \imath \omega \alpha_r  \beta_r 
  \left[
    \beta_r  \tilde{H}_y
  \right]
 +
  \imath \omega
    \alpha_r  \beta_c *
  \left[
    \beta_r  \tilde{H}_y
  \right]
\label{eqn-defs-noframes-daE}
\end{eqnarray}

Adding and subtracting eqns. (\ref{eqn-defs-noframes-dbH}, 
\ref{eqn-defs-noframes-daE}), and using the fourier transform of
the definitions for $G^\pm = \alpha_r E_x \pm \beta_r H_y$, we get 
~
\begin{eqnarray}
  \frac{\partial }{\partial z'}
  \tilde{G}^\pm
&=&
 \pm
  \imath \omega \alpha_r  \beta_r \tilde{G}^\pm
 ~~
 +
  \imath \omega
    \alpha_r  \beta_c * \beta_r \tilde{H}_y
 ~~
 \pm
  \imath \omega
  \beta_r \alpha_c * \alpha_r \tilde{E_x}
\\
&=&
 \pm
  \imath \omega \alpha_r  \beta_r \tilde{G}^\pm
 ~~
 +
  \imath \omega
    \frac{\alpha_r }
         {2}
    \beta_c
   *
    \left( \tilde{G}^+ - \tilde{G}^- \right)
 ~~
 \pm
  \imath \omega
    \frac{\beta_r}
         {2}
   \alpha_c
   *
    \left( \tilde{G}^+ + \tilde{G}^- \right)
\label{eqn-stationaryGpropagation}
\end{eqnarray}

The ``non-magnetic'' limit (strictly, no magnetic corrections to the frame), 
where $\beta_c=0$ is
~
\begin{eqnarray}
  \frac{\partial }{\partial z'}
  \tilde{G}^\pm
&=&
 \pm
  \imath \omega \alpha_r \beta_r 
  \tilde{G}^\pm
 \pm
  \imath \omega  
    \frac{\beta_r }
         {2}
   \alpha_c
   *
  \left( \tilde{G}^+ + \tilde{G}^- \right)
\end{eqnarray}

The reference-only limit of the above equations (with $\alpha_1=0; \beta_1=0$) 
is exactly as you would expect (see eqn.(\ref{eqn-defs-frames-vGplusminus})) --
~
\begin{eqnarray}
  \frac{\partial }{\partial z'}
  \tilde{G}^\pm
&=&
 \pm
  \imath \omega \alpha_r  \beta_r G^\pm
\label{eqn-defs-frames-vGplusminus}
\end{eqnarray}

Clearly, even if $\alpha_c \ll \alpha_r$, both $G^+$ and $G^-$ evolve
on similar scales, as specified by the leading RHS term.  It is 
interesting to compare Kolesik et.al.'s eqn(KMM~7) 
(i.e. (\ref{eqn-Kolesik-7})) 
with my eqn (\ref{eqn-stationaryGpropagation}) above.  Of course their
equation has the $\tilde{G}^\mp$ terms projected out, but otherwise
the two are similar in form.

% --------- --------- ---------

\subsection{Vector stationary frame (Authoritative) }\label{Ss-1storderE-gstationaryvector}

Here I avoid calculations using the $x, y$, and $z$ components of the 
field, and retain a fully vector description.
Using the definitions in subsection \ref{S-medium}, I can 
proceed with the calculation --
~
\begin{eqnarray}
\nabla \times \vec{H} (t)
&=&
  +\partial_t \epsilon \ast \vec{E} (t)
  + \vec{J}
,
\nonumber
\\
\nabla \times \vec{E} (t)
&=&
  -\partial_t  \mu \ast \vec{H} (t)
\\
~
~
\textrm{into temporal-frequency space}
\longrightarrow
~~~~ ~~~~
\nabla \times \vec{H} (\omega)
&=&
  - \imath \omega 
    ~ \alpha (\omega) ^2  
    * \vec{E} (\omega)
  + \vec{J} (\omega)
,
\nonumber
\\
\nabla \times \vec{E} (\omega)
&=&
  + \imath \omega  
    ~ \beta (\omega)^2  
    * \vec{H} (\omega)
\\
~
~
\textrm{swap transverse components}
\longrightarrow
~~~~ ~~~~
\vec{u} \times \left( \nabla \times \vec{H} \right)
&=&
  - \imath \omega   
    ~\alpha^2 
    *
    \left(
      \vec{u} ~ \times \vec{E}
    \right)
  + \vec{u} \times \vec{J}
,
\nonumber
\\
\nabla \times \vec{E}
&=&
   + \imath \omega   
    ~\beta^2  
    * 
    \vec{H}
\\
~
~
\textrm{premultiply}
\longrightarrow
~~~~ ~~~~
\vec{u} \times \left( \nabla \times \beta_r \vec{H} \right)
&=&
  - \imath \omega   
    ~\beta_r \alpha^2   
    *
    \left(
      \vec{u} ~ \times \vec{E}
    \right)
  + \vec{u} \times  \beta_r \vec{J}
,
\nonumber
\\
\nabla \times \alpha_r \vec{E}
&=&
   + \imath \omega   
    ~\alpha_r 
    \beta^2   
    * 
    \vec{H}
\\
~
~
\textrm{sum-and-difference}
\longrightarrow
~~~~ ~~~~
  \nabla \times \alpha_r \vec{E}
 ~~
 \pm
 ~~
  \vec{u} \times \left( \nabla \times \beta_r \vec{H} \right)
&=&
 +
  \imath \omega   
    ~\alpha_r 
    \beta^2
    *
    \vec{H}
 ~~
 \mp 
 ~~
  \imath \omega   
    ~\beta_r 
    \alpha^2   
    *
    \left(
      \vec{u} ~ \times \vec{E}
    \right)
 ~~
\nonumber
\\
&& 
~~~~ 
~~~~
 \pm 
 ~~
  \vec{u} \times  \beta_r \vec{J}
,
\end{eqnarray}

The vector $\vec{G}^\pm$ fields 
 (defined in eqn (\ref{eqn-S-defs-Gvector}))
are ~
\begin{eqnarray}
  \vec{G}^{\pm} 
&=&
  \alpha_r \vec{E} + \vec{u} \times \beta_r \vec{H}
\end{eqnarray}

This means I need to convert both 
the second term on the LHS of the sum-and-difference equation above, 
as well as the RHS.  
It is most imortant for the LHS to be simple, 
because this will define the type of propagation specified by the RHS.  
To convert the LHS, we need two vector identities,
as used in subsection \ref{S-vectorid} to get 
eqn.(\ref{eqn-vectorid-combined}). So using
~
\begin{eqnarray}
  \vec{u} \times \left( \nabla \times \vec{H} \right)
 -
  \nabla \left( \vec{u} \cdot \vec{H} \right)
&=&
  \nabla \times \left( \vec{u} \times \vec{H} \right)
\end{eqnarray}

If I retain the $\vec{u} \cdot \beta_r \vec{H}$ term , which 
will be zero for transverse $H$ fields (THF), then --
~
\begin{eqnarray}
  \nabla \times \alpha_r \vec{E}
 ~~
 \pm
 ~~
  \vec{u} \times \left( \nabla \times \beta_r \vec{H} \right)
&=&
 +
  \imath \omega \alpha_r \beta^2 
  * \vec{H}
 ~~
 \mp 
 ~~
  \imath \omega \beta_r \alpha^2 
  * \left( \vec{u} \times \vec{E} \right)
 ~~
 \pm 
 ~~
  \vec{u} \times  \beta_r \vec{J}
,
\\
  \nabla \times \alpha_r \vec{E}
 ~~
 \pm
 ~~
  \nabla \times \left( \vec{u} \times \beta_r \vec{H} \right)
 \pm
  \nabla \left( \vec{u} \cdot \beta_r \vec{H} \right)
&=&
 +
  \imath \omega \alpha_r \beta^2 
  * \vec{H}
 ~~
 \mp 
 ~~
  \imath \omega \beta_r \alpha^2 
  * \left( \vec{u} \times \vec{E} \right)
 ~~
 \pm 
 ~~
  \vec{u} \times  \beta_r \vec{J}
\\
  \nabla \times 
  \left[
    \alpha_r \vec{E}
    ~~
    \pm
    ~~
    \left( \vec{u} \times \beta_r \vec{H} \right)
  \right]
&=&
 +
  \imath \omega \alpha_r \beta^2 
  * \vec{H}
 ~~
 \mp 
 ~~
  \imath \omega \beta_r \alpha^2 
  * \left( \vec{u} \times \vec{E} \right)
 \mp
  \nabla \left( \vec{u} \cdot \beta_r \vec{H} \right)
 ~~
 \pm 
 ~~
  \vec{u} \times  \beta_r \vec{J}
~~~~ ~~~~
\\
  \nabla \times \vec{G}^{\pm}
&=&
  \imath \omega 
  \left\{
    \alpha_r \beta^2 
    * \vec{H}
   ~~
   \mp 
   ~~
    \beta_r \alpha^2 
    * \left( \vec{u} \times \vec{E} \right)
  \right\}
 \mp
  \nabla \left( \vec{u} \cdot \beta_r \vec{H} \right)
 ~~
 \pm 
 ~~
  \vec{u} \times  \beta_r \vec{J}
\end{eqnarray}

Now note that eqn.(\ref{eqn-vectorid-doublecross-u}) means that
~
\begin{eqnarray}
     \vec{u} \times 
     \left[
       \vec{u} \times \vec{H}
     \right]
&=& 
  \left[ \vec{u} \cdot \vec{H} \right] \vec{u} 
 -
  \vec{H} 
\\
\textrm{and so} 
~~~~ ~~~~
  \vec{u} \times \vec{G}^{\pm}
&=&
  \vec{u} \times \alpha_r \vec{E}
 \pm
  \vec{u} \times 
    \left[ 
      \vec{u} \times \beta_r \vec{H} 
    \right]
\\
&=&
  \vec{u} \times \alpha_r \vec{E}
 \mp
  \beta_r \vec{H} 
 ~~
 \pm
  \left[ \vec{u} \cdot \beta_r \vec{H} \right] \vec{u} 
\end{eqnarray}

So,
~
\begin{eqnarray}
  \nabla \times \vec{G}^{\pm}
&=&
  \imath \omega 
  \left\{
    \alpha_r \beta^2 
    *
    \left[
      \vec{u} \cdot \vec{H}
    \right]
    \vec{u}
   ~~
   -
    \alpha_r \beta^2 
    *
    \left(
      \vec{u} \times \left[ \vec{u} \times \vec{H} \right]
    \right)
   ~~
   \mp 
    \beta_r \alpha^2 
    *
    \left(
      \vec{u} \times \vec{E} 
    \right)
  \right\}
 \mp
  \nabla \left( \vec{u} \cdot \beta_r \vec{H} \right)
 ~~
 \pm 
 ~~
  \vec{u} \times  \beta_r \vec{J}
\\
&=&
 \mp 
  \imath \omega 
  \left\{
    \beta_r \alpha^2 
    *
    \left(
      \vec{u} \times \vec{E}
    \right)
   \pm
    \alpha_r \beta^2 
    *
    \left(
      \vec{u} \times \left[ \vec{u} \times \vec{H} \right]
    \right)
  \right\}
 ~~
 +
  \imath \omega 
    \alpha_r \beta^2 
    *
    \left[
      \vec{u} \cdot \vec{H}
    \right]
    \vec{u}
 ~~
 \mp
  \nabla \left( \vec{u} \cdot \beta_r \vec{H} \right)
 ~~
 \pm 
 ~~
  \vec{u} \times  \beta_r \vec{J}
.
~~~~
~~~~
\end{eqnarray}

I now separate the interaction parts (depending on $\alpha_c$, $\beta_c$) 
from the vacuum-like parts (depending on $\alpha_r$, $\beta_r$), and then
substitute (as far as possible) expressions containing $G$ rather than $E$
or $H$, 
by referring to eqns (\ref{eqn-S-defs-Evector}) 
and (\ref{eqn-S-defs-Hvector}).  
Hence 
~
\begin{eqnarray}
  \nabla \times \vec{G}^{\pm}
&=&
 \mp 
  \imath \omega 
  \alpha_r \beta_r 
  \left\{
    \vec{u} \times \alpha_r \vec{E}
   \pm
    \vec{u} \times \left[ \vec{u} \times \beta_r \vec{H} \right]
  \right\}
 ~~
 +
  \imath \omega 
    \alpha_r \beta_r
    \left[
      \vec{u} \cdot \beta_r \vec{H}
    \right]
    \vec{u}
 ~~
 \mp
  \nabla \left( \vec{u} \cdot \beta_r \vec{H} \right)
\nonumber 
\\
&&
 ~~~~
 \mp 
  \imath \omega 
  \beta_r \alpha_c 
  *
  \left(
    \vec{u} \times \alpha_r \vec{E}
  \right)
 -
  \imath \omega 
  \alpha_r \beta_c
  *
  \left(
    \vec{u} \times \left[ \vec{u} \times \beta_r \vec{H} \right]
  \right)
 ~~
 +
  \imath \omega 
    \alpha_r \beta_c
    *
    \left[
      \vec{u} \cdot \beta_r \vec{H}
    \right]
    \vec{u}
 ~~
 \pm 
 ~~
  \vec{u} \times  \beta_r \vec{J}
\label{eqn-1stfleck-vectorstat}
\\
&=&
 \mp 
  \imath \omega 
  \alpha_r \beta_r 
  ~
  \vec{u} \times \vec{G}^\pm
 ~~
 +
  \imath \omega 
    \alpha_r \beta_r
    \vec{u} ~ {G}^{\circ}
 ~~
 \mp
  \nabla {G}^{\circ}
\nonumber 
\\
&&
 ~~~~
 \mp 
  \frac{\imath \omega \beta_r}
       {2}
  \alpha_c 
  *
  \left(
    \vec{u} \times 
    \left[ \vec{G}^+ + \vec{G}^- \right]
  \right)
 ~~
 -
  \frac{\imath \omega \alpha_r}
       {2} 
  \beta_c
  *
  \left(
    \vec{u} \times \left[ \vec{G}^+ - \vec{G}^- \right]
  \right)
 ~~
 +
  \imath \omega 
    \alpha_r \beta_c
  *
  \left(
    \vec{u} ~ {G}^{\circ} 
  \right)
 ~~
 \pm 
 ~~
  \vec{u} \times  \beta_r \vec{J}
.
~~~~
~~~~
\label{eqn-1storderE-statframe}
\end{eqnarray}

Here the transverse (THF) and longitudinal parts decouple, 
 since $\vec{G}^\pm$ is guaranteed magnetically transverse, 
 and $\vec{u} {G}'^{\circ}$ is magnetically longitudinal.
The two decoupled equations are --
~
\begin{eqnarray}
  \nabla \times \vec{G}^{\pm}
&=&
 \mp 
  \imath \omega 
  \alpha_r \beta_r 
  ~
  \vec{u} \times \vec{G}^\pm
  ~~
 \mp 
  \frac{\imath \omega \beta_r}
       {2}
  \alpha_c 
  *
  \left(
    \vec{u} \times 
    \left[ \vec{G}^+ + \vec{G}^- \right]
  \right)
 ~~
 -
  \frac{\imath \omega \alpha_r}
       {2}
  \beta_c
  *
  \left(
    \vec{u} \times \left[ \vec{G}^+ - \vec{G}^- \right]
  \right)
 ~~
 \pm 
 ~~
  \vec{u} \times  \beta_r \vec{J}
~~~~ ~~~~
~~~~ ~~~~
\label{eqn-1storderE-statframe-transverse}
\\
 \pm
  \nabla {G}^{\circ}
&=&
 +
  \imath \omega 
    \alpha_r \beta_r
    \vec{u} ~ {G}^{\circ}
 ~~
 +
  \imath \omega 
    \alpha_r \beta_c
   *
   \left(
    \vec{u} ~ {G}^{\circ} 
   \right)
.
\label{eqn-1storderE-statframe-longitudinal}
\end{eqnarray}

We see from the following \ref{Sss-1storderE-longitudinal} that these
two equations are sufficient to describe the fields as long as there
is no longitudinal $\vec{E}$ component (i.e. $\vec{G}^\pm$ is transverse) 
or longitudinal $\vec{J}$
(also with reference to the divergence calculation in 
\ref{Sss-definitions-vector-divergence}).

% --------- 
\subsubsection{Longitudinal Equation}\label{Sss-1storderE-longitudinal}

Taking the $\vec{u} \times$ of the $\nabla \times \vec{H}$ equation
 removed a part of the field dynamics lying in the direction
 of the propagation vector $\vec{u}$.   
This part also include the response to longitudinal currents.
To rectify this omission, 
 we take the dot product with $\vec{u}$.
Using the standard vector identity eqn.(\ref{eqn-vectorid-I-7}): 
~
\begin{eqnarray}
\nabla \cdot \left(
    \vec{A}  \times \vec{B} \right) 
&=&
  \vec{A} 
  \cdot 
  \left(
    \nabla  \times \vec{B} 
  \right)
 -
  \vec{B} 
  \cdot 
  \left(
    \nabla  \times \vec{A} 
  \right)
,
\end{eqnarray}
and using $\nabla  \times \vec{u} = 0$; we get
~
\begin{eqnarray}
  \vec{u} 
  \cdot
  \left(
    \nabla \times \vec{H}
  \right)
&=&
  -
  \imath \omega
  \alpha^2
 *
 \left(
  \vec{u} 
  \cdot
  \vec{E} 
 \right)
 +
  \vec{u} 
  \cdot
  \vec{J} 
\\
\textrm{vector identity:} ~~~~ ~~~~
  \nabla 
  \cdot
  \left(
    \vec{u}  \times \vec{H}
  \right)
&=&
  -
  \imath \omega
  \alpha^2
 *
 \left(
  \vec{u} 
  \cdot
  \vec{E} 
 \right)
 +
  \vec{u} 
  \cdot
  \vec{J} 
\\
\textrm{premultiply:} ~~~~ ~~~~
  \alpha_r 
  \nabla 
  \cdot
  \left(
    \vec{u} \times \beta_r \vec{H}
  \right)
&=&
  -
  \imath \omega
  \beta_r \alpha^2 
 *
 \left(
  \vec{u} 
  \cdot
  \alpha_r \vec{E} 
 \right)
 +
  \alpha_r \beta_r
  \vec{u} 
  \cdot
  \vec{J} 
\\
  \alpha_r 
  \nabla 
  \cdot
  \left(
    \vec{G}^{+} - \vec{G}^{-} 
  \right)
&=&
  -
  \imath \omega
  \beta_r \alpha^2
 *
 \left(
  \vec{u} 
  \cdot
    \left[
       \vec{G}^{+} + \vec{G}^{-} 
    \right]
  \right)
 +
  2
  \alpha_r \beta_r
  \vec{u} 
  \cdot
  \vec{J}
\label{eqn-1storderE-statframe-longitgraddot}
\end{eqnarray}

This is the same as the source-free divergence-difference equation 
 seen in \ref{Sss-definitions-vector-divergence}; 
 where we see that both parts of the LHS are zero
 if  $ \vec{u} 
  \cdot
  \left(
    \vec{G}^{+} + \vec{G}^{-} 
  \right) = 0$
and $\vec{u} \cdot \vec{J} = 0$.

% --------- 
\subsubsection{Simple Cases}\label{Sss-1storderE-simple}

In the non-magnetic ($\beta_c=0$) case, 
~
\begin{eqnarray}
  \nabla \times \vec{G}^\pm
&=&
 \mp 
  \imath \omega 
  \alpha_r \beta_r 
  ~
  \vec{u} \times \vec{G}^\pm
 ~~
 +
  \imath \omega 
    \alpha_r \beta_r
    \vec{G}^{\circ} 
 ~~
 \mp 
  \frac{\imath \omega \beta_r}
       {2}
   \alpha_c
  *
  \left(
    \vec{u} \times 
    \left[ \vec{G}^+ + \vec{G}^- \right]
  \right)
 ~~
 \pm 
 ~~
  \vec{u} \times  \beta_r \vec{J}
\label{eqn-1storderE-statframe-nobetaX-transverse}
~~~~ ~~~~ 
~~~~ ~~~~ 
\\
 \pm
  \nabla {G}^{\circ}
&=&
 +
  \imath \omega 
    \alpha_r \beta_r
    \vec{u} {G}^{\circ} 
\label{eqn-1storderE-statframe-nobetaX-longitudinal}
\end{eqnarray}

In the time domain, eqn.(\ref{eqn-1storderE-statframe-nobetaX-transverse}) is 
~
\begin{eqnarray}
  \nabla \times \vec{G}^{\pm}
&=&
 \pm
  \partial_t
  \left[
    \left(
      \alpha_r \ast \beta_r 
    \right)
    ~\ast~
    \left(
      \vec{u} \times \vec{G}^\pm
    \right)
  \right]
%\nonumber
%\\
%&&
 ~
 \pm
  \partial_t
  \left[
    \left(
      \frac{\alpha_c \ast \beta_r}
           {2}
    \right)
    ~\ast~
    \left(
      \vec{u} \times 
      \left[ \vec{G}^+ + \vec{G}^- \right]
    \right)
  \right]
 \pm 
 ~~
  \vec{u} \times  \beta_r \ast \vec{J}
.
~~~~
\label{eqn-1storderE-propagation-approx-time}
\end{eqnarray}

\noindent{\bf Plane Polarized \& Magnetic ($G_x^\pm  \sim E_x \pm H_y$):} 

The $\hat{y}$ components of the LHS and RHS contain the $G_x$ 
values -- as a result of the curl and cross-products respectively.
Compare the previous stand-alone calculation in 
eqn.(\ref{eqn-stationaryGpropagation}) to 
~
\begin{eqnarray}
  \partial_z G_x^{\pm}
&=&
 \mp 
  \imath \omega 
  \alpha_r \beta_r 
  ~
  G_x^\pm
 ~~
 \mp 
  \frac{\imath \omega \beta_r}
       {2}
  \alpha_c
 *
  \left[ G_x^+ + G_x^- \right]
 ~~
 -
  \frac{\imath \omega \alpha_r}
       {2}
  \beta_c
 *
  \left[ G_x ^+ - G_x^- \right]
\label{eqn-1storderE-propagation-planepol}
\end{eqnarray}

This contains a single sign discrepancy compared to 
eqn (\ref{eqn-stationaryGpropagation}); the 
RHS has the opposite sign.

NB: recheck curl/cross-product component signs? (checked OK 20041028)

\noindent{\bf Vector Moving Frame, Transverse, Magnetic:} 

It is worth transforming to the moving frame at this end point,
rather than at the beginning as in 
subsection \ref{Ss-1storderE-gmovingvector}.  
Since eqn.(\ref{eqn-vector-frametranslation}) tells us that 
$\nabla \vec{Q} = 
\nabla' \vec{Q}- \alpha_f \beta_f \vec{u} \times \partial_t' \vec{Q}$, with
$\xi = \alpha_f \beta_f / \alpha_r \beta_r$,
we get 
~
\begin{eqnarray}
  \nabla' \times \vec{G}^{\pm}
  ~~
 +
  \imath \omega \alpha_f \beta_f
  ~
  \vec{u} \times \vec{G}^{\pm}
&=&
 \mp 
  \imath \omega 
  \alpha_r \beta_r 
  ~
  \vec{u} \times \vec{G}^\pm
% ~~
% +
%  \imath \omega 
%    \alpha_r \beta_r
%    \left[
%      \vec{u} \cdot \beta_r \vec{H}
%    \right]
%    \vec{u}
% ~~
% \mp
%  \nabla \left( \vec{u} \cdot \beta_r \vec{H} \right)
%\nonumber 
%\\
%&&
 ~~~~
 \mp 
  \frac{\imath \omega \beta_r}
       {2}
  \alpha_c 
  *
  \left(
    \vec{u} \times 
    \left[ \vec{G}^+ + \vec{G}^- \right]
  \right)
 ~~
 -
  \frac{\imath \omega \alpha_r}
       {2}
  \beta_c
  *
  \left(
    \vec{u} \times \left[ \vec{G}^+ - \vec{G}^- \right]
  \right)
 ~~
% +
%  \imath \omega 
%    \alpha_r \beta_c
%    \left[
%      \vec{u} \cdot \beta_r \vec{H}
%    \right]
%    \vec{u}
~~~~ 
~~~~
\\
  \nabla' \times \vec{G}^{\pm}
&=&
 \mp 
  \imath \omega 
  \alpha_r \beta_r 
  \left( 1 \mp \xi \right)
  ~
  \vec{u} \times \vec{G}^\pm
% ~~
% +
%  \imath \omega 
%    \alpha_r \beta_r
%    \left[
%      \vec{u} \cdot \beta_r \vec{H}
%    \right]
%    \vec{u}
% ~~
% \mp
%  \nabla \left( \vec{u} \cdot \beta_r \vec{H} \right)
\nonumber 
\\
&&
 ~~~~
 \mp 
  \frac{\imath \omega \beta_r}
       {2}
  \alpha_c
  *
  \left(
    \vec{u} \times 
    \left[ \vec{G}^+ + \vec{G}^- \right]
  \right)
 ~~
 -
  \frac{\imath \omega \alpha_r}
       {2}
  \beta_c
  *
  \left(
    \vec{u} \times \left[ \vec{G}^+ - \vec{G}^- \right]
  \right)
% ~~
% +
%  \imath \omega 
%    \alpha_r \beta_c
%    \left[
%      \vec{u} \cdot \beta_r \vec{H}
%    \right]
%    \vec{u}
~~~~ 
~~~~
\end{eqnarray}

This agrees with the standalone vector moving frame calculation in 
 the transverse case (see subsection \ref{Ss-1storderE-gmovingvector}); 
 which culminates in eqn (\ref{eqn-Ss-1storderE-movingvector}).  
I use the transverse case (not the general) to avoid 
 the moving-frame altered $\nabla$
 that occurs in one of the terms.

\noindent{\bf Plane Polarized Moving Frame, Magnetic:} 

This is just a special case of the above vector moving frame equation --
~
\begin{eqnarray}
  \partial_{z'} G_x^{\pm}
&=&
 \mp 
  \imath \omega 
  \alpha_r \beta_r 
  \left( 1 \mp \xi \right)
  ~
  G_x^{\pm}
 ~~
 \mp 
  \frac{\imath \omega \beta_r}
       {2}
   \alpha_c
  *
    \left[ G_x^+ + G_x^- \right]
 ~~
 -
  \frac{\imath \omega \alpha_r}
       {2}
    \beta_c
  *
    \left[ G_x^+ - G_x^- \right]
\end{eqnarray}

This has sign conflicts with the stand-alone calculation in 
subsection \ref{Ss-1storderE-gmoving} which 
culminates in eqn (\ref{eqn-movingGpropagation}).

\noindent{\bf Beltrami-like form:} 

This might be written
~
\begin{eqnarray}
  \left\{
    \left(
      \nabla \times 
    \right)
   \pm
    \imath \omega 
    \alpha_r \beta_r 
    \left(
      \vec{u} \times 
    \right)
  \right\}
  \vec{G}^{\pm}
&=&
  \vec{W}^{\pm}
,
\end{eqnarray}
where $\vec{W}^{\pm}$ is a ``source'' term, but note that  actually it
will be some complicated function of $\vec{G}^\pm$.

% --------- 
\subsubsection{Comments}\label{Sss-1storderE-comments}

Note the $\omega \alpha_r \beta_r = \omega/c_f$ prefactors on the RHS's -- 
these trivially convert ($\omega/c_r \rightarrow \omega / \omega k_r^{-1} 
\rightarrow k_r$) to a wavevector prefactor $k_r$, as would be expected 
for this kind of spatially propagating description.

Inspection of the equations shows that even if (e.g.) $G^+$ starts
 identically equal to zero, 
 that it will pick up some value due to the interaction of the $G^-$ components 
 with the non-vacuum part of the permittivity.  
However, so as long as $\alpha_c \ll \alpha_r$, 
 the transfer will be slow, 
 and because of the leading $2\imath \omega$ term, 
 $G^+$ will rapidly oscillate.  
As a result, 
 we can assume (in an appropriate limit) that
 any $G^+$ contribution to the evolution of $G^-$ will spatially average to
 zero in a kind of rotating wave approximation (RWA), 
 although we should properly define criteria for
 ``slow transfer'' and ``RWA''

One approach that would avoid throwing away the backward propagating terms
would be to describe the $G^+$ field with the reverse frame transformation. 
Whilst this would result in some inconvenience, when simulating numerically it
could be useful in that fields propagating in the unimportant direction could
be thrown away only once they had ceased overlapping with the range covered by
the pulse travelling in the important direction.

% --------- --------- ---------

\subsection{Plane-polarized moving frame (Example only)}
\label{Ss-1storderE-gmoving}

\noindent
NOTE: In this example calculation, 
 the  convolutions required for the nonlinearity are not explicitly 
 included in the notation.  
This does not affect the result, 
 as long as they are reinserted correctly in the final equations.

Here I use the (non relativistic) frame translation defined 
by eqn.(\ref{eqn-z-frametranslation}) in subsection \ref{Ss-movingframe}.
Starting with the source-free Maxwells' equations, we have --
~
\begin{eqnarray}
\frac{\partial H_y (t) }{\partial z}
&=&
  -\frac{\partial }{\partial t} \epsilon \ast E_x (t)
,
\\
\frac{\partial E_x (t)}{\partial z}
&=&
  -\frac{\partial }{\partial t}  \mu \ast H_y (t)
;
\\
\Longrightarrow ~~~~ ~~~~
\left[
  \frac{\partial }{\partial z'}
 +
  \alpha_f \beta_f \frac{\partial }{\partial t'}
\right]
H_y
&=&
  -\frac{\partial }{\partial t'} \epsilon \ast E_x
,
\\
\left[
  \frac{\partial }{\partial z'}
 +
  \alpha_f \beta_f \frac{\partial }{\partial t'}
\right]
E_x
&=&
  -\frac{\partial }{\partial t'} \mu \ast H_y
;
\\
\Longrightarrow ~~~~ ~~~~
  \frac{\partial }{\partial z'}
H_y
&=&
  -
  \frac{\partial}{\partial t'}
  \left[
    \epsilon \ast E_x
   +
    \alpha_f \beta_f  H_y
  \right]
,
\\
  \frac{\partial }{\partial z'}
E_x
&=&
  -
  \frac{\partial}{\partial t'}
  \left[
    \mu \ast  H_y 
   +
    \alpha_f \beta_f  E_x
  \right]
\end{eqnarray}

Now, 
 because the convolutions $\epsilon \ast E_x$ and $\mu \ast  H_y$ 
 complicate the analysis, 
 I will fourier transform into the frequency domain where the
 linear response becomes a simple product. 
Using the definitions in subsection \ref{S-medium}, 
 I can proceed with the calculation --
~
\begin{eqnarray}
  \frac{\partial }{\partial z'}
  \tilde{H}_y (\omega)
&=&
  \imath \omega
  \left[
    \alpha (\omega)^2 \tilde{E_x} (\omega)
   +
    \alpha_f  (\omega)\beta_f  (\omega) ~ \tilde{H}_y (\omega)
  \right]
,
\\
  \frac{\partial }{\partial z'}
  \tilde{E_x} (\omega)
&=&
  \imath \omega
  \left[
    \beta^2  \tilde{H}_y (\omega)
   +
    \alpha_f  (\omega)\beta_f  (\omega) ~  \tilde{E_x} (\omega)
  \right]
\\
\Longrightarrow ~~~~ ~~~~
  \frac{\partial }{\partial z'}
  \tilde{H}_y
&=&
  \imath \omega
  \left[
    \left( \alpha_r^2 +  \alpha_r \alpha_c \right)^2 \tilde{E_x}
   +
    \alpha_f \beta_f \tilde{H}_y
  \right]
,
\\
  \frac{\partial }{\partial z'}
  \tilde{E_x}
&=&
  \imath \omega
  \left[
    \left( \beta_r^2 + \beta_r \beta_c \right)^2  \tilde{H}_y
   +
    \alpha_f \beta_f  \tilde{E_x}
  \right]
\\
\Longrightarrow ~~~~ ~~~~
  \frac{\partial }{\partial z'}
  \beta_r \tilde{H}_y
&=&
  \imath \omega \alpha_r \beta_r 
  \left[
    \alpha_r \tilde{E_x}
   +
     \xi \beta_r \tilde{H}_y
  \right]
 +
  \imath \omega
  \beta_r \alpha_c . \alpha_r \tilde{E_x}
\label{eqn-defs-frames-dbH}
,
\\
  \frac{\partial }{\partial z'}
  \alpha_r  \tilde{E_x}
&=&
  \imath \omega \alpha_r  \beta_r 
  \left[
    \beta_r  \tilde{H}_y
   +
    \xi \alpha_r \tilde{E_x}
  \right]
 +
   \imath \omega
    \alpha_r  \beta_c . \beta_r \tilde{H}_y
\label{eqn-defs-frames-daE}
\end{eqnarray}

Adding and subtracting eqns. (\ref{eqn-defs-frames-dbH}, 
\ref{eqn-defs-frames-daE}), and using the fourier transform of
the definitions for $G^\pm = \alpha_r E_x \pm \beta_r H_y$, we get 
~
\begin{eqnarray}
  \frac{\partial }{\partial z'}
  \tilde{G}^\pm
&=&
 \pm
  \left(1 \pm \xi \right) 
  \imath \omega \alpha_r  \beta_r \tilde{G}^\pm
 +
  \imath \omega
    \alpha_r  \beta_c . \beta_r \tilde{H}_y
 \pm
  \imath \omega
  \beta_r \alpha_c . \alpha_r \tilde{E_x}
\\
&=&
 \pm
  \left(1 \pm \xi \right) 
  \imath \omega \alpha_r  \beta_r \tilde{G}^\pm
 ~~
 +
  \imath \omega
  \frac{ \alpha_r  \beta_c }
       {2}
    \left( \tilde{G}^+ - \tilde{G}^- \right)
 ~~
 \pm
  \imath \omega
  \frac{ \beta_r \alpha_c }
       {2}
    \left( \tilde{G}^+ + \tilde{G}^- \right)
\label{eqn-movingGpropagation}
\end{eqnarray}

The ``non-magnetic'' limit, where $\beta_1=0$ is
~
\begin{eqnarray}
  \frac{\partial }{\partial z'}
  \tilde{G}^\pm
&=&
 \pm
  \left(1 \pm \xi \right) 
  \frac{\imath \omega}{c_f} \tilde{G}^+
 ~~
 \pm
  \imath \omega
  \frac{ \beta_r \alpha_c }
       {2}
    \left( \tilde{G}^+ + \tilde{G}^- \right)
\end{eqnarray}

The reference limit of the above equations (with $\alpha_1=0; \beta_1=0$) 
when the frame is matched to the reference $c_f = c_r$
is exactly as you would expect, given that the new frame is moving in 
the direction of the $G^-$ component, and since the pulse
does not evolve (relative to its centre)  in a vacuum, $\partial_{z'} G^-$ is
zero (see eqn.(\ref{eqn-defs-frames-vGminus})).  Conversely, the $G^+$
component is moving the opposite way to our frame at $c$, thus relative to the
frame it moves at $2c$, hence the derivative is double that in a stationary
frame (see eqn.(\ref{eqn-defs-frames-vGplus})).
~
\begin{eqnarray}
  \frac{\partial }{\partial z'}
  \tilde{G}^+
&=&
 \pm
  2 \frac{\imath \omega}{c_f} G^+
\label{eqn-defs-frames-vGplus}
\\
  \frac{\partial }{\partial z'}
  \tilde{G}^-
&=&
0
\label{eqn-defs-frames-vGminus}
\end{eqnarray}

% --------- --------- ---------

\subsection{Vector moving frame (Example only)}\label{Ss-1storderE-gmovingvector}

\noindent
NOTE: In this example calculation, 
 the nonlinear convolutions are not explicitly 
 included in the notation.  
This does not affect the result, 
 as long as they are reinserted correctly in the final equations.

Note that is better to follow the (authoritative) vector stationary
 frame derviation (see \ref{Ss-1storderE-gstationaryvector}) instead;
 where the frame translation is applied to the wave equation.
I use the (non relativistic) frame translation defined 
 by eqn.(\ref{eqn-vector-frametranslation}) in subsection \ref{Ss-movingframe};
 and using the definitions in section \ref{S-medium}, 
 I can proceed with the calculation --
~
\begin{eqnarray}
  \nabla \times \vec{H} (t)
&=&
  +\partial_t \epsilon \ast \vec{E} (t)
,
\nonumber
\\
  \nabla \times \vec{E} (t)
&=&
  -\partial_t  \mu \ast \vec{H} (t)
\\
~
~
\textrm{into moving frame}
\longrightarrow
~~~~ ~~~~
  \nabla' \times \vec{H}
&=&
 +
  \partial_{t'} \epsilon \ast \vec{E} 
 -
  \alpha_f \beta_f ~  \vec{u} \times \partial_{t'} \vec{H}
,
\nonumber
\\
  \nabla' \times \vec{E}
&=&
 -
  \partial_{t'}  \mu \ast \vec{H} 
 -
  \alpha_f \beta_f ~  \vec{u} \times \partial_{t'} \vec{E}
\\
~
~
\textrm{into temporal-frequency space}
\longrightarrow
~~~~ ~~~~
  \nabla' \times \vec{H} (\omega)
&=&
 -
  \imath \omega 
  \left(
    \alpha (\omega)^2 \vec{E} (\omega)
   -
    \alpha_f (\omega) \beta_f (\omega) ~ \vec{u} \times \vec{H} (\omega)
  \right)
,
\nonumber
\\
  \nabla' \times \vec{E}
&=&
 +
  \imath \omega 
  \left(
    \beta (\omega)^2 \vec{H} (\omega)
   +
    \alpha_f (\omega) \beta_f (\omega) ~ \vec{u} \times \vec{E} (\omega)
  \right)
\\
~
~
\textrm{swap transverse}
\longrightarrow
~~~~ ~~~~
  \vec{u} \times \left( \nabla' \times \vec{H} \right)
&=&
 -
  \imath \omega 
  \left(
    \alpha^2 ~ \vec{u} \times \vec{E}
   -
    \alpha_f \beta_f ~ \vec{u} \times 
      \left[ \vec{u} \times \vec{H} \right]
  \right)
,
\nonumber
\\
  \nabla' \times \vec{E}
&=&
 +
  \imath \omega 
  \left(
    \beta^2 \vec{H}
   +
    \alpha_f \beta_f ~ \vec{u} \times \vec{E}
  \right)
\\
~
~
\textrm{premultiply}
\longrightarrow
~~~~ ~~~~
  \vec{u} \times \left( \nabla' \times \beta_r \vec{H} \right)
&=&
 -
  \imath \omega
  \left(
    \beta_r \alpha^2 ~ \vec{u} \times \vec{E}
   -
    \alpha_f \beta_f \beta_r ~ \vec{u} \times 
      \left[ \vec{u} \times \vec{H} \right]
  \right)
,
\nonumber
\\
  \nabla' \times \alpha_r \vec{E}
&=&
 +
  \imath \omega
  \left(
    \alpha_r \beta^2 \vec{H}
   +
    \alpha_f \beta_f \alpha_r ~ \vec{u} \times \vec{E}
  \right)
\\
~
~
\textrm{sum \& diff}
\longrightarrow
~~~~ 
  \nabla' \times \alpha_r \vec{E}
 ~~
 \pm
 ~~
  \vec{u} \times \left( \nabla' \times \beta_r \vec{H} \right)
&=&
 +
 \imath \omega
  \left(
    \alpha_r \beta^2 \vec{H}
   +
    \alpha_f \beta_f \alpha_r ~ \vec{u} \times \vec{E}
  \right)
\nonumber
\\
&&
 ~~
 \mp 
 ~~
  \imath \omega 
  \left(
    \beta_r \alpha^2 ~ \vec{u} \times \vec{E}
   -
    \alpha_f \beta_f \beta_r ~ \vec{u} \times 
       \left[ \vec{u} \times \vec{H} \right]
  \right)
,
\\
~
~
\textrm{collect terms}
\longrightarrow
~~~~ ~~~~
&=&
  \mp
   \imath \omega
   \left(
     \alpha^2 \beta_r
    \mp
     \alpha_f \beta_f \alpha_r 
   \right)
   \vec{u} \times \vec{E}
\nonumber
\\
&&
 ~~
 +
  \imath \omega 
   \left(
     \alpha_r \beta^2 
     \vec{H}
    ~
    \pm
     \alpha_f \beta_f \beta_r
     ~
     \vec{u} \times 
     \left[
       \vec{u} \times \vec{H}
     \right]
   \right)
,
\end{eqnarray}

The vector $\vec{G}^\pm$ fields (defined in eqn (\ref{eqn-S-defs-Gvector}))
are ~
\begin{eqnarray}
  \vec{G}^{\pm} 
&=&
  \alpha_r \vec{E} + \vec{u} \times \beta_r \vec{H}
\end{eqnarray}

This means I need to convert both 
the second term on the LHS of the sum-and-difference equation above, 
as well as the RHS.  
It is most imortant for the LHS to be simple, 
because this will define the type of propagation specified by the RHS.  
To convert the LHS, we need two vector identities,
as used in subsection \ref{S-vectorid} to get 
eqn.(\ref{eqn-vectorid-combined}). So using
~
\begin{eqnarray}
  \vec{u} \times \left( \nabla \times \vec{H} \right)
 -
  \nabla \left( \vec{u} \cdot \vec{H} \right)
&=&
  \nabla \times \left( \vec{u} \times \vec{H} \right)
\end{eqnarray}

\noindent
WARNING! 
These are moving frame $\nabla$'s arriving from the vector identities!!

If I retain the $\vec{u} \cdot \beta_r \vec{E}$ term, which 
would be zero for strictly transverse $E$ fields (TEF), then --
~
\begin{eqnarray}
  \nabla' \times \alpha_r \vec{E}
 ~~
 \pm
 ~~
  \vec{u} \times \left( \nabla' \times \beta_r \vec{H} \right)
&=&
  \mp
   \imath \omega
   \left(
     \alpha^2 \beta_r
    \mp
     \alpha_f \beta_f \alpha_r
   \right)
   \vec{u} \times \vec{E}
\nonumber
\\
&&
 ~~
 +
  \imath \omega 
   \left(
     \alpha_r \beta^2 
     \vec{H}
    ~
    \pm
     \alpha_f \beta_f \beta_r
     ~
     \vec{u} \times 
     \left[
       \vec{u} \times \vec{H}
     \right]
   \right)
,
\\
\longrightarrow
~~~~ ~~~~
  \nabla' \times \alpha_r \vec{E}
 ~~
 \pm
 ~~
  \nabla' \times \left( \vec{u} \times \beta_r \vec{H} \right)
 \pm
  \nabla' \left( \vec{u} \cdot \beta_r \vec{H} \right)
&=&
  \mp
   \imath \omega
   \left(
     \alpha^2 \beta_r
    \mp
     \alpha_f \beta_f \alpha_r
   \right)
   \vec{u} \times \vec{E}
\nonumber
\\
&&
 ~~
 +
  \imath \omega 
   \left(
     \alpha_r \beta^2 
     \vec{H}
    ~
    \pm 
     \alpha_f \beta_f \beta_r
     ~
     \vec{u} \times 
     \left[
       \vec{u} \times \vec{H}
     \right]
   \right)
\\
(C) 
\longrightarrow
~~~~
  \nabla' \times 
  \left[
    \alpha_r \vec{E}
    ~~
    \pm
    ~~
    \left( \vec{u} \times \beta_r \vec{H} \right)
  \right]
&=&
  \mp
   \imath \omega
   \left(
     \alpha^2 \beta_r
    \mp
     \alpha_f \beta_f \alpha_r
   \right)
   \vec{u} \times \vec{E}
\nonumber
\\
&&
 ~~
 +
  \imath \omega 
   \left(
     \alpha_r \beta^2 
     \vec{H}
    ~
    \pm
     \alpha_f \beta_f \beta_r
     ~
     \vec{u} \times 
     \left[
       \vec{u} \times \vec{H}
     \right]
   \right)
\nonumber
\\
&&
 ~~
 \mp
  \nabla' \left( \vec{u} \cdot \beta_r \vec{H} \right)
\end{eqnarray}

Now note that eqn.(\ref{eqn-vectorid-doublecross-u}) means that
~
\begin{eqnarray}
     \vec{u} \times 
     \left[
       \vec{u} \times \vec{H}
     \right]
&=& 
  \left[ \vec{u} \cdot \vec{H} \right] \vec{u} 
 -
  \vec{H} 
\\
\textrm{and so} 
~~~~ ~~~~
  \vec{u} \times \vec{G}^{\pm}
&=&
  \vec{u} \times \alpha_r \vec{E}
 \pm
  \vec{u} \times 
    \left[ 
      \vec{u} \times \beta_r \vec{H} 
    \right]
\\
&=&
  \vec{u} \times \alpha_r \vec{E}
 \mp
  \beta_r \vec{H} 
 ~~
 \pm
  \left[ \vec{u} \cdot \beta_r \vec{H} \right] \vec{u} 
\end{eqnarray}

So, 
~
\begin{eqnarray}
  \nabla' \times \vec{G}^{\pm}
&=&
  \mp
   \imath \omega
   \left(
     \alpha^2 \beta_r
    \mp
     \alpha_f \beta_f \alpha_r
   \right)
   \vec{u} \times \vec{E}
 ~~
 +
  \imath \omega 
   \left(
     \alpha_r \beta^2 
     \vec{H}
    ~
    \pm
     \alpha_f \beta_f \beta_r
     ~
     \vec{u} \times 
     \left[
       \vec{u} \times \vec{H}
     \right]
   \right)
 ~~
 \mp
  \nabla' \left( \vec{u} \cdot \beta_r \vec{H} \right)
\\
&=&
  \mp
   \imath \omega
   \left(
     \alpha^2 \beta_r
    \mp
     \alpha_f \beta_f \alpha_r
   \right)
   \vec{u} \times \vec{E}
 ~~
 +
  \imath \omega 
   \left(
     \alpha_r \beta^2 
     \left[
       \vec{u} \cdot \vec{H} 
     \right]
     \vec{u}  
    ~
    -
     \alpha_r \beta^2 
     \vec{u} \times 
     \left[
       \vec{u} \times \vec{H}
     \right]
    ~
    \pm
     \alpha_f \beta_f \beta_r
     ~
     \vec{u} \times 
     \left[
       \vec{u} \times \vec{H}
     \right]
   \right)
\nonumber
\\
&&
 ~~~~
 ~~~~
 \mp
  \nabla' \left( \vec{u} \cdot \beta_r \vec{H} \right)
\\
&=&
  \mp
   \imath \omega
   \left(
     \alpha^2 \beta_r
    \mp
     \alpha_f \beta_f \alpha_r
   \right)
   \vec{u} \times \vec{E}
 ~~
 -
  \imath \omega 
   \left(
     \alpha_r \beta^2 
    \mp
     \alpha_f \beta_f \beta_r
   \right)
     \vec{u} \times
     \left[
       \vec{u} \times \vec{H} 
     \right]
 ~~
 +
  \imath \omega 
     \alpha_r \beta^2 
     ~
     \left[
       \vec{u} \cdot \vec{H} 
     \right]
     \vec{u}
 ~~
 \mp
  \nabla' \left( \vec{u} \cdot \beta_r \vec{H} \right)
.
~~~~
~~~~
\end{eqnarray}

I now separate the interaction parts (depending on $\alpha_c$, $\beta_c$) 
from the reference parts (depending on $\alpha_r$, $\beta_r$), and then
substitute (as far as possible) expressions containing $G$ rather than $E$
or $H$, 
by referring to eqns (\ref{eqn-S-defs-Evector}) 
and (\ref{eqn-S-defs-Hvector}).   
Hence 
~
\begin{eqnarray}
  \nabla \times \vec{G}^{\pm}
&=&
  \mp
   \imath \omega \alpha_r \beta_r
   \left(
     1
    \mp
     \xi
   \right)
   \vec{u} \times \alpha_r\vec{E}
 ~~
 -
  \imath \omega \alpha_r \beta_r
   \left(
     1
    \mp
     \xi
   \right)
     \vec{u} \times
     \left[
       \vec{u} \times \beta_r\vec{H} 
     \right]
 ~~
 +
  \imath \omega 
     \alpha_r \beta_r
     ~
     \left[
       \vec{u} \cdot \beta_r\vec{H} 
     \right]
     \vec{u}
 ~~
 \mp
  \nabla' \left( \vec{u} \cdot \beta_r \vec{H} \right)
\nonumber
\\
&&
 ~~~~
 \mp 
   \imath \omega \alpha_c \beta_r 
   \vec{u} \times \alpha_r \vec{E}
 -
  \imath \omega \alpha_r \beta_c
     \vec{u} \times
     \left[
       \vec{u} \times \beta_r \vec{H} 
     \right]
 +
  \imath \omega 
     \alpha_r \beta_c 
     ~
     \left[
       \vec{u} \cdot \beta_r \vec{H} 
     \right]
     \vec{u}
\\
&=&
  \mp
   \imath \omega \alpha_r \beta_r
   \left(
     1
    \mp
     \xi
   \right)
   \left\{
     \vec{u} \times \alpha_r\vec{E}
    \pm
     \vec{u} \times
     \left[
       \vec{u} \times \beta_r\vec{H} 
     \right]
   \right\}
 ~~
 +
  \imath \omega 
     \alpha_r \beta_r
     ~
     \left[
       \vec{u} \cdot \beta_r\vec{H} 
     \right]
     \vec{u}
 ~~
 \mp
  \nabla' \left( \vec{u} \cdot \beta_r \vec{H} \right)
\nonumber
\\
&&
 ~~~~
 \mp 
   \imath \omega \alpha_c \beta_r 
   \vec{u} \times \alpha_r \vec{E}
 -
  \imath \omega \alpha_r \beta_c
     \vec{u} \times
     \left[
       \vec{u} \times \beta_r \vec{H} 
     \right]
 +
  \imath \omega 
     \alpha_r \beta_c 
     ~
     \left[
       \vec{u} \cdot \beta_r \vec{H} 
     \right]
     \vec{u}
\\
&=&
  \mp
   \imath \omega \alpha_r \beta_r
   \left(
     1
    \mp
     \xi
   \right)
   \vec{u} \times \vec{G}^\pm
 ~~
 +
  \imath \omega 
     \alpha_r \beta_r
     ~
     \left[
       \vec{u} \cdot \beta_r\vec{H} 
     \right]
     \vec{u}
 ~~
 \mp
  \nabla' \left( \vec{u} \cdot \beta_r \vec{H} \right)
\nonumber
\\
&&
 ~~~~
 \mp 
   \imath \omega \alpha_c \beta_r 
   \vec{u} \times \alpha_r \vec{E}
 -
  \imath \omega \alpha_r \beta_c
     \vec{u} \times
     \left[
       \vec{u} \times \beta_r \vec{H} 
     \right]
 +
  \imath \omega 
     \alpha_r \beta_c 
     ~
     \left[
       \vec{u} \cdot \beta_r \vec{H} 
     \right]
     \vec{u}
\end{eqnarray}

Note that the non-transverse corrections depend only on the 
 {\em Magnetic} field $\vec{H}$.  
I assume this is a curiousity of the chosen $G$ definition, 
 as (I think) choosing $G^\pm \sim u \times E \pm H $
 would produce non-transverse corrections depending on the electric field.
As yet I do not have a simple vectorised form for calculating $H$ purely
 in terms of $G$'s without these non-transverse corrections occuring.

So using 
 $\vec{E} = ( \vec{G}^+ + \vec{G}^- ) / 2 \alpha_r$; 
 $\vec{u} \times \vec{H} = ( \vec{G}^+ - \vec{G}^- ) / 2 \beta_r$; 
 $\vec{H} = \vec{u} \times  (\vec{u} \times \vec{H}) 
                 - (\vec{u} \cdot \vec{H}) \vec{u}$
~
\begin{eqnarray}
  \nabla' \times \vec{G}^{\pm}
&=&
  \mp
   \imath \omega
     \alpha_r \beta_r 
   \left(
     1 \mp \xi
   \right)
   \vec{u} \times \vec{G}^\pm
  ~~~~
 +
  \imath \omega 
     \alpha_r \beta_r
     ~
     \left[
       \vec{u} \cdot \beta_r \vec{H} 
     \right]
     \vec{u}
 ~~
 \mp
  \nabla' \left( \vec{u} \cdot \beta_r \vec{H} \right)
\nonumber 
\\
&&
~~~~
  \mp
   \frac{\imath \omega \alpha_c \beta_r}
        {2}
   \vec{u} \times \left[ \vec{G}^+ + \vec{G}^- \right]
 ~~
 -
   \frac{\imath \omega \alpha_r \beta_c}
        {2}
  \left[
    \vec{u} \times \left( \vec{G}^+ - \vec{G}^- \right)
  \right]  
 ~~
 +
   \imath \omega
   \alpha_r  \beta_c
   \left[ \vec{u} \cdot \beta_r \vec{H} \right]
   \vec{u}
\label{eqn-Ss-1storderE-movingvector}
\\
~
~
\textrm{non-magnetic}
\rightarrow
~~~~ 
&=&
  \mp
   \imath \omega
     \alpha_r \beta_r 
   \left(
     1 \mp \xi
   \right)
   \vec{u} \times \vec{G}^\pm
  ~~
  \mp
   \frac{\imath \omega \alpha_c \beta_r}
        {2}
   \vec{u} \times \left[ \vec{G}^+ + \vec{G}^- \right]
  ~~
 +
  \imath \omega 
     \alpha_r \beta_r
     ~
     \left[
       \vec{u} \cdot \beta_r \vec{H} 
     \right]
     \vec{u}
 ~~
 \mp
  \nabla' \left( \vec{u} \cdot \beta_r \vec{H} \right)
~~~~ ~~~~ ~~~~
\\
~
~
\textrm{transverse}
\rightarrow
~~~~ 
&=&
  \mp
   \imath \omega
     \alpha_r \beta_r 
   \left(
     1 \mp \xi
   \right)
   \vec{u} \times \vec{G}^\pm
  ~~
  \mp
   \frac{\imath \omega \alpha_c \beta_r}
        {2}
   \vec{u} \times \left[ \vec{G}^+ + \vec{G}^- \right]
\end{eqnarray}

\newpage
% --------- --------- ---------
\subsection{Simulations: a step by step guide}
\label{Ss-1storderE-sims}

% --------- 
\subsubsection{The medium}
\label{Sss-1storderE-sims-medium}

Consider a plane polarized field propagating through a non-magnetic medium.
I choose a stationary frame,
  and a dispersionless linear reference 
 (so $\alpha_r, \beta_r$ are constants).
The permeability of the medium is the vacuum value $\mu_0$, and the 
permittivity has three contributions: 
a vacuum (reference) $\epsilon_0$, 
a linear dispersion $\epsilon_D(t)$,
and an instantaneous nonlinearity $\epsilon_{NL}$.
The medium properties are therefore broken down in the 
following fashion -- 
~
\begin{eqnarray}
%  \epsilon (t)
%&=&
%  \epsilon_0  \delta(t)
% + 
%  \epsilon_D(t)
% +
%  \epsilon_{NL}  \delta(t)
%\\
  \tilde{\epsilon}
&=&
  \epsilon_0
 +
  \tilde{\epsilon}_c^D (\omega)
 +
  \epsilon_c^{NL}
\\
  \tilde{\epsilon}
&=&
  \tilde{\alpha}_r^2 
 +
  \tilde{\epsilon}_c^D  \tilde{\alpha}_r
 +
  \tilde{\epsilon}_c^{NL}  \tilde{\alpha}_r
\\
\textrm{so}
~~~~
~~~~
  \tilde{\epsilon}_c^D 
&=&
  \tilde{\epsilon}_c^D (\omega) / \tilde{\alpha}_r
\\
  \tilde{\epsilon}_c^{NL} 
&=&
  \tilde{\epsilon}_c^{NL} (\omega) / \tilde{\alpha}_r
.
\end{eqnarray}

We can see here that choosing a dispersive reference medium
 would modify both the dispersive correction and the 
 nonlinear term.
Im the nonlinear case, 
 this adds a time dependence which did not previously exist,
 which would increase the computational effort needed if 
 the nonlinearity $\tilde{\epsilon}_c^{NL}$ was instantaneous.

% --------- 
\subsubsection{The full wave equation}
\label{Sss-1storderE-sims-waveequation}

The first order evolution equation, using the parameter definitions
at eqns.(\ref{eqn-defs-alphaX}, \ref{eqn-defs-betaX})), is
~
\begin{eqnarray}
  \partial_z G_x^{\pm}
&=&
 \mp 
  \imath \omega 
  \tilde{\alpha}_r \tilde{\beta}_r 
       \left( 1 \mp \xi \right)
  ~
  G_x^\pm
 ~~
 \mp 
  \frac{\imath \omega \tilde{\alpha}_c \tilde{\beta}_r}
       {2}
  \left[ G_x^+ + G_x^- \right]
\\
&=&
 \mp 
  \imath \omega 
  \tilde{\alpha}_r \tilde{\beta}_r 
       \left( 1 \mp \xi \right)
  ~
  G_x^\pm
 ~~
 \mp 
  \frac{\imath \omega \tilde{\alpha}_c^D \tilde{\beta}_r}
       {2}
  \left[ G_x^+ + G_x^- \right]
 ~~
 \mp 
  \frac{\imath \omega \tilde{\beta}_r}
       {2}
  \tilde{\alpha}_c^{NL} 
  *
  \left[ G_x^+ + G_x^- \right]
\end{eqnarray}

This has three terms -- 
a reference carrier-oscillation-like term ($\sim \tilde{\alpha}_r$), 
a linear dispersion term ($\sim \tilde{\alpha}_c^D$), 
and a nonlinear polarization term ($\sim \tilde{\alpha}_c^{NL}$).
The straightforward way of solving this using a split step method, 
where each contribution is considered to be relatively weak, 
hence we solve the above in three steps.  
Each one integrates forward a distance $\delta z$, 
but since we do one after the other (rather than  doing them simultaneously), 
the procedure is only accurate to first order -- 
but for small enough $\delta z$, 
the error can be assured negligible.

% --------- 
\subsubsection{The reference evolution}
\label{Sss-1storderE-sims-reference}

The reference term just applies a
complex rotation to the field in frequency space, which can be
calculated exactly --
~
\begin{eqnarray}
  G_{x1}^{\pm}
&=&
  G_x^{\pm} (z) 
 \times
  \exp
    \left[
     \mp 
       \imath \omega 
       \tilde{\alpha}_r \tilde{\beta}_r 
       \left( 1 \mp \xi \right)
      .
      \delta z
    \right]
\label{eqn-1storderE-example1-G1}
.
\end{eqnarray}

This will have no effect if we choose the frame velocity to be the same as the
phase velocity of the reference medium.  
This might seem to differ from the usual $E$ field approaches, 
where choosing the frame velocity to match the group
velocity gives the best cancellation of propagation terms.  
However, remember that the way we have chosen our reference medium 
(in this example) means that group velocity corrections 
appear in the second RHS term as part of $\tilde{\alpha}_c^D$.
Choosing a frame velocity equal to the group velocity here will
leave a residual reference-like term, but this can then cancel with part of
the group-velocity-like contribution from the dispersion term, 
which in most cases would give the best cancellation -- 
just as in the usual $E$ field approaches.  
If we wanted, we could easily rearrange
eqn.(\ref{eqn-1storderE-example1-G1}) to incorporate such a cancellation, and
then solve the equation appropriately.

% --------- 
\subsubsection{The dispersive correction}
\label{Sss-1storderE-sims-dispersive}

The next  step is solving for the linear dispersion 
$\tilde{\alpha}_c^D= \tilde{\epsilon}_c^D / \tilde{\alpha}_r$ 
in the frequency domain.  Fortunately, this part of the equation is (also)
trivial to solve exactly in the frequency domain, 
and results in the following exponential 
solution --
~
\begin{eqnarray}
  G_{xD}^{\pm}(z+\delta z)
&=&
  G_x^{\pm} (z) 
 \times
  \exp
    \left[
     \mp 
       \imath k
     ~
       G_x^\pm
       .
       \delta z
     ~~
      \mp 
      \frac{\imath \omega \tilde{\epsilon}_c^D(\omega)}
           {2}
      \sqrt{\frac{\mu_0}{\epsilon_0}}
      \left[ G_x^+(z) + G_x^-(z) \right]
      .
      \delta z
    \right]
.
\end{eqnarray}

Although both reference and dispersion steps can be solved exactly 
using exponentials, there is an important difference.  
The reference evolution of $G^+$ depends only on $G^+$, 
whereas the dispersion evolution depends on the sum $G^++G^-$, 
since the dispersion acts on the electric field.  
In the forward-only approximation, of course, 
$G^-=0$ and the two steps can be trivally combined -- 
as is automagically done in most approaches
solving for the propagation of optical pulses.

% --------- 
\subsubsection{The nonlinear correction}
\label{Sss-1storderE-sims-nonlinear}

The final (third) step is to transform into the time domain 
 and solve for the $n$-th order nonlinear effects.  
Since our reference $\alpha_r, \beta_r$ are constants, 
 $\tilde{\alpha}_c^{NL} = \tilde{\epsilon}_c^{NL} / \tilde{\alpha}_r$ 
 is trivial to calculate and will be just 
 $\tilde{\alpha}_c^{NL} = \chi^{(n)} E^{n-1} / \tilde{\alpha}_r$, 
so (for a simple Euler method integration)
~
\begin{eqnarray}
  G_{x}^{\pm}(z+\delta z)
&=&
  G_{xD}^{\pm}(z+\delta z)
 ~~
 \pm
  \frac{\chi^{(n)}}
       {2}
      \sqrt{\frac{\mu_0}{\epsilon_0}}
  \frac{d}{dt}
  \left[ G_{xD}^+(z+\delta z) + G_{xD}^-(z+\delta z) \right]^n
  \delta z
\end{eqnarray}

For a narrow-band field, the time derivative would be dominated by 
(proportional to)  its centre frequency; 
and indeed in most envelope theories 
we only see a factor of (carrier frequency) $\omega_0$ 
in the analogous expression.  
For a wider-band field, there would be corrections, 
as found for the GFEA approach \cite{Kinsler-N-2003pra}.

In an envelope theory, the time derivative would be dominated by the 
carrier frequency $\omega_0$, hence we would only see a factor of 
$\omega_0$ rather than the differential here. A few-cycle envelope 
theory would give us corrections to the $\omega_0$ prefactor.
If you wanted to, the derivative used in this nonlinear step could 
be calculated in the spectral domain -- you'd have to FT $G$ into the 
time domain, raise it to the $n$-th power, and FT back...

% --------- 
\subsubsection{Initial conditions: matching a pulse to the medium}
\label{Sss-1storderE-sims-initialc}

Here I describe how to make the ``best matched'' initial conditions 
for $G^\pm$ describing a pulse
propagating {\em only} in the ``+'' direction.  Note that
it may be helpful to read section \ref{S-interpretations} if you are
confused by the use of terms like ``pseudo-reflection'' or by the 
meaning of the co-propagating $G^-$ component.

The procedure is:

\noindent
(1) Pick a suitable electric field $E(\omega)$

\noindent
(2) Find out the full dispersive properties of the medium, as 
described by $\epsilon_i$ and $\mu_i$.  
You can even put the nonlinear or any other properities into $\epsilon_i$
and $\mu_i$ as well -- but only if you will still 
know how get a solution for steps (3) and (5) below with those added 
complications.

\noindent
(3) Calculate the $H(\omega)$ corresponding to $E(\omega)$ for a 
``+'' propagating pulse:
~
\begin{eqnarray}
  0 
&=& 
  \sqrt{\epsilon_i(\omega)} E(\omega) 
 +
  \sqrt{\mu_i(\omega)} H(\omega)
\\
  \Rightarrow
  ~~~~
  H(\omega)
&=&
  \sqrt{\frac{\epsilon_i(\omega)}{\mu_i(\omega)}} E(\omega)
.
\end{eqnarray}

\noindent
(4) Choose our reference medium parameters 
$\epsilon_r(\omega)$ and $\mu_r(\omega)$.

\noindent
(5) Calculate our initial $G^\pm$ for the chosen reference medium, 
given our initial $E(\omega)$ and $H(\omega)$ fields.
~
\begin{eqnarray}
  G^\pm
&=&
  \sqrt{\epsilon_r(\omega)} E(\omega) \pm \sqrt{\mu_r(\omega)} H(\omega)
\\
&=&
  \sqrt{\epsilon_r(\omega)} E(\omega) 
 \pm 
  \sqrt{\mu_r(\omega)} 
  \sqrt{\frac{\epsilon_i(\omega)}{\mu_i(\omega)}} 
  E(\omega)
\\
&=&
  \left[
    \sqrt{\epsilon_r(\omega)} 
   \pm 
    \sqrt{\mu_r(\omega)} 
    \sqrt{\frac{\epsilon_i(\omega)}{\mu_i(\omega)}} 
  \right]
  E(\omega)
.
\end{eqnarray}

These $G^\pm$ initial conditions are the best guess we can make for 
a ``+'' propagating pulse in our medium.

~

Note that in step (3) above, I have (in effect) used the definition of $G^\pm$ 
with a ``reference'' equal to the full dispersive properties
of the medium, and achieved a ``+'' propagating pulse by setting
$G^-$ to zero.  In the case where we intend to use these full 
dispersive properties as or simulation reference, of course step
(5) becomes trivial.

Because our reference $\epsilon_r, \mu_r$ will not (in general) include 
all the dispersion $\epsilon_i(\omega), \mu_i(\omega)$,
 our initial conditions will contain both a $G^+$ and a $G^-$ part.  
This initial $G^-$ part will co-propagate along with the $G^+$, 
the combination being a (hopefully) good match to the medium.

If our ``full'' dispersion $\epsilon_i(\omega), \mu_i(\omega)$ are not
a perfect match to the simulated medium, 
then there will be an initial pseudo-reflection.
This pseudo-reflection will consist of a reverse propagating $G^-$, 
whose size will be equivalent to that of the reflection from 
an interface between the dispersion used to construct the initial conditions
and the simulated medium (with $\epsilon_r+\epsilon_c, \mu_r+\mu_c$).
Mostly this kind of pseudo-reflection can be
eliminated with appropriate choice of $\epsilon_i(\omega), \mu_i(\omega)$; 
although nonlinear corrections will usually be impossible to treat, 
and so will still cause pseudo-reflections -- albiet hopefully very small ones.

For example, if we construct our initial conditions using a vacum 
$\epsilon_i(\omega)=\epsilon_0, \mu_i(\omega)=\mu_0$, but simulate a medium
with some refractive index $n$, this will be equivalent to simulating 
a $\epsilon_i, \mu_i \rightarrow \epsilon_r, \mu_r$ interface.  Such
an interface will cause a (pseudo) reflection, which will 
consist of a reverse propagating $G^-$, 
whose size will be equivalent to that of the reflection from a
vacuum/$n$ interface.  A mirror image of the pseudo-reflection will
also appear, and co-propagate with the $G^+$ pulse -- to ensure that the 
tramsmitted $E$ field has reduced appropriately for such an interface.

Note that the ``forward (co) propagating'' $G^-$ component 
still has a Poynting vector directed {\em backwards}.

% --------- 
\subsubsection{Comparing alternate references}
\label{Sss-1storderE-sims-cfaltref}

Here I do the simplest possible comparison between two different 
implementations of the same physical system; and see that (within
the limits set by the approximaions) that they are identical.
In the uncoupled stationary frame ($G^+$ only) case, the wave equation 
with only a reference $\alpha'_0$ term is 
~
\begin{eqnarray}
  \partial_z G_x^{\pm}
&=&
 \mp 
  \imath \omega 
  {\alpha'}_0^2
  \beta_r 
  ~
  G_x^+
\\
&=&
 \mp 
  \imath \omega 
  \left[
    \alpha_r^2
   +
    \alpha_r \alpha_L
  \right]^{1/2}
  \beta_r 
  ~
  G_x^+
\\
&=&
 \mp 
  \imath \omega 
  \alpha_r \beta_r 
  \left[
    1
   +
    \alpha_L / \alpha_r
  \right]^{1/2}
  ~
  G_x^+
\\
&\approx&
 \mp 
  \imath \omega 
  \alpha_r \beta_r 
  \left[
    1
   +
    \frac{1}
         {2}
    \alpha_L / \alpha_r
  \right]
  ~
  G_x^+
\\
&=&
 \mp 
  \imath \omega 
  \left[
    \alpha_r \beta_r 
   +
    \frac{1}
         {2}
    \alpha_L \beta_r
  \right]
  ~
  G_x^+
\\
&=&
 \mp 
  \imath \omega 
  \alpha_r \beta_r 
  ~
  G_x^+
 ~~
 \mp 
  \frac{\imath \omega \alpha_L \beta_r}
       {2}
  G_x^+
\end{eqnarray}

This is identical to the equivalent wave equation 
 with both a reference and correction term, 
 with the two sets of $\alpha$ parameters linked by 
 the appropriate relation 
 $\epsilon = {\alpha'}_0^2 = \alpha_r^2 + \alpha_r \alpha_L$.
The uncoupled approximation 
 (no $G^-$ i.e. $\left| G^-\right| \ll \left| G^+ \right|$) 
 and small correction  ($1 \gg \alpha_L / \alpha_r$) 
 are true in the same regime.

%\end{section}

\newpage
% ----------------------------------------------------------------------

\section{First order evolution equations (Primary $\vec{H}$)}\label{S-1storderH}

If we were more interested in $\vec{H}$ than $\vec{E}$, 
 we might instead define the field variables 
 as in  \ref{Ss-definitions-vectorH},
 using eqn (\ref{eqn-S-defs-Gpvector}, \ref{eqn-S-defs-EvectorGp}, 
           \ref{eqn-S-defs-HvectorGp}).
So
~
\begin{eqnarray}
  \vec{G}'^{\pm} (\omega)
&=&
  \beta_r (\omega)
  ~ \vec{H}  (\omega)
 \pm
   \vec{u} \times 
  ~ \alpha_r  (\omega)
  ~ \vec{E}  (\omega)
,
\\
\textrm{so} ~~~~ ~~~~
  \vec{u} \times \alpha_r \vec{E} 
&=&
  \frac{1}{2} \left[ \vec{G}'^+ - \vec{G}'^- \right]
\\
\textrm{and} ~~~~ ~~~~
  \beta_r \vec{H} 
&=& 
  \frac{1}{2} \left[ \vec{G}'^+ + \vec{G}'^- \right]
.
\end{eqnarray}

% --------- --------- ---------

\subsection{Vector stationary frame (Authoritative) }\label{Ss-1storderH-gstationaryvector}

Here I avoid calculations using the $x, y$, and $z$ components of the 
field, and retain a fully vector description.
Using the definitions in subsection \ref{S-medium}, I can 
proceed with the calculation --
~
\begin{eqnarray}
\nabla \times \vec{H} (t)
&=&
  +\partial_t \epsilon \ast \vec{E} (t)
  + \vec{J}
,
\nonumber
\\
\nabla \times \vec{E} (t)
&=&
  -\partial_t  \mu \ast \vec{H} (t)
\\
~
~
\textrm{into temporal-frequency space}
\longrightarrow
~~~~ ~~~~
\nabla \times \vec{H} (\omega)
&=&
  - \imath \omega 
    ~ \alpha (\omega) ^2  
    * \vec{E} (\omega)
  + \vec{J} (\omega)
,
\nonumber
\\
\nabla \times \vec{E} (\omega)
&=&
  + \imath \omega  
    ~ \beta (\omega)^2  
    * \vec{H} (\omega)
\\
~
~
\textrm{swap transverse components}
\longrightarrow
~~~~ ~~~~
\nabla \times \vec{H}
&=&
  - \imath \omega   
    ~\alpha^2 * \vec{E}
  + \vec{J}
,
\nonumber
\\
\vec{u} \times \left( \nabla \times \vec{E} \right)
&=&
   + \imath \omega   
    ~\beta^2  
    *
    \left(
     \vec{u} \times \vec{H}
    \right)
\\
~
~
\textrm{premultiply}
\longrightarrow
~~~~ ~~~~
\nabla \times \beta_r \vec{H}
&=&
  - \imath \omega   
    ~\beta_r \alpha^2   
    *
    \vec{E}
  + \beta_r \vec{J}
,
\nonumber
\\
\vec{u} \times \left( \nabla \times \alpha_r \vec{E}  \right)
&=&
   + \imath \omega   
    ~\alpha_r \beta^2   
    *
    \left(  
      \vec{u} \times \vec{H}
    \right)
\\
~
~
\textrm{sum-and-difference}
\longrightarrow
~~~~ ~~~~
  \nabla \times \beta_r \vec{H}
 ~~
 \pm
 ~~
  \vec{u} \times \left( \nabla \times \alpha_r \vec{E} \right)
&=&
 -
  \imath \omega   
    ~\beta_r \alpha^2   
    *
     \vec{E}
 ~~
 \pm
 ~~
  \imath \omega   
    ~\alpha_r \beta^2   
    *
    \left(  
      \vec{u} \times \vec{H}
    \right)
 ~~
 +
 ~~
  \beta_r \vec{J}
,
\end{eqnarray}

The vector $\vec{G}'^\pm$ fields (defined in eqn (\ref{eqn-S-defs-Gpvector}))
are ~
\begin{eqnarray}
  \vec{G}'^{\pm} 
&=&
  \beta_r \vec{H} \pm \vec{u} \times \alpha_r \vec{E} 
\end{eqnarray}

This means I need to convert both 
the second term on the LHS of the sum-and-difference equation above, 
as well as the RHS.  
It is most imortant for the LHS to be simple, 
because this will define the type of propagation specified by the RHS.  
To convert the LHS, we need two vector identities,
as used in subsection \ref{S-vectorid} to get 
eqn.(\ref{eqn-vectorid-combined}). So using
~
\begin{eqnarray}
  \vec{u} \times \left( \nabla \times \vec{H} \right)
 -
  \nabla \left( \vec{u} \cdot \vec{H} \right)
&=&
  \nabla \times \left( \vec{u} \times \vec{H} \right)
\end{eqnarray}

If I retain the $\vec{u} \cdot \alpha_r \vec{D}$ term, which 
will be zero for strictly transverse $D$ fields, then --
~
\begin{eqnarray}
  \nabla \times \beta_r \vec{H}
 ~~
 \pm
 ~~
  \vec{u} \times \left( \nabla \times \alpha_r \vec{E} \right)
&=&
 -
  \imath \omega \beta_r \alpha^2 
  * \vec{E}
 ~~
 \pm
 ~~
  \imath \omega \alpha_r \beta^2 * \left( \vec{u} \times \vec{H} \right)
 ~~
 +
 ~~
  \beta_r \vec{J}
,
\\
  \nabla \times \beta_r \vec{H}
 ~~
 \pm
 ~~
  \nabla \times \left( \vec{u} \times \alpha_r \vec{E} \right)
 \pm
  \nabla \left( \vec{u} \cdot \alpha_r \vec{E} \right)
&=&
 - 
  \imath \omega \beta_r \alpha^2 
  * \vec{E}
 ~~
 \pm
  \imath \omega \alpha_r \beta^2 
  *  \left( \vec{u} \times \vec{H} \right)
 ~~
 + 
 ~~
  \beta_r \vec{J}
\\
  \nabla \times 
  \left[
    \beta_r \vec{H} 
    ~~
    \pm
    ~~
    \left( \vec{u} \times \alpha_r \vec{E} \right)
  \right]
&=&
 - 
  \imath \omega \beta_r \alpha^2 
  * \vec{E}
 ~~
 \pm
  \imath \omega \alpha_r \beta^2 
  *  \left( \vec{u} \times \vec{H} \right)
 \mp
  \nabla \left( \vec{u} \cdot \alpha_r \vec{E} \right)
 ~~
 +
 ~~
  \beta_r \vec{J}
~~~~ ~~~~
\\
  \nabla \times \vec{G}'^{\pm}
&=&
 -
  \imath \omega 
  \left\{
    \beta_r \alpha^2 
    * \vec{E} 
   \mp
    \alpha_r \beta^2 * \left( \vec{u} \times \vec{H} \right)
  \right\}
 \mp
  \nabla \left( \vec{u} \cdot \alpha_r \vec{E} \right)
 ~~
 + 
 ~~
  \beta_r \vec{J}
\end{eqnarray}

Now note that eqn.(\ref{eqn-vectorid-doublecross-u}) means that
~
\begin{eqnarray}
     \vec{u} \times 
     \left[
       \vec{u} \times \vec{H}
     \right]
&=& 
  \left[ \vec{u} \cdot \vec{H} \right] \vec{u} 
 -
  \vec{H} 
\\
\textrm{and so} 
~~~~ ~~~~
  \vec{u} \times \vec{G}'^{\pm}
&=&
      \vec{u} \times \beta_r \vec{H} 
 \pm
  \vec{u} \times 
    \left[ 
  \vec{u} \times \alpha_r \vec{E}
    \right]
\\
&=&
  \vec{u} \times \beta_r \vec{H} 
 \mp
  \alpha_r \vec{E} 
 ~~
 \pm
  \left[ \vec{u} \cdot \alpha_r \vec{E} \right] \vec{u} 
\end{eqnarray}

So,
~
\begin{eqnarray}
  \nabla \times \vec{G}'^{\pm}
&=&
  -
  \imath \omega 
  \left\{
    \beta_r \alpha^2 
    *
    \left[
      \vec{u} \cdot \vec{E}
    \right]
    \vec{u}
   ~~
   -
    \beta_r \alpha^2 
    *
    \left(
      \vec{u} \times \left[ \vec{u} \times \vec{E} \right]
    \right)
   ~~
   \mp 
    \alpha_r \beta^2 
    *
    \left[ \vec{u} \times \vec{H}  \right]
  \right\}
 \mp
  \nabla \left( \vec{u} \cdot \alpha_r \vec{E} \right)
 ~~
 +
 ~~
  \beta_r \vec{J}
\\
&=&
 +
  \imath \omega 
  \left\{
    \beta_r \alpha^2 
    *
    \vec{u} \times \left[ \vec{u} \times \vec{E} \right]
   \pm
    \alpha_r \beta^2 
    *
    \left[ \vec{u} \times \vec{H} \right]
  \right\}
 ~~
 -
  \imath \omega 
    \beta_r \alpha^2 
    *
    \left[
      \vec{u} \cdot \vec{E}
    \right]
    \vec{u}
 ~~
 \mp
  \nabla \left( \vec{u} \cdot \alpha_r \vec{E} \right)
 ~~
 +
 ~~
  \beta_r \vec{J}
\end{eqnarray}

I now separate the interaction parts (depending on $\alpha_c$, $\beta_c$) 
from the reference parts (depending on $\alpha_r$, $\beta_r$), and then
substitute (as far as possible) expressions containing $G'^\pm$ rather than $E$
or $H$, 
by referring to eqns (\ref{eqn-S-defs-EvectorGp}) 
and (\ref{eqn-S-defs-HvectorGp}).  
Hence 
~
\begin{eqnarray}
  \nabla \times \vec{G}'^{\pm}
&=&
 +
  \imath \omega 
  \alpha_r \beta_r 
  \left\{
    \vec{u} \times \left[ \vec{u} \times \alpha_r \vec{E} \right]
   \pm
    \left[ \vec{u} \times \beta_r \vec{H} \right]
  \right\}
 ~~
 -
  \imath \omega 
    \alpha_r \beta_r
    \left[
      \vec{u} \cdot \alpha_r \vec{E}
    \right]
    \vec{u}
 ~~
 \mp
  \nabla \left( \vec{u} \cdot \alpha_r \vec{E} \right)
\nonumber 
\\
&&
 ~~~~
 +
  \imath \omega 
  \beta_r  \alpha_c
  *
  \left(
    \vec{u} \times \left[ \vec{u} \times \alpha_r \vec{E} \right]
 \right)
 ~~
   \pm
  \imath \omega 
  \alpha_r \beta_c 
  *
    \left[ \vec{u} \times \beta_r \vec{H} \right]
 ~~
 -
  \imath \omega 
    \beta_r \alpha_c
    *
    \left[
      \vec{u} \cdot \alpha_r \vec{E}
    \right]
    \vec{u}
 ~~
 + 
  \beta_r \vec{J}
\label{eqn-1storderH-vectorstat}
\\
&=&
 \pm
  \imath \omega 
  \alpha_r \beta_r 
  \vec{u} \times \vec{G}'^\pm
 ~~
 -
  \imath \omega 
    \alpha_r \beta_r
    \vec{u} ~
    {G'}^\circ
 ~~
 \mp
  \nabla {G'}^\circ
\nonumber 
\\
&&
 ~~~~
 +
  \frac{\imath \omega \beta_r }
       {2}
    \alpha_c
   *
   \left(
    \vec{u} \times 
    \left[ \vec{G}'^+ + \vec{G}'^- \right]
   \right)
 ~~
   \pm
  \frac{\imath \omega \alpha_r }
       {2}
    \beta_c
   *
   \left(
    \vec{u} \times 
    \left[ \vec{G}'^+ - \vec{G}'^- \right]
   \right)
 ~~
 -
  \imath \omega 
    \beta_r \alpha_c
   *
   \left(
    \vec{u} ~
    {G'}^\circ
   \right)
 ~~
 + 
  \beta_r \vec{J}
.
~~~~
\label{eqn-1storderH-statframe}
\end{eqnarray}

Here the transverse (TEF) and longitudinal parts decouple, 
 since $\vec{G}'^\pm$ is guaranteed electrically transverse, 
 and $\vec{u} {G}'^{\circ}$ is guaranteed electrically longitudinal.
In any case, the two decoupled equations are --
~
\begin{eqnarray}
  \nabla \times \vec{G}'^{\pm}
&=&
 \pm
  \imath \omega 
  \alpha_r \beta_r 
  ~
  \vec{u} \times \vec{G}'^\pm
  ~~
 +
  \frac{\imath \omega \beta_r}
       {2}
    \alpha_c
   *
   \left(
    \vec{u} \times 
    \left[ \vec{G}'^+ + \vec{G}'^- \right]
   \right)
 ~~
 \pm
  \frac{\imath \omega \alpha_r}
       {2}
    \beta_c
   *
   \left(
    \vec{u} \times \left[ \vec{G}'^+ - \vec{G}'^- \right]
   \right)
 ~~
 +
 ~~
  \beta_r \vec{J}
~~~~ ~~~~
~~~~ ~~~~
\label{eqn-1storderH-statframe-transverse}
\\
 \pm
  \nabla {G'}^{\circ}
&=&
 -
  \imath \omega 
    \alpha_r \beta_r
    \vec{u} ~ {G'}^{\circ}
 ~~
 -
  \imath \omega 
    \alpha_r \beta_c
   *
   \left(
    \vec{u} ~ {G'}^{\circ} 
   \right)
.
\label{eqn-1storderH-statframe-longitudinal}
\end{eqnarray}

We see from the following \ref{Sss-1storderH-longitudinal} that these
 two equations are sufficient to describe the fields as long as there
 is no longitudinal $\vec{H}$ component (i.e. $\vec{G}'^\pm$ is transverse);
 (also with reference to the divergence calculation in 
 \ref{Sss-definitions-vectorH-divergence}).

% --------- 
\subsubsection{Longitudinal Equation}\label{Sss-1storderH-longitudinal}

Taking the $\vec{u} \times$ of the $\nabla \times \vec{E}$ equation
 removed a part of the field dynamics lying in the direction
 of the propagation vector $\vec{u}$.   
To rectify this omission, 
 we take the dot product with $\vec{u}$.
Using the standard vector identity eqn.(\ref{eqn-vectorid-I-7}):
~
\begin{eqnarray}
\nabla \cdot \left(
    \vec{A}  \times \vec{B} \right) 
&=&
  \vec{A} 
  \cdot 
  \left(
    \nabla  \times \vec{B} 
  \right)
 -
  \vec{B} 
  \cdot 
  \left(
    \nabla  \times \vec{A} 
  \right)
,
\end{eqnarray}
and using $\nabla  \times \vec{u} = 0$; we get 
~
\begin{eqnarray}
  \vec{u} 
  \cdot
  \left(
    \nabla \times \vec{E}
  \right)
&=&
  -
  \imath \omega
  \beta^2
 *
 \left(
  \vec{u} 
  \cdot
  \vec{H} 
 \right)
\\
\textrm{vector identity:} ~~~~ ~~~~
  \nabla 
  \cdot
  \left(
    \vec{u}  \times \vec{E}
  \right)
&=&
  -
  \imath \omega
  \beta^2
 *
 \left(
  \vec{u} 
  \cdot
  \vec{H} 
 \right)
\\
\textrm{premultiply:} ~~~~ ~~~~
  \beta_r
  \nabla 
  \cdot
  \left(
    \vec{u} \times \alpha_r \vec{E} 
  \right)
&=&
  -
  \imath \omega
  \alpha_r \beta^2
 *
 \left(
  \vec{u} 
  \cdot
  \beta_r \vec{H} 
 \right)
\\
  \alpha_r 
  \nabla 
  \cdot
  \left(
    \vec{G}'^{+} - \vec{G}'^{-} 
  \right)
&=&
  -
  \imath \omega
  \alpha_r \beta^2
 *
 \left(
  \vec{u} 
  \cdot
  \left[
    \vec{G}'^{+} + \vec{G}'^{-} 
  \right]
 \right)
\end{eqnarray}

This is the same as the divergence-difference equation 
 seen in \ref{Sss-definitions-vectorH-divergence}; 
 where we see that both parts of the LHS are zero
 if  $ \vec{u} 
  \cdot
  \left(
    \vec{G}'^{+} + \vec{G}'^{-} 
  \right) = 0$.

% --------- 
\subsubsection{Simple Cases}\label{Sss-1storderH-simple}

In the non-magnetic ($\beta_c=0$) case, 
~
\begin{eqnarray}
  \nabla \times \vec{G}'^{\pm}
&=&
 \pm
  \imath \omega 
  \alpha_r \beta_r 
  ~
  \vec{u} \times \vec{G}'^\pm
  ~~
 +
  \frac{\imath \omega \beta_r}
       {2}
    \alpha_c
   *
   \left(
    \vec{u} \times 
    \left[ \vec{G}'^+ + \vec{G}'^- \right]
   \right)
 ~~
 +
 ~~
  \beta_r \vec{J}
~~~~ ~~~~
~~~~ ~~~~
\label{eqn-1storderH-statframe-nobetaX-transverse}
\\
 \pm
  \nabla {G'}^{\circ}
&=&
 -
  \imath \omega 
    \alpha_r \beta_r
    \vec{u} ~ {G'}^{\circ}
.
\label{eqn-1storderH-statframe-nobetaX-longitudinal}
\end{eqnarray}

In the time domain, eqn.(\ref{eqn-1storderH-statframe-nobetaX-transverse}) is 
~
\begin{eqnarray}
  \nabla \times \vec{G}'^{\pm}
&=&
 \mp
  \partial_t
  \left[
    \left(
      \alpha_r \ast \beta_r 
    \right)
    ~\ast~
    \left(
      \vec{u} \times \vec{G}'^\pm
    \right)
  \right]
%\nonumber
%\\
%&&
 ~
 -
  \partial_t
  \left[
    \left(
      \frac{\beta_r}
           {2}
    \right)
   *
    \alpha_c
   *
    \left(
      \vec{u} \times 
      \left[ \vec{G}^+ + \vec{G}^- \right]
    \right)
  \right]
 \pm 
 ~~
  \vec{u} \times  \beta_r \ast \vec{J}
.
~~~~
\label{eqn-1storderH-propagation-approx-time}
\end{eqnarray}

\noindent{\bf Plane Polarized \& Magnetic ($G_x^\pm  \sim E_x \pm H_y$):} 

The $\hat{y}$ components of the LHS and RHS contain the $G_x$ 
values -- as a result of the curl and cross-products respectively.
Compare the previous stand-alone calculation in 
eqn.(\ref{eqn-stationaryGpropagation}) to 
~
\begin{eqnarray}
  \partial_z {G_x'}^{\pm}
&=&
 \mp 
  \imath \omega 
  \alpha_r \beta_r 
  ~
  {G_x'}^\pm
 ~~
 \mp 
  \frac{\imath \omega \beta_r}
       {2}
  \alpha_c
 *
  \left[ {G_x'}^+ + {G_x'}^- \right]
 ~~
 -
  \frac{\imath \omega \alpha_r}
       {2}
  \beta_c
 *
  \left[ {G_x'} ^+ - {G_x'}^- \right]
\end{eqnarray}

\noindent{\bf Vector Moving Frame, Transverse, Magnetic:} 

It is worth transforming to the moving frame at this end point,
 rather than at the beginning.  
Since eqn.(\ref{eqn-vector-frametranslation}) tells us that 
 $\nabla \vec{Q} = 
 \nabla' \vec{Q}- \alpha_r \beta_r \vec{u} \times \partial_t' \vec{Q}$, with
 $\xi = \alpha_r \beta_r / \alpha_r \beta_r$,
 we get 
~
\begin{eqnarray}
  \nabla' \times \vec{G}'^{\pm}
  ~~
 +
  \imath \omega \alpha_r \beta_r
  ~
  \vec{u} \times \vec{G}'^{\pm}
&=&
 \pm
  \imath \omega 
  \alpha_r \beta_r 
  ~
  \vec{u} \times \vec{G}^\pm
 ~~~~
 +
  \frac{\imath \omega \beta_r}
       {2}
    \alpha_c
   *
   \left(
    \vec{u} \times 
    \left[ \vec{G}'^+ + \vec{G}'^- \right]
   \right)
 ~~
 \mp
  \frac{\imath \omega \alpha_r}
       {2}
   \beta_c
   *
   \left(
    \vec{u} \times \left[ \vec{G}'^+ - \vec{G}'^- \right]
   \right)
 ~~
~~~~ 
~~~~
\\
  \nabla' \times \vec{G}^{\pm}
&=&
 \pm
  \imath \omega 
  \alpha_r \beta_r 
  \left( 1 \mp \xi \right)
  ~
  \vec{u} \times \vec{G}'^\pm
 ~~~~
 + 
  \frac{\imath \omega \beta_r}
       {2}
   \alpha_c
   *
   \left(
    \vec{u} \times 
    \left[ \vec{G}'^+ + \vec{G}'^- \right]
   \right)
 ~~
 \mp
  \frac{\imath \omega \alpha_r}
       {2}
    \beta_c
   *
   \left(
    \vec{u} \times \left[ \vec{G}'^+ - \vec{G}'^- \right]
   \right)
~~~~ 
~~~~
\end{eqnarray}

\noindent{\bf Plane Polarized Moving Frame, Magnetic:} 

This is just a special case of the above vector moving frame equation --
~
\begin{eqnarray}
  \partial_{z'} {G_x'}^{\pm}
&=&
 \pm
  \imath \omega 
  \alpha_r \beta_r 
  \left( 1 \mp \xi \right)
  ~
  G_x^{\pm}
 ~~
 +
  \frac{\imath \omega \beta_r}
       {2}
   \alpha_c
  *
    \left[ {G_x'}^+ + {G_x'}^- \right]
 ~~
 \mp
  \frac{\imath \omega \alpha_r}
       {2}
    \beta_c
   *
    \left[ {G_x'}^+ - {G_x'}^- \right]
\end{eqnarray}

\newpage

% ----------------------------------------------------------------------

\section{First order evolution equations (primary $\vec{D}$)}\label{S-1storderD}

Follows the derivation in the previous section, 
 but uses the ``Primary $\vec{D}$'' (or $\vec{F}^\pm$) form, 
 as defined in \ref{Ss-definitions-vector},
 with
 eqn. (\ref{eqn-S-defs-Fvector}, \ref{eqn-S-defs-EvectorF}, 
     \ref{eqn-S-defs-BvectorF}). 
So (dropping the tildes out of laziness)
~
\begin{eqnarray}
  \vec{F}^{\pm} (\omega)
&=&
  \alpha_r^{-1}(\omega)
  ~ \vec{D} (\omega)
 \pm 
  \vec{u} \times 
  ~ \beta_r^{-1}(\omega)
  ~ \vec{B} (\omega)
,
\\
\textrm{so} ~~~~ ~~~~
  \alpha_r^{-1}
  ~ \vec{D} 
&=&
  \frac{1}{2} 
  \left[ \vec{F}^+ + \vec{F}^- \right]
\\
\textrm{and} ~~~~ ~~~~
  \vec{u} \times 
  ~ \beta_r^{-1}
  ~ \vec{B} 
&=& 
  \frac{1}{2} \left[ \vec{F}^+ - \vec{F}^- \right]
.
\end{eqnarray}

We would do this so that the displacement field (and magnetic fields) 
 could be easily reconstructed from $ \vec{F}^\pm$; 
 and, in analogy to the THF and TEF mentioned above, 
 we would use a transverse magnetic induction $B$ approximation (TBF).
This this form differs from the primary $E$ form 
 that defines $G^{\pm}$ in that it contains
 $\alpha_r^{-1} D =  \alpha_r^{-1} \alpha^2 E$ not $\alpha_r E$; 
 i.e. the electric-like contribution differs only by 
 $\epsilon/\epsilon_r = \alpha^2/\alpha_r^2$.
A similar comparison holds for the magnetic components.

% --------- --------- ---------

\subsection{Vector stationary frame (Authoritative) }\label{Ss-1storderD-gstationaryvector}

Here I avoid calculations using the $x, y$, and $z$ components of the 
field, and retain a fully vector description.
Using the definitions in subsection \ref{S-medium}, I can 
proceed with the calculation --
~
\begin{eqnarray}
\nabla \times \vec{H} (t)
&=&
  +\partial_t \vec{D} (t)
  + \vec{J}
,
\nonumber
\\
\nabla \times \vec{E} (t)
&=&
  -\partial_t  \vec{B} (t)
\\
~
~
\textrm{into temporal-frequency space}
\longrightarrow
~~~~ ~~~~
\nabla \times \vec{H} (\omega)
&=&
  - \imath \omega 
    ~ \vec{D} (\omega)
  + \vec{J} (\omega)
,
\nonumber
\\
\nabla \times \vec{E} (\omega)
&=&
  + \imath \omega  
    ~ \vec{B} (\omega)
\\
~
~
\textrm{swap transverse components}
\longrightarrow
~~~~ ~~~~
\vec{u} \times \left( \nabla \times \vec{H} \right)
&=&
  - \imath \omega 
    \vec{u}   
    ~\times \vec{D}
  + \vec{u} \times \vec{J}
,
\nonumber
\\
\nabla \times \vec{E}
&=&
   + \imath \omega   
    ~\vec{B}
\\
~
~
\textrm{premultiply}
\longrightarrow
~~~~ ~~~~
\vec{u} \times \left( \nabla \times \alpha_r \beta^2 * \vec{H} \right)
&=&
  - \imath \omega   
    ~\alpha_r  \beta^2 
    *
    \left(
     \vec{u} \times \vec{D}
    \right)
  + \vec{u} \times \alpha_r  \beta^2  * \vec{J}
,
\nonumber
\\
\nabla \times \beta_r \alpha^2 * \vec{E}
&=&
   + \imath \omega   
    ~\beta_r \alpha^2 
    *
    \vec{B}
\\
~
~
\textrm{simplify}
\longrightarrow
~~~~ ~~~~
\vec{u} \times \left( \nabla \times \alpha_r \vec{B} \right)
&=&
  - \imath \omega   
    ~\alpha_r  \beta^2 
    *
    \left(
      \vec{u} \times \vec{D}
    \right)
  + \vec{u} \times \alpha_r  \beta^2 * \vec{J}
,
\nonumber
\\
\nabla \times \beta_r \vec{D}
&=&
   + \imath \omega   
    ~\beta_r \alpha^2 
    *
     \vec{B}
\\
~
~
\textrm{sum-and-difference}
\longrightarrow
~~~~ ~~~~
  \nabla \times \beta_r \vec{D}
 ~~
 \pm
 ~~
  \vec{u} \times \left( \nabla \times \alpha_r \vec{B} \right)
&=&
 +
  \imath \omega   
    ~\beta_r \alpha^2 
    *
     \vec{B}
 ~~
 \mp 
 ~~
  \imath \omega   
    ~\alpha_r  \beta^2  
    *
    \left(
      \vec{u} \times \vec{D}
    \right)
 ~~
\nonumber
\\
&&
~~~~ 
~~~~
 \pm 
 ~~
  \vec{u} \times \alpha_r  \beta^2 * \vec{J}
,
\end{eqnarray}

The vector $\vec{F}^\pm$ fields (defined in eqn (\ref{eqn-S-defs-Fvector}))
are ~
\begin{eqnarray}
  \vec{F}^{\pm} 
&=&
  \alpha_r^{-1} \vec{D} \pm \vec{u} \times \beta_r^{-1} \vec{B}
\end{eqnarray}

This means I need to convert both 
the second term on the LHS of the sum-and-difference equation above, 
as well as the RHS.  
It is most imortant for the LHS to be simple, 
because this will define the type of propagation specified by the RHS.  
To convert the LHS, we need two vector identities,
as used in subsection \ref{S-vectorid} to get 
eqn.(\ref{eqn-vectorid-combined}). So using
~
\begin{eqnarray}
  \vec{u} \times \left( \nabla \times \vec{H} \right)
 -
  \nabla \left( \vec{u} \cdot \vec{H} \right)
&=&
  \nabla \times \left( \vec{u} \times \vec{H} \right)
\end{eqnarray}

If I retain the $\vec{u} \times \beta_r \vec{H}$ term in the above, which 
will be zero for strictly transverse fields, then --
~
\begin{eqnarray}
  \nabla \times \beta_r \vec{D}
 ~~
 \pm
 ~~
  \vec{u} \times \left( \nabla \times \alpha_r \vec{B} \right)
&=&
 +
  \imath \omega   
    ~\beta_r \alpha^2 
    *
     \vec{B}
 ~~
 \mp 
 ~~
  \imath \omega   
    ~\alpha_r  \beta^2 
    *
    \left( \vec{u} \times \vec{D} \right)
 ~~
 \pm 
 ~~
  \vec{u} \times \alpha_r  \beta^2 * \vec{J}
\\
  \nabla \times \beta_r \vec{D}
 ~~
 \pm
 ~~
  \nabla \times \left( \vec{u} \times \alpha_r \vec{B} \right)
 \pm
  \nabla \left( \vec{u} \cdot \alpha_r \vec{B} \right)
&=&
 +
  \imath \omega   
    ~\beta_r \alpha^2 
    *
    \vec{B}
 ~~
 \mp 
 ~~
  \imath \omega   
    ~\alpha_r  \beta^2 
    *
    \left( \vec{u} \times \vec{D} \right)
 ~~
 \pm 
 ~~
  \vec{u} \times \alpha_r  \beta^2 * \vec{J}
\\
  \nabla \times 
  \left[
    \beta_r \vec{D}
    ~~
    \pm
    ~~
    \left( \vec{u} \times \alpha_r \vec{B} \right)
  \right]
&=&
 +
  \imath \omega   
    ~\beta_r \alpha^2 
    *
    \vec{B}
 ~~
 \mp 
 ~~
  \imath \omega   
    ~\alpha_r  \beta^2 
    *
    \left( \vec{u} \times \vec{D} \right)
 ~~
 \mp
  \nabla \left( \vec{u} \cdot \alpha_r \vec{B} \right)
\nonumber
\\
&&
~~~~ 
~~~~ 
 ~~
 \pm 
  \vec{u} \times \alpha_r  \beta^2 * \vec{J}
~~~~ ~~~~
\\
  \nabla \times \vec{F}^{\pm}
&=&
  \imath \omega 
  \left\{
    \beta_r \alpha^2 * \vec{B}
   ~~
   \mp 
   ~~
    \alpha_r \beta^2 
    * 
    \left( \vec{u} \times \vec{D} \right)
  \right\}
 ~~
 \mp
  \nabla \left( \vec{u} \cdot \alpha_r \vec{B} \right)
 ~~
 \pm 
  \vec{u} \times \alpha_r  \beta^2 * \vec{J}
~~~~ ~~~~
\end{eqnarray}

Now note that eqn.(\ref{eqn-vectorid-doublecross-u}) means that
~
\begin{eqnarray}
     \vec{u} \times 
     \left[
       \vec{u} \times \vec{H}
     \right]
&=& 
  \left[ \vec{u} \cdot \vec{H} \right] \vec{u} 
 -
  \vec{H} 
\\
\textrm{and so} 
~~~~ ~~~~
  \left(
    \alpha_r \beta_r
  \right)
  ~
  \vec{u} \times \vec{F}^{\pm}
&=&
  \vec{u} \times \beta_r \vec{D}
 \pm
  \vec{u} \times 
    \left[ 
      \vec{u} \times \alpha_r \vec{B} 
    \right]
\\
&=&
  \vec{u} \times \beta_r \vec{D}
 \mp
  \alpha_r \vec{B} 
 ~~
 \pm
  \left[ \vec{u} \cdot \alpha_r \vec{B} \right] \vec{u} 
\end{eqnarray}

So,
~
\begin{eqnarray}
  \left(
    \alpha_r \beta_r
  \right)
  ~
  \nabla \times \vec{F}^{\pm}
&=&
  \imath \omega 
  \left\{
    \beta_r \alpha^2 
    *
    \left[
      \vec{u} \cdot \vec{B}
    \right]
    \vec{u}
   ~~
   -
    \beta_r \alpha^2 
    *
    \vec{u} \times \left[ \vec{u} \times \vec{B} \right]
   ~~
   \mp 
    \alpha_r \beta^2 
    *
    \left(
      \vec{u} \times \vec{D} 
    \right)
  \right\}
 ~~
 \mp
  \nabla \left( \vec{u} \cdot \alpha_r \vec{B} \right)
\nonumber
\\
&&
~~~~ 
~~~~ 
 ~~
 \pm 
 ~~
  \vec{u} \times \alpha_r  \beta^2 * \vec{J}
\\
&=&
 \mp 
  \imath \omega 
  \left\{
    \alpha_r \beta^2 
    ~
    \vec{u} \times \vec{D}
   \pm
    \beta_r \alpha^2 
    \vec{u} \times \left[ \vec{u} \times \vec{B} \right]
  \right\}
 ~~
 +
  \imath \omega 
     \beta_r \alpha^2 
    \left[
      \vec{u} \cdot \vec{B}
    \right]
    \vec{u}
 ~~
 \mp
  \nabla \left( \vec{u} \cdot \alpha_r \vec{B} \right)
\nonumber
\\
&&
~~~~ 
~~~~ 
 ~~
 \pm 
 ~~
  \vec{u} \times \alpha_r  \beta^2 * \vec{J}
\end{eqnarray}

I now separate the interaction parts (depending on $\alpha_c$, $\beta_c$) 
 from the reference parts (depending on $\alpha_r$, $\beta_r$), 
 and then substitute (as far as possible) expressions containing $F$ 
 rather than $D$ or $B$, 
 by referring to eqns (\ref{eqn-S-defs-EvectorF})
 and (\ref{eqn-S-defs-BvectorF}).  
Hence 
~
\begin{eqnarray}
  \nabla \times \vec{F}^{\pm}
&=&
 \mp 
  \imath \omega 
  \alpha_r \beta_r 
  \left\{
    \vec{u} \times \alpha_r^{-1} \vec{D}
   \pm
    \vec{u} \times \left[ \vec{u} \times \beta_r^{-1} \vec{B} \right]
  \right\}
 ~~
 +
  \imath \omega 
    \alpha_r \beta_r
    \left[
      \vec{u} \cdot \beta_r^{-1} \vec{B}
    \right]
    \vec{u}
 ~~
 \mp
  \nabla \left( \vec{u} \cdot \beta_r^{-1} \vec{B} \right)
\nonumber 
\\
&&
 ~~~~
 \mp 
  \imath \omega 
  \alpha_r \beta_c 
  *
  \left(
    \vec{u} \times \alpha_r^{-1} \vec{D}
  \right)
 -
  \imath \omega 
  \beta_r \alpha_c
  *
  \left(
    \vec{u} \times \left[ \vec{u} \times \beta_r^{-1} \vec{B} \right]
  \right)
 ~~
 +
  \imath \omega 
   \beta_r  \alpha_c
   *
    \left[
      \vec{u} \cdot \beta_r^{-1} \vec{B}
    \right]
    \vec{u}
\nonumber
\\
&&
~~~~
~~~~
 \pm 
 ~~
  \vec{u} \times  \beta_r^{-1} 
  \left( \beta_r^2 + \beta_r \beta_c \right)
  * \vec{J}
\label{eqn-1storderD-vectorstat}
\\
&=&
 \mp 
  \imath \omega 
  \alpha_r \beta_r 
  ~
  \vec{u} \times \vec{F}^\pm
 ~~
 +
  \imath \omega 
    \alpha_r \beta_r
    \vec{u} ~ 
    {F}^{\circ}
 ~~
 \mp
  \nabla {F}^{\circ}
\nonumber 
\\
&&
 ~~~~
 \mp 
  \frac{\imath \omega \alpha_r}
       {2}
    \beta_c
  *
  \left(
    \vec{u} \times 
    \left[ \vec{F}^+ + \vec{F}^- \right]
  \right)
 ~~
 -
  \frac{\imath \omega \beta_r}
       {2}
    \alpha_c
  *
  \left(
    \vec{u} \times \left[ \vec{F}^+ - \vec{F}^- \right]
  \right)
 ~~
 +
  \imath \omega 
    \beta_r \alpha_c
  *
  \left(
    \vec{u} ~ 
    {F}^{\circ}
  \right)
\nonumber
\\
&&
~~~~
~~~~
 \pm 
 ~~
  \vec{u} \times  
  \left( \beta_r + \beta_c \right)
  * \vec{J}
\label{eqn-1storderD-statframe}
\end{eqnarray}

Here the transverse (TDF) and longitudinal parts decouple, 
 since $\vec{F}^\pm$ is guaranteed displacment transverse, 
 and $\vec{u} {F}^{\circ}$ is displacment longitudinal.
Note that in this {\em  transverse/ longitudinal split}, 
 the role of the current $\vec{J}$ has not been checked.
In any case, the two decoupled equations are --
~
\begin{eqnarray}
  \nabla \times \vec{F}^{\pm}
&=&
 \mp 
  \imath \omega 
  \alpha_r \beta_r 
  ~
  \vec{u} \times \vec{F}^\pm
 ~~
 \mp 
  \frac{\imath \omega \alpha_r}
       {2}
    \beta_c
   *
   \left(
    \vec{u} \times 
    \left[ \vec{F}^+ + \vec{F}^- \right] 
  \right)
 ~~
 -
  \frac{\imath \omega \beta_r}
       {2}
    \alpha_c
   *
   \left(
    \vec{u} \times \left[ \vec{F}^+ - \vec{F}^- \right]
  \right)
 ~~
 \pm 
 ~~
  \vec{u} \times  
  \left( \beta_r + \beta_c \right)
  * \vec{J}
~~~~ ~~~~
~~~~ ~~~~
\label{eqn-1storderD-statframe-transverse}
\\
 \pm
  \nabla {F}^{\circ}
&=&
 +
  \imath \omega 
    \alpha_r \beta_r
    \vec{u} ~ 
    {F}^{\circ}
 ~~
 +
  \imath \omega 
    \beta_r \alpha_c
   *
   \left(
    \vec{u} ~ 
    {F}^{\circ}
  \right)
.
\label{eqn-1storderD-statframe-longitudinal}
\end{eqnarray}

We see from the following \ref{Sss-1storderD-longitudinal} that these
 two equations are sufficient to describe the fields as long as there
 is no longitudinal $\vec{D}$ component (i.e. $\vec{F}^\pm$ is transverse)
 or longitudinal $\vec{J}$
 (also with reference to the divergence calculation in 
 \ref{Sss-definitions-vectorD-divergence}).

% --------- 
\subsubsection{Longitudinal Equation}\label{Sss-1storderD-longitudinal}

Taking the $\vec{u} \times$ of the $\nabla \times \vec{H}$ equation
 removed a part of the field dynamics lying in the direction
 of the propagation vector $\vec{u}$.   
This part also include the response to longitudinal currents.
To rectify this omission, 
 we take the dot product with $\vec{u}$.
Using the standard vector identity eqn.(\ref{eqn-vectorid-I-7}):
~
\begin{eqnarray}
\nabla \cdot \left(
    \vec{A}  \times \vec{B} \right) 
&=&
  \vec{A} 
  \cdot 
  \left(
    \nabla  \times \vec{B} 
  \right)
 -
  \vec{B} 
  \cdot 
  \left(
    \nabla  \times \vec{A} 
  \right)
,
\end{eqnarray}
and using $\nabla  \times \vec{u} = 0$; we get
~
\begin{eqnarray}
  \vec{u} 
  \cdot
  \left(
    \nabla \times \vec{H}
  \right)
&=&
  -
  \imath \omega
  \vec{u} 
  \cdot
  \vec{D} 
 +
  \vec{u} 
  \cdot
  \vec{J} 
\\
\textrm{vector identity:} ~~~~ ~~~~
  \nabla 
  \cdot
  \left(
    \vec{u}  \times \vec{H}
  \right)
&=&
  -
  \imath \omega
  \vec{u} 
  \cdot
  \vec{D} 
 +
  \vec{u} 
  \cdot
  \vec{J} 
\\
\textrm{premultiply:} ~~~~ ~~~~
  \alpha_r^{-1}
  \nabla 
  \cdot
  \left(
    \vec{u} \times \beta_r^{-1} \vec{B}
  \right)
&=&
  -
  \imath \omega
  \beta_r^{-1} \beta^2
  *
  \left(
    \vec{u} 
    \cdot
    \alpha_r^{-1} \vec{E} 
  \right)
 +
  \alpha_r^{-1} \beta_r^{-1} \beta^2
  *
  \left(
    \vec{u} 
    \cdot
    \vec{J} 
  \right)
\\
  \nabla 
  \cdot
  \left(
    \vec{F}^{+} - \vec{F}^{-} 
  \right)
&=&
  -
  \imath \omega
  \alpha_r 
  \left( \beta_r + \beta_c \right)
  *
  \left(
    \vec{u} 
    \cdot
    \left[
      \vec{F}^{+} + \vec{F}^{-} 
    \right]
  \right)
 +
  \left( \beta_r + \beta_c \right)
  *
  \left(
    \vec{u} 
    \cdot
    \vec{J} 
  \right)
\label{eqn-1storderD-statframe-divergence}
\end{eqnarray}

This is the same as the divergence-difference equation 
 seen in \ref{Sss-definitions-vectorD-divergence}; 
 where we see that both parts of the LHS are zero
 if  $ \vec{u} 
  \cdot
  \left(
    \vec{F}^{+} + \vec{F}^{-} 
  \right) = 0$
and $\vec{u} \cdot \vec{J} = 0$.

% --------- 
\subsubsection{Simple Cases}\label{Sss-1storderD-simple}

In the non-magnetic ($\beta_c=0$) case, 
~
\begin{eqnarray}
  \nabla \times \vec{F}^{\pm}
&=&
 \mp 
  \imath \omega 
  \alpha_r \beta_r 
  ~
  \vec{u} \times \vec{F}^\pm
 ~~
 -
  \frac{\imath \omega \beta_r}
       {2}
    \alpha_c
   *
   \left(
    \vec{u} \times \left[ \vec{F}^+ - \vec{F}^- \right]
   \right)
 ~~
 \pm 
 ~~
  \vec{u} \times  \left(\beta_r + \beta_c \right) * \vec{J}
~~~~ ~~~~
~~~~ ~~~~
\label{eqn-1storderD-statframe-nobetaX-transverse}
\\
 \pm
  \nabla {F}^{\circ}
&=&
 +
  \imath \omega 
    \alpha_r \beta_r
    \vec{u} ~ 
    {F}^{\circ}
 ~~
 +
  \imath \omega 
    \beta_r \alpha_c
   *
   \left(
    \vec{u} ~ 
    {F}^{\circ}
   \right)
.
\label{eqn-1storderD-statframe-nobetaX-longitudinal}
\end{eqnarray}

In the time domain, eqn.(\ref{eqn-1storderD-statframe-nobetaX-transverse}) is 
~
\begin{eqnarray}
  \nabla \times \vec{F}^{\pm}
&=&
 \pm
  \partial_t
  \left[
    \left(
      \alpha_r \ast \beta_r 
    \right)
    ~\ast~
    \left(
      \vec{u} \times \vec{F}^\pm
    \right)
  \right]
%\nonumber
%\\
%&&
 ~
 +
  \partial_t
  \left[
    \left(
      \frac{\beta_r}
           {2}
    \right)
    *
    \alpha_c 
    *
    \left(
      \vec{u} \times 
      \left[ \vec{F}^+ - \vec{F}^- \right]
    \right)
  \right]
 \pm 
 ~~
  \vec{u} \times 
  \left(\beta_r + \beta_c \right)  \ast \vec{J}
.
~~~~
\label{eqn-1storderD-propagation-approx-time}
\end{eqnarray}

\noindent{\bf Plane Polarized \& Magnetic ($F_x^\pm  \sim _x \pm B_y$):} 

The $\vec{y}$ components of the LHS and RHS contain the $F_x$ 
values -- as a result of the curl and cross-products respectively.
Compare the previous stand-alone calculation in 
eqn.(\ref{eqn-stationaryGpropagation}) to 
~
\begin{eqnarray}
  \partial_z F_x^{\pm}
&=&
 \mp 
  \imath \omega 
  \alpha_r \beta_r 
  ~
  F_x^\pm
 ~~
 - 
  \frac{\imath \omega \beta_r}
       {2}
  \alpha_c
  *
  \left[ F_x^+ - F_x^- \right]
 ~~
 \mp
  \frac{\imath \omega \alpha_r}
       {2}
  \beta_c
  *
  \left[ F_x ^+ + F_x^- \right]
\end{eqnarray}

\noindent{\bf Vector Moving Frame, Transverse, Magnetic:} 

It is worth transforming to the moving frame at this end point,
 rather than at the beginning.  
Since eqn.(\ref{eqn-vector-frametranslation}) tells us that 
 $\nabla \vec{Q} = 
 \nabla' \vec{Q}- \alpha_r \beta_r \vec{u} \times \partial_t' \vec{Q}$, with
 $\xi = \alpha_r \beta_r / \alpha_r \beta_r$,
 we get 
~
\begin{eqnarray}
  \nabla' \times \vec{F}^{\pm}
  ~~
 +
  \imath \omega \alpha_r \beta_r
  ~
  \vec{u} \times \vec{F}^{\pm}
&=&
 \mp 
  \imath \omega 
  \alpha_r \beta_r 
  ~
  \vec{u} \times \vec{F}^\pm
 ~~~~
 \mp 
  \frac{\imath \omega \alpha_r}
       {2}
   \beta_c
   *
   \left(
    \vec{u} \times 
    \left[ \vec{F}^+ + \vec{F}^- \right]
   \right)
 ~~
 -
  \frac{\imath \omega \beta_r}
       {2}
   \alpha_c
   *
   \left(
    \vec{u} \times \left[ \vec{F}^+ - \vec{F}^- \right]
   \right)
 ~~
~~~~ 
~~~~
\\
  \nabla' \times \vec{F}^{\pm}
&=&
 \mp 
  \imath \omega 
  \alpha_r \beta_r 
  \left( 1 \mp \xi \right)
  ~
  \vec{u} \times \vec{F}^\pm
\nonumber
\\
&&
 ~~~~
 \mp 
  \frac{\imath \omega \alpha_r}
       {2} 
   \beta_c
   *
   \left(
    \vec{u} \times 
    \left[ \vec{F}^+ + \vec{F}^- \right]
   \right)
 ~~
 -
  \frac{\imath \omega \beta_r}
       {2}
   \alpha_c
   *
   \left(
    \vec{u} \times \left[ \vec{F}^+ - \vec{F}^- \right]
   \right)
~~~~ 
~~~~
\end{eqnarray}

\noindent{\bf Plane Polarized Moving Frame, Magnetic:} 

This is just a special case of the above vector moving frame equation --
~
\begin{eqnarray}
  \partial_{z'} F_x^{\pm}
&=&
 \mp 
  \imath \omega 
  \alpha_r \beta_r 
  \left( 1 \mp \xi \right)
  ~
  F_x^{\pm}
 ~~
 \mp 
  \frac{\imath \omega \alpha_r}
       {2}
   \beta_c
   *
   \left(
    \left[ F_x^+ + F_x^- \right]
   \right)
 ~~
 -
  \frac{\imath \omega \beta_r}
       {2}
   \alpha_c
   *
   \left(
    \left[ F_x^+ - F_x^- \right]
   \right)
\end{eqnarray}

\noindent{\bf Beltrami-like form:} 

This might be written
~
\begin{eqnarray}
  \left\{
    \left(
      \nabla \times 
    \right)
   \pm
    \imath \omega 
    \alpha_r \beta_r 
    \left(
      \vec{u} \times 
    \right)
  \right\}
  \vec{F}^{\pm}
&=&
  \vec{W}^{\pm}
,
\end{eqnarray}
where $\vec{W}^{\pm}$ is a ``source'' term, but note that  actually it
will be some complicated function of $\vec{G}^\pm$.

\newpage
% ----------------------------------------------------------------------

\section{First order evolution equations (Primary $\vec{B}$)}\label{S-1storderB}

Follows the derivation in the previous sections, 
 but uses the ``Primary $\vec{B}$'' (or $\vec{F}'^\pm$) form, 
 as defined in \ref{Ss-definitions-vectorB},
 with
 eqns. (\ref{eqn-S-defs-Fpvector}, \ref{eqn-S-defs-DvectorFp}, 
     \ref{eqn-S-defs-BvectorFp}). 

\noindent
{\bf NOTE:}
Not yet calculated, 
 but it will have many similarities to the 
 preceeding three derivations in 
 sections \ref{S-1storderE}, \ref{S-1storderH}, \ref{S-1storderD}.

% ----------------------------------------------------------------------

\section{The Energy density and the Poynting vector}\label{S-eflux}

Here I discuss the role and calculation of 
 the energy densities and Poynting vector for the EM field 
 described in terms of the Primary $\vec{E}$ ($\vec{G}^\pm$) fields.  
While it turns out that the THF
 Poynting vector is always a simple function of the $\vec{G}^\pm$ fields, 
 the energy density is more complicated and depends on the degree of
 mismatch between the reference parameters and the true material parameters.
Because of this difficulty of reconstructing $\vec{H}$ from $\vec{G}^\pm$,
 all these calculations currently resort to restricting themselves 
 to the THF case only (i.e. $G^\circ = 0$, 
 see \ref{Ss-definitions-vector}).

In practise it might well be simpler to just reconstuct the 
 $\vec{E}$ and $\vec{B}$ fields from known $\vec{G}^\pm$, 
 and use the traditional forms for energy density and Poynting vector
 rather than evaluate the complicated expressions derived in this section.
Nevertheless, 
 these complicated expression may provide insight into the nature
 of the $\vec{G}^\pm$ fields, 
 even if they a not used for computation.

Here I will restrict mself to the non-dispersive reference case,
 where $\epsilon, \mu$ are constants independent of frequency.  
Without this, 
 getting $\vec{E}(t)$ from $\vec{G}^\pm(t)$ requires a deconvolution
 to get the frequency-dependance of $\epsilon, \mu$ out of $\vec{G}^\pm(t)$.
This restriction can be avoided entirely by taking the formulae 
 below to hold in the frequency (rather than time) domain.

I define $\odot$, denoting a convolving dot product, and
 $\otimes$, denoting a convolving cross product, so
~
\begin{eqnarray} 
  \vec{A} \odot \vec{B}
&=&
  \int d\omega' 
    ~~
    \vec{A} (\omega') 
   \cdot
    \vec{B}(\omega-\omega');
\\
  \vec{A} \otimes \vec{B}
&=&
  \int d\omega' 
    ~~
    \vec{A} (\omega') 
   \times 
    \vec{B}(\omega-\omega')
.
\end{eqnarray}

I've used
~
\begin{eqnarray} 
  \left[ A(t) \ast \vec{B}(t) \right] \cdot \vec{C}(t) 
\leftrightarrow 
  \left[ A(\omega) \vec{B}(\omega) \right] \odot \vec{C}(\omega)
\\
  \left[ A(t) \ast \vec{B}(t) \right] \times \vec{C}(t) 
\leftrightarrow 
  \left[ A(\omega) \vec{B}(\omega) \right] \otimes \vec{C}(\omega)
.
\end{eqnarray}

To avoid cluttered equations, I will also use this shorthand notation 
~
\begin{eqnarray} 
\vec{G}^{\pm}_\epsilon = \vec{G}^{\pm}/\alpha_r^{1/2},
~~~~ &~~~~& ~~~~
\vec{G}^{\pm}_\mu =  \vec{G}^{\pm}/\beta_r^{1/2}
.
\label{eqn-eflux-shorthand-epsGmuG}
\end{eqnarray}

% --------- --------- --------- ---------
\subsection{Energy Density}\label{S-eflux-energy}

The standard expression for bulk energy density of the EM field in
 terms of the electric and magnetic fields can be easily 
 converted into one in terms of $\vec{G}^\pm$ in the THF case.
In the following, 
 I explicitly allow for the possibility of 
 mismatched reference and material parameters 
 (i.e. $\epsilon \neq \epsilon_r$ etc).

% --------- 
\subsubsection{Miscellaneous Identities}\label{Sss-eflux-energy-vidents}

Where, at the transverse approximation:
~
\begin{eqnarray} 
  \left( \vec{u} \times \vec{A} \right) 
 \cdot 
  \left( \vec{u} \times \vec{B} \right)
&=&
 \vec{A}' \cdot \left( \vec{u} \times \vec{B} \right)
~~~~ 
=
 -\vec{A}' \cdot \left( \vec{B} \times \vec{u} \right)
~~~~ 
=
 -\vec{B} \cdot \left( \vec{u} \times \vec{A}' \right)
\\
&=&
 -\vec{B} \cdot \left( \vec{u} \times  \vec{u} \times \vec{A} \right)
~~~~ 
=
  -\vec{B} 
 \cdot 
   \left( \vec{u} \times  
     \left[ 
       \vec{u} \times \vec{A} 
     \right]
   \right)
\\
&=&
  +\vec{B} 
 \cdot 
   \left( 
     \left[ 
       \vec{u} \times \vec{A} 
     \right]
    \times  
     \vec{u} 
   \right)
~~~~ 
=
  +\vec{B} 
 \cdot 
   \left( 
     \left[ 
       \vec{u} \cdot \vec{u} 
     \right]
     \vec{A} 
    -
     \left[ 
       \vec{u} \cdot \vec{A} 
     \right]
     \vec{u} 
   \right)
\\
&=&
 +\vec{B} \cdot \vec{A}
 - 
  \left[ \vec{B} \cdot \vec{u} \right]
  \left[ \vec{u} \cdot \vec{A} \right]
\\
\textrm{transverse.} \Rightarrow 
~~~~ ~~~~
  \left( \vec{u} \times \vec{A} \right) 
 \cdot 
  \left( \vec{u} \times \vec{B} \right)
&\approx&
 +\vec{B} \cdot \vec{A}
~~~~ ~~~~
=
 \vec{A} \cdot \vec{B}
\label{eqn-vectorid-UxAdXxB-transverse}
\end{eqnarray}

The following brief calculation illustrates the non-trivial nature
of getting a simple expression in the frequency domain --
~
\begin{eqnarray} 
  \mathscr{U}_a
&=&
  \epsilon \ast E \cdot E
\\
&=&
  \left[
    \int 
      \epsilon(t') E(t-t') 
      dt'
   \right]
\cdot E(t)
\\
  \tilde{\mathscr{U}}_a
&=&
  \int
    e^{-\imath \omega t}
    \left[
      \int 
        \epsilon(t') E(t-t') 
        dt'
     \right]
    \cdot E(t)
    dt
\\
&=&
  \int
    \epsilon(t') 
    \left[
      \int 
        e^{-\imath \omega t}
        E(t-t') 
        \cdot E(t)
        dt
     \right]
    dt'
\\
&=&
  \int
    \epsilon(t') 
    \tilde{F}(\omega, t')
    dt'
\\
\textrm{where}
~~~~ ~~~~
  \tilde{F}(\omega, t')
&=&
       \int 
        e^{-\imath \omega t}
        E(t-t') 
        \cdot E(t)
        dt 
\end{eqnarray}

% --------- --------- 
\subsubsection{Transverse Case}\label{Sss-eflux-energy-tmf}

The time domain energy density calculation is 
(making use of eqn.(\ref{eqn-eflux-shorthand-epsGmuG})),
~
\begin{eqnarray} 
  \mathscr{U}
&=&
  \frac{1}{2}
  \epsilon \ast \vec{E} \cdot \vec{E} 
 +
  \frac{1}{2}
  \mu \ast \vec{H} \cdot \vec{H} 
\\
\textrm{freq. domain} \Rightarrow
~~~~ ~~~~
  2 
  ~
  \mathscr{U} (\omega)
&=&
  \left(
    \epsilon \vec{E} 
  \right)
  \odot 
  \vec{E} 
 +
  \left(
    \mu \vec{H} 
  \right)
  \odot 
    \vec{H} 
\label{eqn-eflux-Udispless}
\\
  8
  ~
  \mathscr{U} (\omega)
&=&
  %\frac{1}{\epsilon_r}
  \left(
    \epsilon 
    \left[
      \vec{G}^{+}_\epsilon
     +
      \vec{G}^{-}_\epsilon
    \right]
  \right)
  \odot 
    \left[
      \vec{G}^{+}_\epsilon
     +
      \vec{G}^{-}_\epsilon
    \right]
 +
  %\frac{1}{\mu_r}
  \left(
    \mu ~
    \vec{u} \times 
    \left[
      \vec{G}^{+}_\mu
     +
      \vec{G}^{-}_\mu
    \right]
  \right)
  \odot 
  \left(
    \vec{u} \times 
    \left[
      \vec{G}^{+}_\mu
     +
      \vec{G}^{-}_\mu
    \right]
  \right)
~~~~ ~~~~
\\
  8 
  ~
  \mathscr{U} (\omega)
&=&
  %\mu_r
  \left(
    \epsilon 
    \vec{G}^{+}_\epsilon
  \right)
  \odot 
    \vec{G}^{+}_\epsilon
 +
  %\mu_r
  \left(
    \epsilon 
    \vec{G}^{+}_\epsilon
  \right)
  \odot 
    \vec{G}^{-}_\epsilon
 +
  %\mu_r
  \left(
    \epsilon 
    \vec{G}^{-}_\epsilon
  \right)
  \odot 
    \vec{G}^{+}_\epsilon
 +
  %\mu_r
  \left(
    \epsilon 
    \vec{G}^{-}_\epsilon
  \right)
  \odot 
    \vec{G}^{-}_\epsilon
\nonumber
\\
&&
~~~~
 +
  %\epsilon_r
  \left(  
    \vec{u} \times \mu \vec{G}^{+}_\mu
  \right)
  \odot 
  \left(
    \vec{u} \times \vec{G}^{+}_\mu
  \right)
 -
  %\epsilon_r
  \left(  
    \vec{u} \times \mu \vec{G}^{+}_\mu
  \right)
  \odot 
  \left(
    \vec{u} \times \vec{G}^{-}_\mu
  \right)
\nonumber
\\
&&
~~~~ ~~~~
 -
  %\epsilon_r
  \left(  
    \vec{u} \times \mu \vec{G}^{-}_\mu
  \right)
  \odot 
  \left(
    \vec{u} \times \vec{G}^{+}_\mu
  \right)
 +
  %\epsilon_r
  \left(  
    \vec{u} \times \mu \vec{G}^{-}_\mu
  \right)
  \odot 
  \left(
    \vec{u} \times \vec{G}^{-}_\mu
  \right)
\\
\textrm{transverse approx eqn(\ref{eqn-vectorid-UxAdXxB-transverse}).} \Rightarrow
~~~~ ~~~~
&\approx&
  %\mu_r
  \left(
    \epsilon 
    \vec{G}^{+}_\epsilon 
  \right)
  \odot 
    \vec{G}^{+}_\epsilon 
 +
  %\mu_r
  \left(
    \epsilon 
    \vec{G}^{+}_\epsilon 
  \right)
  \odot 
    \vec{G}^{-}_\epsilon 
 +
  %\mu_r
  \left(
    \epsilon 
    \vec{G}^{-}_\epsilon 
  \right)
  \odot 
    \vec{G}^{+}_\epsilon 
 +
  %\mu_r
  \left(
    \epsilon 
    \vec{G}^{-}_\epsilon
  \right)
  \odot 
    \vec{G}^{-}_\epsilon
\nonumber
\\
&&
~~~~
 +
  %\epsilon_r
  \left(  
    \mu \vec{G}^{+}_\mu
  \right)
  \odot 
    \vec{G}^{+}_\mu
 -
  %\epsilon_r
  \left(  
    \mu \vec{G}^{+}_\mu
  \right)
  \odot 
    \vec{G}^{-}_\mu
\nonumber
\\
&&
~~~~ ~~~~
 -
  %\epsilon_r
  \left(  
    \mu \vec{G}^{-}_\mu
  \right)
  \odot 
    \vec{G}^{+}_\mu
 +
  %\epsilon_r
  \left(  
    \mu \vec{G}^{-}_\mu
  \right)
  \odot 
    \vec{G}^{-}_\mu
\\
&=&
  %\mu_r
  \left(
    \epsilon 
    \vec{G}^{+}_\epsilon
  \right)
  \odot 
    \vec{G}^{+}_\epsilon
 +
  %\epsilon_r
  \left(  
    \mu \vec{G}^{+}_\mu
  \right)
  \odot 
    \vec{G}^{+}_\mu
~~~~
 +
  %\mu_r
  \left(
    \epsilon 
    \vec{G}^{-}_\epsilon
  \right)
  \odot 
    \vec{G}^{-}_\epsilon
 +
  %\epsilon_r
  \left(  
    \mu \vec{G}^{-}_\mu
  \right)
  \odot 
    \vec{G}^{-}_\mu
\nonumber
\\
&&
~~~~ 
 +
  %\mu_r
  \left(
    \epsilon 
    \vec{G}^{+}_\epsilon
  \right)
  \odot 
    \vec{G}^{-}_\epsilon
 -
  %\epsilon_r
  \left(  
    \mu \vec{G}^{+}_\mu
  \right)
  \odot 
    \vec{G}^{-}_\mu
~~~~ 
 +
  %\mu_r
  \left(
    \epsilon 
    \vec{G}^{-}_\epsilon
  \right)
  \odot 
    \vec{G}^{+}_\epsilon
 -
  %\epsilon_r
  \left(  
    \mu \vec{G}^{-}_\mu
  \right)
  \odot 
    \vec{G}^{+}_\mu
\\
  8
  ~
  \mathscr{U} (\omega)
&=&
  \left(
    \left[
      \frac{\epsilon}{\alpha_r}
      \vec{G}^{+}
    \right]
   \odot 
    \left[
      \frac{1}{\alpha_r}
      \vec{G}^{+}
    \right]
  +
    \left[
      \frac{\mu}{\beta_r}
      \vec{G}^{+}
    \right]
  \odot 
    \left[
      \frac{1}{\beta_r}
      \vec{G}^{+}
    \right]
  \right)
\nonumber
\\
&&
~~~~
 +
  \left(
    \left[
      \frac{\epsilon}{\alpha_r}
      \vec{G}^{-}
    \right]
   \odot 
    \left[
      \frac{1}{\alpha_r}
      \vec{G}^{-}
    \right]
  +
    \left[
      \frac{\mu}{\beta_r}
      \vec{G}^{-}
    \right]
  \odot 
    \left[
      \frac{1}{\beta_r}
      \vec{G}^{-}
    \right]
  \right)
\nonumber
\\
&&
~~~~ ~~~~ 
 +
  \left(
    \left[
      \frac{\epsilon}{\alpha_r}
      \vec{G}^{+}
    \right]
   \odot 
    \left[
      \frac{1}{\alpha_r}
      \vec{G}^{-}
    \right]
  -
    \left[
      \frac{\mu}{\beta_r}
      \vec{G}^{+}
    \right]
  \odot 
    \left[
      \frac{1}{\beta_r}
      \vec{G}^{-}
    \right]
  \right)
\nonumber
\\
&&
~~~~ ~~~~  ~~~~ 
 +
  \left(
    \left[
      \frac{\epsilon}{\alpha_r}
      \vec{G}^{-}
    \right]
   \odot 
    \left[
      \frac{1}{\alpha_r}
      \vec{G}^{+}
    \right]
  -
    \left[
      \frac{\mu}{\beta_r}
      \vec{G}^{-}
    \right]
  \odot 
    \left[
      \frac{1}{\beta_r}
      \vec{G}^{+}
    \right]
  \right)
\\
\textrm{time domain.} \Rightarrow
~~~~ ~~~~
  8
  ~
  \mathscr{U} (t)
&=&
  \left(
    \left[
      \frac{\epsilon}{\alpha_r}
     \ast
      \vec{G}^{+}
    \right]
   \cdot 
    \left[
      \frac{1}{\alpha_r}
     \ast
      \vec{G}^{+}
    \right]
  +
    \left[
      \frac{\mu}{\beta_r}
     \ast
      \vec{G}^{+}
    \right]
  \cdot 
    \left[
      \frac{1}{\beta_r}
     \ast
      \vec{G}^{+}
    \right]
  \right)
\nonumber
\\
&&
~~~~
 +
  \left(
    \left[
      \frac{\epsilon}{\alpha_r}
     \ast
      \vec{G}^{-}
    \right]
   \cdot 
    \left[
      \frac{1}{\alpha_r}
     \ast
      \vec{G}^{-}
    \right]
  +
    \left[
      \frac{\mu}{\beta_r}
     \ast
      \vec{G}^{-}
    \right]
  \cdot 
    \left[
      \frac{1}{\beta_r}
     \ast
      \vec{G}^{-}
    \right]
  \right)
\nonumber
\\
&&
~~~~ ~~~~ 
 +
  \left(
    \left[
      \frac{\epsilon}{\alpha_r}
     \ast
      \vec{G}^{+}
    \right]
   \cdot 
    \left[
      \frac{1}{\alpha_r}
     \ast
      \vec{G}^{-}
    \right]
  -
    \left[
      \frac{\mu}{\beta_r}
     \ast
      \vec{G}^{+}
    \right]
  \cdot 
    \left[
      \frac{1}{\beta_r}
     \ast
      \vec{G}^{-}
    \right]
  \right)
\nonumber
\\
&&
~~~~ ~~~~  ~~~~ 
 +
  \left(
    \left[
      \frac{\epsilon}{\alpha_r}
     \ast
      \vec{G}^{-}
    \right]
   \cdot 
    \left[
      \frac{1}{\alpha_r}
     \ast
      \vec{G}^{+}
    \right]
  -
    \left[
      \frac{\mu}{\beta_r}
     \ast
      \vec{G}^{-}
    \right]
  \cdot 
    \left[
      \frac{1}{\beta_r}
     \ast
      \vec{G}^{+}
    \right]
  \right)
\label{eqn-energydensity-dispersivetransverse}
.
\end{eqnarray}

The first two lines of eqn.(\ref{eqn-energydensity-dispersivetransverse})
 are the $G^{+2}$ and $G^{-2}$ terms 
 that might naturally be expected to occur; 
 the final two lines are $G^{+} \cdot G^{-}$ terms 
 caused by interference between the two fields.
In the dispersionless reference case, 
 eqn.(\ref{eqn-energydensity-dispersivetransverse}) reduces to
~
\begin{eqnarray} 
\textrm{freq. domain.} \Rightarrow
~~~~ ~~~~
  8
  ~
  \mathscr{U} (\omega)
&=&
  \left(
    \left[
      \frac{\epsilon}{\epsilon_r}
     +
      \frac{\mu}{\mu_r}
    \right]
    \vec{G}^{+}
   \right)
   \odot 
      \vec{G}^{+}
~~~~
 +
  \left(
    \left[
      \frac{\epsilon}{\epsilon_r}
     +
      \frac{\mu}{\mu_r}
    \right]
    \vec{G}^{-}
   \right)
   \odot 
      \vec{G}^{-}
\nonumber
\\
&&
~~~~ ~~~~ 
 +
  \left(
    \left[
      \frac{\epsilon}{\epsilon_r}
     -
      \frac{\mu}{\mu_r}
    \right]
    \vec{G}^{+}
   \right)
   \odot 
      \vec{G}^{-}
~~~~
 +
  \left(
    \left[
      \frac{\epsilon}{\epsilon_r}
     -
      \frac{\mu}{\mu_r}
    \right]
    \vec{G}^{-}
   \right)
   \odot 
      \vec{G}^{+}
\\
\textrm{time domain.} \Rightarrow
~~~~ ~~~~
  8
  ~
  \mathscr{U} (t)
&=&
  \left(
    \left[
      \frac{\epsilon}{\epsilon_r}
     +
      \frac{\mu}{\mu_r}
    \right]
    \ast
    \vec{G}^{+}
   \right)
   \cdot 
      \vec{G}^{+}
~~~~
 +
  \left(
    \left[
      \frac{\epsilon}{\epsilon_r}
     +
      \frac{\mu}{\mu_r}
    \right]
    \ast
    \vec{G}^{-}
   \right)
   \cdot 
      \vec{G}^{-}
\nonumber
\\
&&
~~~~
 +
  \left(
    \left[
      \frac{\epsilon}{\epsilon_r}
     -
      \frac{\mu}{\mu_r}
    \right]
    \ast
    \vec{G}^{+}
   \right)
   \cdot 
      \vec{G}^{-}
~~~~
 +
  \left(
    \left[
      \frac{\epsilon}{\epsilon_r}
     -
      \frac{\mu}{\mu_r}
    \right]
    \ast
    \vec{G}^{-}
   \right)
   \cdot 
      \vec{G}^{+}
\label{eqn-energydensity}
.
\end{eqnarray}

In the fully dispersionless case, 
eqn.(\ref{eqn-energydensity-dispersivetransverse})
reduces to
~
\begin{eqnarray} 
\textrm{freq. domain.} \Rightarrow
~~~~ ~~~~
  8
  ~
  \mathscr{U} (\omega)
&=&
    \left[
      \frac{\epsilon}{\epsilon_r}
     +
      \frac{\mu}{\mu_r}
    \right]
    \vec{G}^{+}
   \odot 
      \vec{G}^{+}
~~~~
 +
    \left[
      \frac{\epsilon}{\epsilon_r}
     +
      \frac{\mu}{\mu_r}
    \right]
    \vec{G}^{-}
   \odot 
      \vec{G}^{-}
\nonumber
\\
&&
~~~~ ~~~~ 
 +
    \left[
      \frac{\epsilon}{\epsilon_r}
     -
      \frac{\mu}{\mu_r}
    \right]
    \vec{G}^{+}
   \odot 
      \vec{G}^{-}
~~~~
 +
    \left[
      \frac{\epsilon}{\epsilon_r}
     -
      \frac{\mu}{\mu_r}
    \right]
    \vec{G}^{-}
   \odot 
      \vec{G}^{+}
\\
\textrm{time domain.} \Rightarrow
~~~~ ~~~~
  8
  ~
  \mathscr{U} (t)
&=&
    \left[
      \frac{\epsilon}{\epsilon_r}
     +
      \frac{\mu}{\mu_r}
    \right]
    ~
    \vec{G}^{+}
   \cdot 
      \vec{G}^{+}
~~~~
 +
    \left[
      \frac{\epsilon}{\epsilon_r}
     +
      \frac{\mu}{\mu_r}
    \right]
    ~
    \vec{G}^{-}
   \cdot 
      \vec{G}^{-}
\nonumber
\\
&&
~~~~
 +
    2
    \left[
      \frac{\epsilon}{\epsilon_r}
     -
      \frac{\mu}{\mu_r}
    \right]
    ~
    \vec{G}^{+}
   \cdot 
      \vec{G}^{-}
\label{eqn-energydensity-displess}
.
\end{eqnarray}

We see here that the energy density takes a more complicated form 
 if we choose to use reference parameters 
 that do not match the true medium ones.
If the reference parameters have a 
 different degree of mismatch between $\epsilon$ and $\mu$, 
 which will in fact normally be the case, 
 then the cross terms dependent on $\vec{G}^{+} \cdot \vec{G}^{-}$ 
 do not vanish.

% --------- --------- 
\subsubsection{Full Vector Case}\label{Sss-eflux-energy-fullvector}

The energy density in the fully vectorised case 
 can be found by retaining the $\left[ \vec{B} \cdot \vec{u} \right]
  \left[ \vec{u} \cdot \vec{A} \right]$ corrections removed by
 the transverse approximation.
Thus the non-dispersive reference energy density, 
 as specified by eqn.(\ref{eqn-energydensity}), becomes 
~
\begin{eqnarray} 
  8
  ~
  \mathscr{U}
&=&
  \left(
    \left[
      \frac{\epsilon}{\epsilon_r}
     +
      \frac{\mu}{\mu_r}
    \right]
    \ast
    \vec{G}^{+}
  \right)
  \cdot 
    \vec{G}^{+}
~~~~
 +
  \left(
    \left[
      \frac{\epsilon}{\epsilon_r}
     +
      \frac{\mu}{\mu_r}
    \right]
    \ast
    \vec{G}^{-}
  \right)
  \cdot 
    \vec{G}^{-}
\nonumber
\\
&&
~~~~ 
  \left(
    \left[
      \frac{\epsilon}{\epsilon_r}
     -
      \frac{\mu}{\mu_r}
    \right]
    \ast
    \vec{G}^{+}
  \right)
  \cdot 
    \vec{G}^{-}
~~~~ 
 +
  \left(
    \left[
      \frac{\epsilon}{\epsilon_r}
     -
      \frac{\mu}{\mu_r}
    \right]
    \ast
    \vec{G}^{-}
  \right)
  \cdot 
    \vec{G}^{+}
\nonumber
\\
&&
~~~~ ~~~~
 + 
      \frac{\mu}{\mu_r}
    \ast 
    \left\{
      \left[ \vec{G}^{+} \cdot \vec{u} \right]
      \left[ \vec{u} \cdot \vec{G}^{+} \right]
     +
      \left[ \vec{G}^{-} \cdot \vec{u} \right]
      \left[ \vec{u} \cdot \vec{G}^{-} \right]
     -
      \left[ \vec{G}^{+} \cdot \vec{u} \right]
      \left[ \vec{u} \cdot \vec{G}^{-} \right]
     -
      \left[ \vec{G}^{-} \cdot \vec{u} \right]
      \left[ \vec{u} \cdot \vec{G}^{+} \right]
   \right\}
\label{eqn-energydensityx}
.
\end{eqnarray}

The fully dispersive version of the corrections is
~
\begin{eqnarray} 
  8
  ~
  \mathscr{U}_{NT}
&=&
 + 
      \left[ 
        \left(
          \frac{\mu}{\beta_r} \ast 
          \vec{G}^{+}  
        \right)
        \cdot \vec{u}
      \right]
      \left[ 
        \vec{u} \cdot 
        \left(
           \frac{1}{\beta_r} \ast 
           \vec{G}^{+} 
        \right)
      \right]
     +
      \left[ 
        \left(
          \frac{\mu}{\beta_r} \ast 
          \vec{G}^{-} 
        \right)
        \cdot \vec{u} 
      \right]
      \left[ 
        \vec{u} \cdot 
        \left(
          \frac{1}{\beta_r} \ast 
          \vec{G}^{-} 
        \right)
      \right]
\nonumber
\\
&&
~~~~ ~~~~
   -
      \left[ 
        \left(
          \frac{\mu}{\beta_r} \ast 
          \vec{G}^{+}  
        \right)
        \cdot \vec{u}
      \right]
      \left[ 
        \vec{u} \cdot 
        \left(
           \frac{1}{\beta_r} \ast 
           \vec{G}^{-} 
        \right)
      \right]
     +
      \left[ 
        \left(
          \frac{\mu}{\beta_r} \ast 
          \vec{G}^{-} 
        \right)
        \cdot \vec{u} 
      \right]
      \left[ 
        \vec{u} \cdot 
        \left(
          \frac{1}{\beta_r} \ast 
          \vec{G}^{+} 
        \right)
      \right]
\label{eqn-energydensity-nontransversecorrections}
.
\end{eqnarray}

% --------- --------- --------- ---------
\subsection{Poynting vector}\label{S-eflux-poynting}

The Poynting vector defines the energy flux, and is
~
\begin{eqnarray}
\vec{P} = \vec{E} \times \vec{H} \longrightarrow P_z = E_x H_y
\end{eqnarray}

We will see that a $G^+$ component always corresponds to a 
 Poynting vector (energy flux) in the $+z$ direction, 
 and the $G^-$ to an flux in the opposite direction (i.e. $-z$).  
This is the basis of the assertion that these $G^\pm$ variable
 are direction, 
 and thus correspond to forward and backward 
 directed field components.

\noindent
{\bf NOTE:}
A critique of Poynting vectors -- 
 see e.g W. Gough, {\em ``Poynting in the wrong direction''}, 
 Eur. J. Phys. {\bf 3}, 83 (1982).

% --------- 
\subsubsection{Vector Identities}\label{Sss-eflux-poynting-vidents}

Because the convolution integrals are independent of the vector properties, 
 we can work in the transverse $\vec{G}^{\pm}$ limit to get
~
\begin{eqnarray} 
 -
  \vec{G}^{+} \otimes \vec{u} \times \vec{G}^{-}
&=&
 -
  \left[ \vec{G}^{+} \odot \vec{G}^{-} \right] \vec{u} 
 +
  \left[ \vec{G}^{+} \cdot \vec{u} \right] \ast \vec{G}^{-}
\\
&=&
 -
  \left[ \vec{G}^{+} \odot \vec{G}^{-} \right] \vec{u} 
\\
  \textrm{and}
  ~~~~ ~~~~ 
  \vec{G}^{-} \otimes \vec{u} \times \vec{G}^{+}
&=&
  \left[ \vec{G}^{-} \odot \vec{G}^{+} \right] \vec{u} 
 -
  \left[ \vec{G}^{-} \cdot \vec{u} \right] \ast \vec{G}^{+}
\\
&=&
 +
  \left[ \vec{G}^{-} \odot \vec{G}^{+} \right] \vec{u} 
\end{eqnarray}

%$G^\alpha$ (NB: $\alpha = \pm$,
%$\bar{\alpha}=\mp$, $\beta=\pm 1$, $\bar{\beta}=\mp 1$ (changed notation!))

% --------- 
\subsubsection{Transverse Case}\label{Ss-eflux-poynting-vector}

The Poynting vector defines the energy flux.  
To get a simple expression, 
 we need to work in the transverse 
 case and specialise to a non-dispersive reference medium.
Some expressions remain in the weaker 
 partially transverse ``THF'' case -- 
 see \ref{Ss-definitions-vector}, 
 where $\vec{u} \cdot \vec{H} = 0$
 and $2 \beta_r \vec{H} 
  = \vec{u} \times \left( \vec{G}^+ - \vec{G}^- \right)$.
Starting from the standard expression, 
 and making use of the notation in eqn.(\ref{eqn-eflux-shorthand-epsGmuG})),
 we get
~
\begin{eqnarray} 
  \vec{S} (t)
&=&
  \vec{E} (t) \times \vec{H}  (t)
\\
\textrm{frequency domain} \Rightarrow
~~~~ ~~~~
  \vec{S} (\omega)
&=&
  \int d\omega' 
    ~~
    \vec{E} (\omega') 
   \times 
    \vec{H}(\omega-\omega');
\\
&=&
  \vec{E} (\omega) 
 ~\otimes~ 
  \vec{H}(\omega);
\\
\textrm{THF approx.} \Rightarrow
~~~~ ~~~~
&=&
 \left\{
  \left(
    \frac{1}{2 \alpha_r}
  \right)
  \left[
    \vec{G}^{+} + \vec{G}^{-}  
  \right]
 \right\}
 \otimes
 \left\{
  \left[
    \vec{u} \times \vec{G}^{+} - \vec{u} \times \vec{G}^{-}  
  \right]
  \left(
    \frac{-1}{2 \sqrt{ \mu_r}}
  \right)
 \right\}
\\
%\textrm{non-dispersive ref.} \Rightarrow
  -4 
 ~
  \vec{S}
&=&
    \vec{G}^{+}_\epsilon \otimes \left[ \vec{u} \times \vec{G}^{+}_\mu \right]
   -
    \vec{G}^{+}_\epsilon \otimes \left[ \vec{u} \times \vec{G}^{-}_\mu \right]
   +
    \vec{G}^{-}_\epsilon \otimes \left[ \vec{u} \times \vec{G}^{+}_\mu \right]
   -
    \vec{G}^{-}_\epsilon \otimes \left[ \vec{u} \times \vec{G}^{-}_\mu \right]
\\
  \textrm{simplify as below (\ref{Sss-eflux-poynting-vidents});} ~~~~ ~~~~
&=& 
    \vec{G}^{+}_\epsilon \otimes \left[ \vec{u} \times \vec{G}^{+} \right]
   -
    \vec{G}^{-}_\epsilon \otimes \left[ \vec{u} \times \vec{G}^{-} \right]
\\
&=& 
    \left[ \vec{G}^{+}_\epsilon \odot \vec{G}^{+}_\mu \right]  \vec{u}
   -
    \left[ \vec{G}^{+}_\epsilon \cdot \vec{u}     \right]  \ast \vec{G}^{+}_\mu 
  -
    \left[ \vec{G}^{-}_\epsilon \odot \vec{G}^{-}_\mu \right]  \vec{u}
   +
    \left[ \vec{G}^{-}_\epsilon  \cdot \vec{u}     \right]  \ast \vec{G}^{-}_\mu 
\\
&=& 
    \left[ \vec{G}^{+}_\epsilon \odot \vec{G}^{+}_\mu \right]  \vec{u}
   -
    0
  -
    \left[ \vec{G}^{-}_\epsilon \odot \vec{G}^{-}_\mu \right]  \vec{u}
   +
    0
\\
&=& 
  \left[ 
    \vec{G}^{+}_\epsilon \odot \vec{G}^{+}_\mu 
   -
    \vec{G}^{-}_\epsilon \odot \vec{G}^{-}_\mu
  \right]  \vec{u}
\\
&=& 
  \left[ 
    \left(
      \alpha_r^{-1} \vec{G}^{+} 
   \right)
   \odot 
    \left(
      \beta_r^{-1} \vec{G}^{+} 
   \right)
   -
    \left(
      \alpha_r^{-1} \vec{G}^{-} 
   \right)
    \odot 
    \left(
      \beta_r^{-1} \vec{G}^{-}
   \right)
  \right]  \vec{u}
\\
  \textrm{time domain;} ~~~~ ~~~~
  -4 
  \vec{S}
&=& 
  \left[ 
    \left(
      \alpha_r^{-1} \ast \vec{G}^{+} 
   \right)
   \cdot 
    \left(
      \beta_r^{-1} \ast \vec{G}^{+} 
   \right)
   -
    \left(
      \alpha_r^{-1} \ast \vec{G}^{-} 
   \right)
    \cdot 
    \left(
      \beta_r^{-1} \ast \vec{G}^{-}
   \right)
  \right]  \vec{u}
\end{eqnarray}

% --------- 
\subsubsection{Minkowski Version}\label{Sss-eflux-poynting-Minkowski}

Starting from the standard expression, I use the shorthand notation 
$\vec{G}^{\pm}_{M\epsilon} = \vec{G}^{\pm} \epsilon . \epsilon_r^{-1/2}$ 
and $\vec{G}^{\pm}_{M\mu} = \vec{G}^{\pm} \mu . \mu_r^{-1/2}$
to avoid cluttered equations; and so
~
\begin{eqnarray} 
  \vec{S}_M (t)
&=&
  \vec{D} (t) \times \vec{B}  (t)
\\
&=&
  \epsilon \ast \vec{E} (t) \times \mu \ast \vec{H}  (t)
\\
\textrm{frequency domain} \Rightarrow
~~~~ ~~~~
  \vec{S}_M (\omega)
&=&
  \int d\omega' 
    ~~
    \epsilon (\omega')  \vec{E} (\omega') 
   \times 
    \mu (\omega-\omega') \ast \vec{H}(\omega-\omega');
\\
&=&
  \epsilon \vec{E}
 ~\otimes~ 
  \mu \vec{H}(\omega);
\\
\textrm{THF approx.} \Rightarrow
~~~~ ~~~~
&=&
 \left\{
  \left(
    \frac{\epsilon}{2 \alpha_r}
  \right)
  \left[
    \vec{G}^{+} + \vec{G}^{-}  
  \right]
 \right\}
 \otimes
 \left\{
  \left[
    \vec{u} \times \vec{G}^{+} - \vec{u} \times \vec{G}^{-}  
  \right]
  \left(
    \frac{-\mu}{2 \sqrt{ \mu_r}}
  \right)
 \right\}
\\
%\textrm{non-dispersive ref.} \Rightarrow
  -4 
 ~
  \vec{S}_M
&=&
    \vec{G}^{+}_{M\epsilon} \otimes \left[ \vec{u} \times \vec{G}^{+}_{M\mu} \right]
   -
    \vec{G}^{+}_{M\epsilon} \otimes \left[ \vec{u} \times \vec{G}^{-}_{M\mu} \right]
   +
    \vec{G}^{-}_{M\epsilon} \otimes \left[ \vec{u} \times \vec{G}^{+}_{M\mu} \right]
   -
    \vec{G}^{-}_{M\epsilon} \otimes \left[ \vec{u} \times \vec{G}^{-}_{M\mu} \right]
\\
  \textrm{simplify as below (\ref{Sss-eflux-poynting-vidents});} ~~~~ ~~~~
&=& 
    \vec{G}^{+}_{M\epsilon} \otimes \left[ \vec{u} \times \vec{G}^{+} \right]
   -
    \vec{G}^{-}_{M\epsilon} \otimes \left[ \vec{u} \times \vec{G}^{-} \right]
\\
&=& 
    \left[ \vec{G}^{+}_{M\epsilon} \odot \vec{G}^{+}_{M\mu} \right]  \vec{u}
   -
    \left[ \vec{G}^{+}_{M\epsilon} \cdot \vec{u}     \right]  \ast \vec{G}^{+}_{M\mu} 
  -
    \left[ \vec{G}^{-}_{M\epsilon} \odot \vec{G}^{-}_{M\mu} \right]  \vec{u}
   +
    \left[ \vec{G}^{-}_{M\epsilon}  \cdot \vec{u}     \right]  \ast \vec{G}^{-}_{M\mu} 
\\
&=& 
    \left[ \vec{G}^{+}_{M\epsilon} \odot \vec{G}^{+}_{M\mu} \right]  \vec{u}
   -
    0
  -
    \left[ \vec{G}^{-}_{M\epsilon} \odot \vec{G}^{-}_{M\mu} \right]  \vec{u}
   +
    0
\\
&=& 
  \left[ 
    \vec{G}^{+}_{M\epsilon} \odot \vec{G}^{+}_{M\mu} 
   -
    \vec{G}^{-}_{M\epsilon} \odot \vec{G}^{-}_{M\mu}
  \right]  \vec{u}
\\
&=& 
  \left[ 
    \left(
      \frac{\epsilon}{\alpha_r}
      \vec{G}^{+} 
   \right)
   \odot 
    \left(
      \frac{\mu}{\beta_r}
      \vec{G}^{+} 
   \right)
   -
    \left(
      \frac{\epsilon}{\alpha_r}
      \vec{G}^{-} 
   \right)
    \odot 
    \left(
      \frac{\mu}{\beta_r}
      \vec{G}^{-}
   \right)
  \right]  \vec{u}
\\
  \textrm{time domain;} ~~~~ ~~~~
  -4 
  \vec{S}_M
&=& 
  \left[ 
    \left(
      \frac{\epsilon}{\alpha_r}
      \ast \vec{G}^{+} 
   \right)
   \cdot 
    \left(
      \frac{\mu}{\beta_r}
      \ast \vec{G}^{+} 
   \right)
   -
    \left(
      \frac{\epsilon}{\alpha_r}
      \ast \vec{G}^{-} 
   \right)
    \cdot 
    \left(
      \frac{\mu}{\beta_r}
      \ast \vec{G}^{-}
   \right)
  \right]  \vec{u}
\end{eqnarray}

%\end{section}

% ----------------------------------------------------------------------

\section{Interpretations}\label{S-interpretations}

The most comprehensive discussions on the interpretation of 
 $G^\pm$ fields are contained in Kinsler et.al. 2005 \cite{Kinsler-RN-2005pra}.
Other important remarks on the behaviour of the propagating fields 
 at interfaces are covered in my report 
 {\em ``Causality in spatially propagated optics''} (tb arXiv).

% ----------------------------------------------------------------------
\newpage 

\section{Second order evolution equations}
\label{S-evolution}

In sections \ref{S-1storderE}, \ref{S-1storderH}, and \ref{S-1storderD}
  I calculated various forms of first-order
 propagation equation for the $G^{\pm}$ field variables.  
However, 
 many pulse propagation theories start from a second-order form for $E$, 
 as in e.g. \cite{Brabec-K-1997prl},
~
\begin{eqnarray}
  \left( \partial_z^2 + \nabla_\bot^2 \right) E(\vec{r},t)
- \frac{1}{c^2}
  \partial_t^2
  \int_{-\infty}^t dt' \epsilon(t-t') E(\vec{r},t')
&=&
  \frac{4\pi}{c^2}
  \partial_t^2
  P(\vec{r},t)
.
\label{startingpoint-copy}
\end{eqnarray}

To address this point of interest,  
 here I follow a similar path and derive a second-order 
 $G^{\pm}$ propagation equation.  
As expected, 
 it looks rather similar to the standard equivalents for $E$, 
 but with some extra derivative (curl) terms.
First I attempt to make eqns.(\ref{scaled-a},\ref{scaled-b}) look similar
 to eqn.(\ref{startingpoint-copy})
 in subsection \ref{SS-evolution-scalar}.
Then in subsection \ref{SS-evolution-vector}
I present a more complete derivation of the vector form in the 
 transverse field case.

% -----------------

\subsection{Scalar Form}
\label{SS-evolution-scalar}

This is intended as a quick, 
 intuitive calculation to get a scalar second-order 
 equation for $G^\pm$ fields.  
It is based on the old Fleck\cite{Fleck-1970prb} definitions, 
 so I leave the linear dispersion 
 (from the permittivity $\epsilon$ or permeability $\mu$)
 inside the polarization $P$ rather than separate it out as 
 an extra term (thus it contains a hidden convolution).  
This approach has been superceeded (I now put everything in $\epsilon, \mu$), 
 but this part is rather old
 and has not been updated accordingly.
To start, 
 I multiply eqn.(\ref{scaled-a}) by the LHS derivative term 
 seen in eqn.(\ref{scaled-b}):
~
\begin{eqnarray}
\left[
  \partial_t 
  -
  \partial_z 
\right]
\left[
 \partial_t 
 +
 \partial_z 
\right]
G^{+}
&=&
\left[
  \partial_t 
  -
  \partial_z 
\right]
\left[
 - \partial_t P
 - \sigma 
   \left( 
       G^{+} + G^{-}
   \right)
\right]
\label{scaled-a21}
\\
~
~
\textrm{(A)} ~~~~ ~~~~
&=&
-
\left[
  \partial_t 
  -
  \partial_z 
\right]
\partial_t P
-
\sigma 
\left[
  \partial_t 
  -
  \partial_z 
\right]
G^{+}
-
\sigma 
\left[
  \partial_t 
  -
  \partial_z 
\right]
G^{-}
\label{scaled-a22}
\\
~
~
\textrm{(B: sub (\ref{scaled-b}) for last RHS term)} ~~~~ ~~~~
&=&
-
\left[
  \partial_t 
  -
  \partial_z 
\right]
\partial_t P
-
\sigma 
\left[
  \partial_t 
  -
  \partial_z 
\right]
G^{+}
-
\sigma 
\left[
 - \partial_t P
 - \sigma G^{+}
 - \sigma G^{-}
\right]
\label{scaled-a23}
\\
~
~
\textrm{(C: iterate)} ~~~~ ~~~~
&=&
-
\left[
  \partial_t 
  -
  \partial_z 
  -
  \sigma 
\right]
\partial_t P
-
\sigma 
\left[
  \partial_t 
  -
  \partial_z 
  -
  \sigma
\right]
G^{+}
+
\sigma^2 G^{-}
\label{scaled-a24}
\\
~
~
\textrm{(D)} ~~~~ ~~~~
&=&
-
\left[
  \partial_t 
  -
  \partial_z 
  +
  \sum_{n=1}^{N}
  \left(-\sigma\right)^n 
\right]
\partial_t P
-
\sigma 
\left[
  \partial_t 
  -
  \partial_z 
  +
  \sum_{n=1}^{N}
  \left(-\sigma\right)^n 
\right]
G^{+}
+
\sigma^{N+1} G^{-}
~~~~
~~~~
\label{scaled-a25}
\\
&=&
-
\left[
  \partial_t 
  -
  \partial_z 
  +
  Q
\right]
\partial_t P
-
\sigma 
\left[
  \partial_t 
  -
  \partial_z 
  +
  Q
\right]
G^{+}
+
\sigma^{N+1} G^{-}
\label{scaled-a26}
,
\\
Q
&=&
\frac{1-\sigma^N}{1-\sigma}
\textrm{, for $\sigma < 1$}
,
\end{eqnarray}
It would now be straightforward to introduce 
 a variety of useful approximations, 
 e.g. $\sigma \ll 1$ or similar.
Alternatively, 
 I could group the $\sigma$ terms with the derivative operators
 in eqns.(\ref{scaled-a},\ref{scaled-b}).  
 I rearrange eqns.(\ref{scaled-a},\ref{scaled-b}) to give
~
\begin{eqnarray}
\left[
 \partial_t 
 +
 \partial_z 
 +
 \sigma
\right]
G^{+}
&=&
 - \partial_t P
 - \sigma 
   G^{-}
\label{scaled-a31}
,
\\
\left[
 \partial_t 
 -
 \partial_z 
 +
 \sigma
\right]
G^{-}
&=&
 - \partial_t P
 - \sigma 
   G^{+}
,
\label{scaled-b31}
\end{eqnarray}
and apply the  differential operator from the LHS of eqn.(\ref{scaled-b31})
to eqn.(\ref{scaled-a31})
~
\begin{eqnarray}
\left[
 \partial_t 
 -
 \partial_z 
 +
 \sigma
\right]
\left[
 \partial_t 
 +
 \partial_z 
 +
 \sigma
\right]
G^{+}
&=&
\left[
 \partial_t 
 -
 \partial_z 
 +
 \sigma
\right]
\left[ - \partial_t P
 - \sigma 
   G^{-}
\right]
\label{scaled-a32}
,
\\
&=&
-
\left[
 \partial_t 
 -
 \partial_z 
 +
 \sigma
\right]
\partial_t P
-
  \sigma
  \left[
   - \partial_t P
   - \sigma 
     G^{+}
  \right]
\label{scaled-a33}
\\
&=&
-
\left[
 \partial_t 
 -
 \partial_z 
\right]
\partial_t P
+
  \sigma^2
     G^{+}
\label{scaled-a34}
\\
\left\{
  \left[
   \partial_t 
   -
   \partial_z 
   +
  \sigma
  \right]
  \left[
   \partial_t 
   +
   \partial_z 
   +
   \sigma
  \right]
-
  \sigma^2
\right\}
G^{+}
&=&-
\left[
 \partial_t 
 -
 \partial_z 
\right]
\partial_t P
\label{scaled-a35}
.
\end{eqnarray}

Now I simplify the term in braces $\left\{ ... \right\}$, 
 but retain the ordering of $\sigma$'s and $\partial$, 
 because $\sigma$ contains $\epsilon$ --
 I might like to make the medium dispersive, 
 even though it is customary to add dispersion at a later stage.
~
\begin{eqnarray}
\left\{
  \left[
   \partial_t 
   -
   \partial_z 
   +
  \sigma
  \right]
  \left[
   \partial_t 
   +
   \partial_z 
   +
   \sigma
  \right]
-
  \sigma^2
\right\}
G^{+}
\\
\left\{
  \left[
    \left(
      \partial_t + \sigma
    \right)
   -
   \partial_z 
  \right]
  \left[
    \left(
      \partial_t + \sigma
    \right)
   +
   \partial_z 
  \right]
-
  \sigma^2
\right\}
G^{+}
\\
\left\{
    \left(
      \partial_t + \sigma
    \right)^2
   +
    \left( \partial_t + \sigma \right) 
    \partial_z
   -
    \partial_z
    \left( \partial_t + \sigma \right) 
   -
   \partial_z^2
-
  \sigma^2
\right\}
G^{+}
\\
\left\{
    \left(
      \partial_t + \sigma
    \right)^2
   +
   \partial_t \partial_z - \partial_z \partial_t
   +
   \sigma \partial_z - \partial_z \sigma
   -
   \partial_z^2
-
  \sigma^2
\right\}
G^{+}
\\
\left\{
    \left(
      \partial_t + \sigma
    \right)^2
   +
   \left( \sigma \partial_z - \partial_z \sigma \right)
   -
   \partial_z^2
-
  \sigma^2
\right\}
G^{+}
\\
\left\{
    \partial_t^2 
   + 
    \sigma \partial_t
   + 
    \partial_t \sigma
   +
   \left( \sigma \partial_z - \partial_z \sigma \right)
   -
   \partial_z^2
\right\}
G^{+}
.
\end{eqnarray}

And for $G^{-}$, 
 the term would be $\left\{ \partial_t^2 + \sigma \partial_t +
  \partial_t \sigma - \left( \sigma \partial_z - \partial_z \sigma \right) -
  \partial_z^2 \right\} G^{-}$.  
Apply this to the full equation, 
 assert the usual case of a uniform material 
 (so $\sigma \partial_z G^{\pm}= \partial_z \sigma  G^{\pm}$), 
 and then duplicate the calculation for $G^-$ ...
~
\begin{eqnarray}
\left\{
    \partial_z^2
   -
    \partial_t^2 
   -
    \sigma \partial_t - \partial_t \sigma
\right\}
G^{+}
&=&
\left[
 \partial_t 
 -
 \partial_z 
\right]
\partial_t P
\label{scaledsecond-a}
,
\\
\left\{
    \partial_z^2
  -
    \partial_t^2 
   - 
    \sigma \partial_t
   - 
    \partial_t \sigma
\right\}
G^{-}
&=&
\left[
 \partial_t 
 +
 \partial_z 
\right]
\partial_t P
,
\label{scaledsecond-b}
\end{eqnarray}

Compare this to the ``standard'' second order wave equation
 eqn.(\ref{startingpoint-copy}):
(1) the $\partial_z^2 G^\pm$ is  analogous to $\partial_z^2 E$;
(2) $\left( \partial_t  + \partial_z \right) \partial_t P$ 
 has an extra $\partial_z \partial_t P$ part 
 which might be expected given the directional nature of $G^\pm$;
(3) the $-\left(\partial_t^2 + \sigma \partial_t + \partial_t \sigma
  \right) G^\pm$ is 
  the analogous term to $-\partial_t^2 \int \epsilon E dt$; 
 remember that $\sigma$ is a conductivity term not usually 
 included in pulse propagation problems.
If we add and subtract eqns.(\ref{scaledsecond-a}) and (\ref{scaledsecond-b}),
 setting $\sigma=0$, 
 we get
~
\begin{eqnarray}
\left\{
    \partial_z^2
   -
    \partial_t^2 
\right\}
\left( G^{+} + G^{-} \right)
&=&
\partial_t^2 P
\label{scaledsecond-E}
,
\\
\left\{
    \partial_z^2
  -
    \partial_t^2 
\right\}
\left( G^{+} - G^{-} \right)
&=&
 \partial_z 
 \partial_t P
.
\label{scaledsecond-H}
\end{eqnarray}
Of course $E_x \propto G^{+} + G^{-}$, and $H_y \propto G^{+} - G^{-}$, 
 so the equivalence between eqn.(\ref{scaledsecond-E}) 
 and eqn.(\ref{startingpoint-copy}) is to be expected.
Presumably there would also be a related equivalence between
 eqn.(\ref{scaledsecond-H}) and the $H_y$ version of
 eqn.(\ref{startingpoint-copy}).

%\end{subsection}

% -----------------

\subsection{Vector Form}
\label{SS-evolution-vector}

This is a vectorised derivation of the second order wave
 equation for $\vec{G}^\pm$ fields.
Unlike the previous scalar derivation,
 it is reasonably complete and without stringent approximation
 (barring its transverse field approx).
The polarization term $P$ remains as part of the dispersion term, 
 thus simplifying the equations; 
 until the end when I split it off for illustrative purposes.
From section \ref{S-definitions}, 
 the unit vector in the propagation direction is $\vec{u}$, 
 and the vectorised version of $G$ is given by 
 eqn.(\ref{eqn-S-defs-Gvector}) --
~
\begin{eqnarray}
  \vec{G}^{\pm} 
&=&
  \tilde{\alpha}_r \vec{E} + \vec{u} \times \tilde{\beta}_r \vec{H}
.
\end{eqnarray}

In frequency space,
~
\begin{eqnarray}
 - \imath \omega 
 \left( \tilde{\epsilon}\vec{E} \right) 
~~~~
=
 - \imath \omega 
 \left( \tilde{\alpha}_r^2 + \tilde{\alpha}_r \tilde{\alpha}_c \right) \vec{E}
&= &
\nabla \times \vec{H} 
-
\vec{J}
,
\\
 - \imath \omega 
 \left(\tilde{\mu} \ast \vec{H} \right) 
~~~~
=
 - \imath \omega 
 \left( \tilde{\beta}_r^2 + \tilde{\beta}_r \tilde{\beta}_c \right) \vec{H}
&=&
-
\nabla \times \vec{E} 
.
\end{eqnarray}

In the following calculation, 
 I omit the convolution symbols ``$*$'' that should appear 
 between the $\tilde{\alpha}_c, \tilde{\beta}_c$ parameters 
 and the $G^+ \pm G^-$ fields; 
 this is in the interests of both my laziness
 and reducing notational clutter.
I split (as before in subsection \ref{S-medium}) 
 the permittivity and permeability into 
 instantaneous and time-dependent parts with 
 $\epsilon(\omega) 
  = \alpha(\omega)^2 = \alpha_r^2 
  + \alpha_r \alpha_c(\omega) $ 
 (and similarly for $\mu$ and $\beta$).
This mimics traditional derivations, 
 where dispersion and nonlinear polarization are handled separately.
Using 
 $\vec{u} \times  \nabla \times \vec{Q} 
  - \nabla \left( \vec{u} \cdot \vec{Q} \right) 
 =  \nabla \times \vec{u} \times \vec{Q}$,
 I take the scaled time derivative of the definition of $\vec{G}^{\pm}$, 
~
\begin{eqnarray}
  \left( - \imath \omega \right)
  \alpha_r \beta_r
  \vec{G}^{\pm}
&=&
  - \imath \omega
  \alpha_r^2 \beta_r
  \vec{E}
\pm
  \left( - \imath \omega \right)
  \alpha_r \beta_r^2
  ~ 
  \vec{u} \times
  \vec{H}
\\
&=&
  \nabla \times 
    \beta_r \vec{H} 
 - 
  \beta_r
  \vec{J} 
 +
  \imath \omega
  \alpha_c 
  \alpha_r \vec{E}
 ~~
 \mp
  \vec{u}\times 
  \left[ \nabla  \times \alpha_r \vec{E} \right]
 \pm
  \imath \omega
  \vec{u} \times
  \beta_c
  \beta_r \vec{H}
\\
&=&
  \nabla \times 
    \beta_r \vec{H} 
 - 
  \beta_r
  \vec{J} 
 +
  \imath \omega 
  \alpha_c
  \alpha_r \vec{E}
 ~~
 \mp
  \nabla \times 
  \left[ \vec{u} \times \alpha_r \vec{E} \right]
 \mp
  \nabla 
  \left[ \vec{u} \cdot \alpha_r \vec{E} \right]
 \pm
  \imath \omega
  \vec{u} \times
  \beta_c
  \beta_r \vec{H}
\\
&=&
  \nabla \times 
    \beta_r \vec{H} 
 ~~
 - 
  \beta_r
  \vec{J} 
 ~~
 \mp
  \nabla \times 
  \left[ \vec{u} \times \alpha_r \vec{E} \right]
 \mp
  \nabla 
  \left[ \vec{u} \cdot \alpha_r \vec{E} \right]
 ~~
 +
  \imath \omega
  \alpha_c 
  \alpha_r \vec{E}
 ~~
 \pm
  \imath \omega
  \vec{u} \times
  \beta_c
  \beta_r \vec{H}
\\
&=&
 \mp
  \nabla \times 
  \left[
    \vec{u} \times 
    \alpha_r
    \vec{E} 
   \mp
    \beta_r
    \vec{H}
  \right]
 ~~
 - 
  \beta_r
  \vec{J} 
 ~~
 \mp
  \nabla 
  \left[ \vec{u} \cdot \alpha_r \vec{E} \right]
 ~~
 +
  \imath \omega
  \alpha_c 
  \alpha_r \vec{E}
 ~~
 \pm
  \imath \omega
  \vec{u} \times
  \beta_c
  \beta_r \vec{H}
\\
&=&
 \mp
  \nabla \times 
  \left[
    \vec{u} \times 
    \alpha_r
    \vec{E} 
   \pm
    \vec{u} \times 
    \vec{u} \times 
    \beta_r
    \vec{H}
  \right]
 \pm
  \nabla \times 
  \left(
    \vec{u} \cdot
    \beta_r
    \vec{H}
  \right)
  \vec{u}
 ~~
 - 
  \beta_r
  \vec{J} 
 ~~
 \mp
  \nabla 
  \left[ \vec{u} \cdot \alpha_r \vec{E} \right]
\nonumber
\\
&&
 ~~~~
 ~~
 +
  \imath \omega
  \alpha_c 
  \alpha_r \vec{E}
 ~~
 \pm
  \imath \omega
  \vec{u} \times
  \beta_c
  \beta_r \vec{H}
.
\end{eqnarray}

Now I combine $\vec{E}$'s and $\vec{H}$'s to form $\vec{G}^{\pm}$'s,
~
\begin{eqnarray}
  - \imath \omega
  \alpha_r \beta_r
  \vec{G}^{\pm}
&=&
 \mp
  \nabla \times 
  \left[
    \vec{u} \times 
    \vec{G}^{\pm}
  \right]
 \pm
  {G}^{\circ}
  \nabla \times 
  \vec{u}
 ~~
 \mp
  \nabla 
  \left[ \vec{u} \cdot \alpha_r \vec{E} \right]
 ~~
 +
  \imath \omega
  \alpha_c 
  \alpha_r \vec{E}
 ~~
 \pm
  \imath \omega
  \vec{u} \times
  \beta_c
  \beta_r \vec{H}
 ~~
 - 
  \beta_r
  \vec{J} 
\\
&=&
 \mp
  \nabla \times
  \left[
    \vec{u} \times 
    \vec{G}^{\pm}
  \right]
 ~~
 +
  \frac{\imath \omega}
       {2}
    \alpha_c 
    \left[
      \vec{G}^{+} + \vec{G}^{-}
    \right]
 ~~
 \pm
  \frac{\imath \omega}
       {2}
    \beta_c 
    \left[
      \vec{G}^{+} - \vec{G}^{-}
    \right]
 ~~
 \mp
  \frac{1}{2}
  \nabla 
  \left[
    \vec{u} \cdot 
    \left(
      \vec{G}^{+} + \vec{G}^{-}
    \right)
  \right]
 ~~
 - 
  \beta_r
  \vec{J} 
\\
&=&
\mp
  \vec{u} \times
  \left[
    \nabla\times 
    \vec{G}^{\pm}
  \right]
 \pm
  \nabla
  \left[
    \vec{u}
    \cdot
    \vec{G}^{\pm}
  \right]
 ~~
 +
  \frac{\imath \omega}
       {2}
    \alpha_c 
    \left[
      \vec{G}^{+} + \vec{G}^{-}
    \right]
 ~~
 \pm
  \frac{\imath \omega}
       {2}
    \beta_c 
    \left[
      \vec{G}^{+} - \vec{G}^{-}
    \right]
\nonumber
\\
&&
 ~~~~
 ~~
 \mp
  \frac{1}{2}
  \nabla 
  \left[
    \vec{u} \cdot 
    \left(
      \vec{G}^{+} + \vec{G}^{-}
    \right)
  \right]
 ~~
 - 
  \beta_r
  \vec{J} 
\\
&=&
\mp
  \vec{u} \times
  \left[
    \nabla\times 
    \vec{G}^{\pm}
  \right]
 ~~
 +
  \frac{\imath \omega}
       {2}
    \alpha_c 
    \left[
      \vec{G}^{+} + \vec{G}^{-}
    \right]
 ~~
 \pm
  \frac{\imath \omega}
       {2}
    \beta_c 
    \left[
      \vec{G}^{+} - \vec{G}^{-}
    \right]
 ~~
 +
  \frac{1}{2}
  \nabla 
  \left[
    \vec{u} \cdot 
    \left(
      \vec{G}^{+} - \vec{G}^{-}
    \right)
  \right]
 ~~
 - 
  \beta_r
  \vec{J} 
, ~~~~
\end{eqnarray}
and apply another scaled time derivative $-\imath \omega \alpha_r \beta_r$,
~
\begin{eqnarray}
 -\omega^2
  \alpha_r^2 \beta_r^2
  \vec{G}^{\pm}
&=&
 \pm
  \imath \omega
  \alpha_r \beta_r
  \vec{u} \times
  \nabla \times
  \vec{G}^{\pm}
 +
  \frac{\omega^2}{2}
  \alpha_r \beta_r \alpha_c
    \left[
      \vec{G}^{+} + \vec{G}^{-}
    \right]
 \mp
  \frac{\omega^2}{2}
  \alpha_r \beta_r \beta_c
    \left[
      \vec{G}^{+} - \vec{G}^{-}
    \right]
\nonumber
\\
&&
 ~~~~
 ~~
 -
  \frac{\imath \omega}{2}
  \alpha_r \beta_r
  \nabla 
  \left[
    \vec{u} \cdot 
    \left(
      \vec{G}^{+} - \vec{G}^{-}
    \right)
  \right]
 ~~
 +
  \imath \omega
  \alpha_r \beta_r^2
  \vec{J} 
\\
\textrm{(B)} ~~~~ ~~~~
&=&
 \mp
  \vec{u} \times
  \nabla \times
  \left(
   -
   \imath \omega
   \alpha_r \beta_r
   \vec{G}^{\pm}
  \right)
 +
  \frac{\omega^2}{2}
  \alpha_r \beta_r \alpha_c 
    \left[
      \vec{G}^{+} + \vec{G}^{-}
    \right]
 \mp
  \frac{\omega^2}{2}
  \alpha_r \beta_r \beta_c 
    \left[
      \vec{G}^{+} - \vec{G}^{-}
    \right]
\nonumber
\\
&&
 ~~~~
 ~~
 -
  \frac{\imath \omega}{2}
  \alpha_r \beta_r
  \nabla 
  \left[
    \vec{u} \cdot 
    \left(
      \vec{G}^{+} - \vec{G}^{-}
    \right)
  \right]
 ~~
 +
  \imath \omega
  \alpha_r \beta_r^2
  \vec{J} 
\\
\textrm{(C)} ~~~~ ~~~~
&=&
  \left(
    \mp
    \vec{u} \times
    \nabla \times
  \right)
  \left\{
\mp
  \vec{u} \times
  \left[
    \nabla\times 
    \vec{G}^{\pm}
  \right]
 ~~
 +
  \frac{\imath \omega}
       {2}
    \alpha_c 
    \left[
      \vec{G}^{+} + \vec{G}^{-}
    \right]
 ~~
 \pm
  \frac{\imath \omega}
       {2}
    \beta_c 
    \left[
      \vec{G}^{+} - \vec{G}^{-}
    \right]
 ~~
 +
  \frac{1}{2}
  \nabla 
  \left[
    \vec{u} \cdot 
    \left(
      \vec{G}^{+} - \vec{G}^{-}
    \right)
  \right]
 ~~
 - 
  \beta_r
  \vec{J} 
  \right\}
\nonumber
\\
 && 
 +
  \frac{\omega^2}{2}
  \alpha_r \beta_r \alpha_c 
    \left[
      \vec{G}^{+} + \vec{G}^{-}
    \right]
 \mp
  \frac{\omega^2}{2}
  \alpha_r \beta_r \beta_c 
    \left[
      \vec{G}^{+} - \vec{G}^{-}
    \right]
 ~~
 -
  \frac{\imath \omega}{2}
  \alpha_r \beta_r
  \nabla 
  \left[
    \vec{u} \cdot 
    \left(
      \vec{G}^{+} - \vec{G}^{-}
    \right)
  \right]
 ~~
 +
  \imath \omega
  \alpha_r \beta_r^2
  \vec{J} 
\\
\textrm{(D)} ~~~~ ~~~~
&=&
    \vec{u} \times
    \nabla \times
  \vec{u} \times
  \left[
    \nabla\times 
    \vec{G}^{\pm}
  \right]
 ~~
 \mp
  \frac{\imath \omega}
       {2}
    \alpha_c 
    \vec{u} \times
    \nabla \times
    \left[
      \vec{G}^{+} + \vec{G}^{-}
    \right]
 ~~
 -
  \frac{\imath \omega}
       {2}
    \beta_c 
    \vec{u} \times
    \nabla \times
    \left[
      \vec{G}^{+} - \vec{G}^{-}
    \right]
\nonumber
\\
 && 
 ~~
 ~~
 \mp
  \frac{1}{2}
  \nabla 
    \vec{u} \times
    \nabla \times
  \left[
    \vec{u} \cdot 
    \left(
      \vec{G}^{+} - \vec{G}^{-}
    \right)
  \right]
 ~~
 \pm
    \vec{u} \times
    \nabla \times
  \beta_r
  \vec{J} 
\nonumber
\\
 && 
 ~~~~
 ~~
 +
  \frac{\omega^2}{2}
  \alpha_r \beta_r \alpha_c 
    \left[
      \vec{G}^{+} + \vec{G}^{-}
    \right]
 \mp
  \frac{\omega^2}{2}
  \alpha_r \beta_r \beta_c 
    \left[
      \vec{G}^{+} - \vec{G}^{-}
    \right]
 ~~
 -
  \frac{\imath \omega}{2}
  \alpha_r \beta_r
  \nabla 
  \left[
    \vec{u} \cdot 
    \left(
      \vec{G}^{+} - \vec{G}^{-}
    \right)
  \right]
 ~~
 +
  \imath \omega
  \alpha_r \beta_r^2
  \vec{J} 
 ~~~~
\\
\textrm{(E)} ~~~~ ~~~~
&=&
    \vec{u} \times
    \nabla \times
  \vec{u} \times
  \left[
    \nabla\times 
    \vec{G}^{\pm}
  \right]
 ~~
 +
  \frac{\omega^2}{2}
  \alpha_r \beta_r \alpha_c 
    \left[
      \vec{G}^{+} + \vec{G}^{-}
    \right]
 ~~
 \mp
  \frac{\imath \omega}
       {2}
    \alpha_c 
    \vec{u} \times
    \nabla \times
    \left[
      \vec{G}^{+} + \vec{G}^{-}
    \right]
\nonumber
\\
 && 
 ~~~~
 ~~
 \mp
  \frac{\omega^2}{2}
  \alpha_r \beta_r \beta_c 
    \left[
      \vec{G}^{+} - \vec{G}^{-}
    \right]
 ~~
 -
  \frac{\imath \omega}
       {2}
    \beta_c 
    \vec{u} \times
    \nabla \times
    \left[
      \vec{G}^{+} - \vec{G}^{-}
    \right]
\nonumber
\\
 && 
 ~~~~ ~~~~
 ~~
 \mp
  \frac{1}{2}
  \nabla 
    \vec{u} \times
    \nabla \times
  \left[
    \vec{u} \cdot 
    \left(
      \vec{G}^{+} - \vec{G}^{-}
    \right)
  \right]
 ~~
 -
  \frac{\imath \omega}{2}
  \alpha_r \beta_r
  \nabla 
  \left[
    \vec{u} \cdot 
    \left(
      \vec{G}^{+} - \vec{G}^{-}
    \right)
  \right]
 ~~
 +
  \imath \omega
  \alpha_r \beta_r^2
  \vec{J} 
 ~~
 \pm
    \vec{u} \times
    \nabla \times
  \beta_r
  \vec{J} 
\\
\textrm{(F)} ~~~~ ~~~~
&=&
    \vec{u} \times
    \nabla \times
  \vec{u} \times
  \left[
    \nabla\times 
    \vec{G}^{\pm}
  \right]
 ~~
 -
  \left(
   - \imath \omega
  \right)
    \frac{1}{2}
  \left[
    \frac{1}{c_r}
    \left(
     - \imath \omega
    \right)
   \mp
    \vec{u} \times
    \nabla \times
  \right]
    \alpha_c
    \left[
      \vec{G}^{+} + \vec{G}^{-}
    \right]
\nonumber
\\
 && 
 ~~~~
 ~~
 \pm
  \left(
   - \imath \omega
  \right)
    \frac{1}{2}
  \left[
    \frac{1}{c_r}
    \left(
     - \imath \omega
    \right)
   \mp
    \vec{u} \times
    \nabla \times
  \right]
    \beta_c
    \left[
      \vec{G}^{+} - \vec{G}^{-}
    \right]
\nonumber
\\
 && 
 ~~~~ ~~~~
 ~~
 +
    \frac{1}{2}
    \nabla
  \left[
    \frac{1}{c_r}
    \left(
     - \imath \omega
    \right)
   \mp
    \vec{u} \times
    \nabla \times
  \right]
    \vec{u} \cdot 
    \left[
      \vec{G}^{+} - \vec{G}^{-}
    \right]
 ~~
 -
  \left[
    \frac{1}{c_r}
    \left(
     - \imath \omega
    \right)
   \mp
    \vec{u} \times
    \nabla \times
  \right]
  \beta_r
  \vec{J} 
\\
\textrm{(G)} ~~~~ ~~~~
&=&
    \vec{u} \times
    \nabla \times
  \vec{u} \times
  \left[
    \nabla\times 
    \vec{G}^{\pm}
  \right]
 ~~
\nonumber
\\
 && 
 ~~~~
 ~~
 -
  \left(
   - \imath \omega
  \right)
    \frac{1}{2}
  \left[
    \frac{1}{c_r}
    \left(
     - \imath \omega
    \right)
   \mp
    \vec{u} \times
    \nabla \times
  \right]
 \left\{
    \alpha_c
    \left[
      \vec{G}^{+} + \vec{G}^{-}
    \right]
   \pm
    \beta_c
    \left[
      \vec{G}^{+} - \vec{G}^{-}
    \right]
   +
  \beta_r
  \vec{J} 
  \right\}
\nonumber
\\
 && 
 ~~~~ ~~~~
 ~~
 +
    \frac{1}{2}
    \nabla
  \left[
    \frac{1}{c_r}
    \left(
     - \imath \omega
    \right)
   \mp
    \vec{u} \times
    \nabla \times
  \right]
    \vec{u} \cdot 
    \left[
      \vec{G}^{+} - \vec{G}^{-}
    \right]
\end{eqnarray}

Time domain (non-dispersive reference) --
~
\begin{eqnarray}
\frac{d^2}{dt^2} {G}^{\pm}
&=&
    \vec{u} \times
    \nabla \times
  \vec{u} \times
  \left[
    \nabla\times 
    \vec{G}^{\pm}
  \right]
 ~~
\nonumber
\\
 && 
 ~~~~
 ~~
 -
  \frac{1}{2}
  \frac{d}{dt}
  \left[
    \frac{1}{c_r}
  \frac{d}{dt}
   \mp
    \vec{u} \times
    \nabla \times
  \right]
 \left\{
    \alpha_c \ast
    \left[
      \vec{G}^{+} + \vec{G}^{-}
    \right]
   \pm
    \beta_c \ast
    \left[
      \vec{G}^{+} - \vec{G}^{-}
    \right]
   +
  \beta_r
  \vec{J} 
  \right\}
\nonumber
\\
 && 
 ~~~~ ~~~~
 ~~
 +
    \frac{1}{2}
    \nabla
  \left[
    \frac{1}{c_r}
    \frac{d}{dt}
   \mp
    \vec{u} \times
    \nabla \times
  \right]
    \vec{u} \cdot 
    \left[
      \vec{G}^{+} - \vec{G}^{-}
    \right]
.
\end{eqnarray}

Since I've already assumed transverse fields, 
 in the source-free case  $\nabla \cdot \vec{E} = 0$) so 
 $\nabla \times \nabla \times \vec{Q} \rightarrow \nabla^2 \vec{Q} $.
Thus,
~
\begin{eqnarray}
  \nabla^2 \vec{G}^{\pm}
 -
  \frac{1}{c^2}
  \frac{d^2 }{dt^2} 
  \vec{G}^{\pm}
 -
    \frac{1}{2}
  \frac{d}{dt}
  \left\{
    \frac{1}{c_r}
    \frac{d}{dt}
   \mp
    \vec{u} \times 
    \nabla \times 
  \right\}
  \left\{
    \alpha_c \ast
    \left[
      \vec{G}^{+} + \vec{G}^{-}
    \right]
   \pm
    \beta_c \ast
    \left[
      \vec{G}^{+} - \vec{G}^{-}
    \right]
  \right\}
&=&
0.
\end{eqnarray}

Note the strong  similarity to
 eqns.(\ref{scaledsecond-a},\ref{scaledsecond-b}).  
The usual polarization term $\vec{P}$ is contained in 
 the $\alpha_c$ (and $\beta_c$) terms.
If I write $\alpha_c = \alpha_c^{D} + \alpha_c^{NL}$
 and $\vec{P} = \alpha_r \alpha_c^{NL} \vec{E}$; 
 dropping the magnetic terms (i.e. $\beta_c=0$) gives us:
~
\begin{eqnarray}
  \nabla^2 \vec{G}^{\pm}
 -
  \frac{1}{c^2}
  \frac{d^2 }{dt^2} 
  \vec{G}^{\pm}
 -
    \frac{1}{2}
  \frac{d}{dt}
  \left\{
    \frac{1}{c_r}
    \frac{d}{dt}
   \mp
    \vec{u} \times 
    \nabla \times 
  \right\}
    \alpha_c^{D} \ast
    \left[
      \vec{G}^{+} + \vec{G}^{-}
    \right]
&=&
 +
    \frac{1}{2\alpha_r}
  \frac{d}{dt}
  \left\{
    \frac{1}{c_r}
    \frac{d}{dt}
   \mp
    \vec{u} \times 
    \nabla \times 
  \right\}
  \vec{P}
.
\end{eqnarray}

The 
 $1/\alpha_r$ prefactor occurs on the RHS because we are relating 
 $G^\pm$ to the polarization $P$, 
 rather than the usual $E$ and $P$.

We could use these second order equations to apply diffraction to a standard 
 first-order equation propagation,
 by split-stepping the diffraction given here $\nabla^2 \vec{G}^\pm$
 with the usual dispersive and nonlinear propagation.

% -----------------

\subsection{GFEA Form}
\label{SS-evolution-gfea}

My {\em ``Few cycle pulse propagation''} 
 detailed calculation \cite{Kinsler-FCPP} has included terms for 
 $G^\pm$ field envelopes; 
 where the GFEA equation is ($g=\pm 1$):
~
\begin{eqnarray}
     q \partial_\xi
  A(\vec{r}_\bot,\xi,\tau) 
&=&
 -
  \frac{ \sigma \mp \sigma'}
       { 1 \pm \imath \sigma' \partial_\tau}
  \partial_\tau A(\vec{r}_\bot,\xi,\tau) 
 +
  \frac{ 1  +  \imath \sigma \partial_\tau}
       { 1 \pm \imath \sigma' \partial_\tau}
    \left( 
    - \frac{ \alpha_r}{\beta_r}
    + \imath  \hat{D}'
    \right)
  A(\vec{r}_\bot,\xi,\tau) 
  + 
    \frac{\imath }
         {2 \beta_r^2 \left( 1 \pm \imath \sigma' \partial_\tau \right)}
      \nabla_\bot^2 
    A(\vec{r}_\bot,\xi,\tau) 
\nonumber
\\
&~&
  +
    \frac{2 \imath \pi }{n_0^2}
      \left[
            1 
          \pm  \imath \partial_\tau 
          \opm \imath g \sigma' \partial_\tau
       \omp
        g
        \left(
            1 
          - 
            \imath q \partial_\xi
        \right)
      \right]
    \frac{
      \left( 
             1 \pm \imath \partial_\tau  
      \right)
    }
    {
             {\left( 1 \pm \imath \sigma' \partial_\tau \right)}
    }
    B(\vec{r}_\bot,\xi,\tau ; A) 
  + 
    \frac{ T_\Upsilon + T_{RHS} }
         { 1 \pm \imath \sigma' \partial_\tau }
\label{exact-BKP-extra}
.
\end{eqnarray}

Here the $\opm$ and $\omp$ symbols act like the usual $\pm, \mp$ ones;
but varying according to which $G$ variable is being described --
i.e. $G^{\opm}$.  The normal $\pm, \mp$ ones are relate to the 
carrier direction.

%\end{section}

% ----------------------------------------------------------------------

\section{Envelope Propagation Equations}\label{S-envelopeprops}

It is common in relatively narrowband cases to solve for 
 the propagation of field envelopes centred around chosen carrier frequencies 
 rather than the full detailed oscillations of the EM field. 
In fact, if sufficient care is taken with approximations, 
 and if the system simulated is well behaved, 
 even quite wideband pulses can be sucessfully modelled using envelopes.
The advantage of the $G^\pm$ wave equations presented in 
 sections \ref{S-1storderE}, \ref{S-1storderH}, and \ref{S-1storderD}
 is that we dispense with the second order form of the wave equation, 
 and generate envelope equations from first order equations.
This greatly reduces the number of approximations without increasing
 the complexity of the theory.
Although a full model requires four envelopes to describe the $G^\pm$,
 just as a full Maxwell theory requires four envelopes,
 (both backward and forward travelling $E$ and $H$), 
 this is rarely a case we are interested in solving, 
 and so in practice we can use just one envelope.

A full expansion of $G^\pm$ into forward and backward envelopes
$G^\pm_f$, $G^\pm_b$
would be
~
\begin{eqnarray}
  {G}^\pm (\omega)
&=&
    \mathscr{G}^{\pm}_f (\omega \mp \omega_0)
    e^{ \pm \imath k_0 z }
   +
    \mathscr{G}^{\pm}_f (\omega_0 \pm \omega)
    e^{ \mp \imath k_0 z }
\nonumber
\\
&&
 +
    \mathscr{G}^{\pm}_b (\omega  \mp \omega_0)
    e^{ \mp  \imath k_0 z }
   +
    \mathscr{G}^{\pm}_b (\omega_0 \mp \omega)
    e^{ \pm \imath k_0 z }
,
~~~~ 
\end{eqnarray}
 where we have suppressed the $z$ argument on the 
 envelope functions for brevity.  
Note that the argument shift ($\omega \rightarrow \omega \pm \omega_0$)
 is how the $-\imath \omega_0 t$ part of the carrier wave is accounted for.
If inserted into the wave equations, 
 this expansion would result in a large number of terms to consider, 
 even for the relatively simple case of a third order  nonlinearity.
However, 
 by specializing to the typical case where we are only interested
 in forward travelling waves, we can instead use
~
\begin{eqnarray}
  {G}^\pm (\omega)
&=&
    \mathscr{G}^{\pm}_f (\omega \mp \omega_0)
    e^{ \pm \imath k_0 z }
   +
    \mathscr{G}^{\pm}_f (\omega_0 \pm \omega)
    e^{ \mp \imath k_0 z }
\\
&=&
    \mathscr{G}^{\pm}_f (\omega \mp \omega_0)
    e^{ \pm  \imath k_0 z }
   +
    \mathscr{G}^{\pm}_f (\omega \mp \omega_0)^*
    e^{ \mp \imath k_0 z }
,
\\
  {G}^\pm (t)
&=&
    \mathscr{G}^{\pm}_f (t)
    e^{ \pm \imath k_0 z }
    e^{ \mp \imath \omega_0 t }
   +
    \mathscr{G}^{\pm}_f (t)^*
    e^{ \mp \imath k_0 z }
    e^{ \pm \imath \omega_0 t }
~~~~
\end{eqnarray}

This has the nice advantage that it eliminates and backward travelling waves, 
 and so we can propagate pulses efficiently in a moving frame.  
This is important, 
 because the backward parts in a moving frame move 
 at {\em twice} the frame speed.  
In a full (non-envelope) simulation, 
 we need to either take special efforts to filter out these backward parts,
 or any hoped-for numerical gains are reversed by 
 the finer $z$-step required to integrate (the backward parts) accurately.

% --------  --------  --------  -------- 
\subsection{Foward travelling envelopes in a first order wave equation}\label{Ss-envelopeprops-firstfwd}

I start with the plane polarized first order wave 
 eqn. (\ref{eqn-1storderE-propagation-planepol})
~
\begin{eqnarray}
  \partial_{z'} G_x^{\pm}
&=&
 \mp 
  \imath \omega 
  \alpha_r \beta_r 
  \left( 1 \mp \xi \right)
  ~
  G_x^{\pm}
 ~~
 \mp 
  \frac{\imath \omega \beta_r}
       {2}
    \alpha_c 
  *
    \left[ G_x^+ + G_x^- \right]
 ~~
 -
  \frac{\imath \omega \alpha_r}
       {2}
    \beta_c
  *
    \left[ G_x^+ - G_x^- \right]
.
\label{eqn-envelope-2398756}
\end{eqnarray}

I insert envelopes using the shorthand notation $\Xi= \imath k_0z$.
Since $\alpha_c$ may contain a dependence on the fields $G^\pm$,
 I split it using $\alpha_c = a_c + a_c^*$. 
In the case of dispersion, 
 $a_c=\alpha_c/2$;
 but for a nonlinearity it will be 
 the appropriately carrier-matched, positive frequency part
 of $\alpha_c$, 
 once the field values it contains have been expanded in terms of the 
 envelope and carrier functions.
I will proceed in the stationary frame case ($\xi=0)$,
 so
~
\begin{eqnarray}
  \partial_{z} %'} 
  \left[
        \mathscr{G}^{\pm}_f
    e^{ \pm \Xi}
   +
    \mathscr{G}^{\pm}_f ~^*
    e^{ \mp \Xi}
  \right]
&=&
 \mp 
  \imath \omega 
  \alpha_r \beta_r 
  % \left( 1 \mp \xi \right)
  ~
  \left[
        \mathscr{G}^{\pm}_f
    e^{ \pm \Xi}
   +
    \mathscr{G}^{\pm}_f ~^*
    e^{ \mp \Xi}
  \right]
\nonumber
\\
&&
 ~~
 \mp 
  \frac{\imath \omega \left( a_c + a_c^* \right) \beta_r}
       {2}
  \left\{
          \mathscr{G}^{\pm}_f
      e^{ \pm \Xi}
     +
      \mathscr{G}^{\pm}_f ~^*
      e^{ \mp \Xi}
   +
          \mathscr{G}^{\mp}_f
      e^{ \mp \Xi}
     +
      \mathscr{G}^{\mp}_f ~^*
      e^{ \pm \Xi}
  \right\}
 ~~~~
\\
    e^{ \pm \Xi}
    \partial_{z} 
      \mathscr{G}^{\pm}_f
   \pm
     \imath k_0
     e^{ \pm \Xi}
     \mathscr{G}^{\pm}_f
  % \mp
  %   \imath \omega_0 c_f^{-1}
  %   e^{ \pm \Xi}
  %  \delta^\pm
%~~~~ ~~~~ ~~~~
%&&
%\\
   +
    e^{ \mp \Xi}
    \partial_{z} 
      \mathscr{G}^{\pm}_f ~^*
   \mp
     \imath k_0
     e^{ \mp \Xi}
      \mathscr{G}^{\pm}_f ~^*
 %  \pm
 %   \imath \omega_0 c_f^{-1}
 %   e^{ \mp \Xi}
 %   \mathscr{G}^{\pm}_f ~^*
&=&
 \mp 
  \imath \omega 
  \alpha_r \beta_r 
  % \left( 1 \mp \xi \right)
  ~
  \left[
        \mathscr{G}^{\pm}_f
    e^{ \pm \Xi}
   +
    \mathscr{G}^{\pm}_f ~^*
    e^{ \mp \Xi}
  \right]
\nonumber
\\
&&
 ~~
 \mp 
  \frac{\imath \omega \left( a_c + a_c^* \right) \beta_r}
       {2}
  \left\{
          \mathscr{G}^{\pm}_f
      e^{ \pm \Xi}
     +
      \mathscr{G}^{\pm}_f ~^*
      e^{ \mp \Xi}
   +
          \mathscr{G}^{\mp}_f
      e^{ \mp \Xi}
     +
      \mathscr{G}^{\mp}_f ~^*
      e^{ \pm \Xi}
  \right\}
 ~~~~
\\
\textrm{split c.c. halves;} ~~~~ ~~~~
%&&
%\nonumber
%\\
    e^{ \pm \Xi}
    \partial_{z'} 
      \mathscr{G}^{\pm}_f
   \pm
     \imath k_0
     e^{ \pm \Xi}
     \mathscr{G}^{\pm}_f
   %\mp
   %  \imath \omega c_f^{-1}
   %  e^{ \pm \Xi}
   %  \mathscr{G}^{\pm}_f
&=&
 \mp 
  \imath \omega 
  \alpha_r \beta_r 
  %\left( 1 \mp \xi \right)
  ~
        \mathscr{G}^{\pm}_f
    e^{ \pm \Xi}
 ~~
 \mp 
  \frac{\imath \omega ~ a_c \beta_r}
       {2}
  \left\{
          \mathscr{G}^{\pm}_f
      e^{ \pm \Xi}
     +
      \mathscr{G}^{\mp}_f ~^*
      e^{ \pm \Xi}
  \right\}
\\
\textrm{cancel} ~~ e^{ \pm \Xi} ; ~~~~ ~~~~
    \partial_{z} 
      \mathscr{G}^{\pm}_f
   \pm
     \imath k_0
     \mathscr{G}^{\pm}_f
  % \mp
  %   \imath \omega c_f^{-1}
  %   \mathscr{G}^{\pm}_f
&=&
 \mp 
  \imath \omega 
  \alpha_r \beta_r 
  % \left( 1 \mp \xi \right)
  ~
        \mathscr{G}^{\pm}_f
 \mp 
  \frac{\imath \omega ~ a_c \beta_r}
       {2}
  \left\{
          \mathscr{G}^{\pm}_f
     +
      \mathscr{G}^{\mp}_f  ~^*
  \right\}
\\
    \partial_{z} 
      \mathscr{G}^{\pm}_f
&=&
 \mp
  \imath 
  \left( \omega \alpha_r \beta_r - k_0 \right)
  ~
        \mathscr{G}^{\pm}_f
 ~~
 \mp
  \frac{\imath \omega ~ a_c \beta_r}
       {2}
  \left\{
          \mathscr{G}^{\pm}_f
     +
      \mathscr{G}^{\mp}_f ~^*
  \right\}
,
\end{eqnarray}

If we do the sensible thing and relate the carrier parameters 
$k_0$ and $\omega_0$ using the reference parameters
(so $k_0 = \omega_0/c_r(\omega_0)$), 
we have 
$\omega \alpha_r \beta_r - k_0 
= (\omega-\omega_0)\alpha_r \beta_r + \omega_0 \alpha_r \beta_r - k_0 
= (\omega-\omega_0)\alpha_r \beta_r$:
~
\begin{eqnarray}
    \partial_{z} 
      \mathscr{G}^{\pm}_f
&=&
 \mp
  \imath 
  \left( \omega  - \omega_0 \right) \alpha_r \beta_r
  ~
        \mathscr{G}^{\pm}_f
 ~~
 \mp
  \frac{\imath \omega ~ a_c \beta_r}
       {2}
  \left\{
          \mathscr{G}^{\pm}_f
     +
      \mathscr{G}^{\mp}_f ~^*
  \right\}
,
\end{eqnarray}

Now transform into the moving frame.  
Note that we will no longer want the frame to move 
at the phase velocity at $\omega_0$, 
because that job has been taken over by the carrier; 
but we could use it to absorb any residual (reference) dispersion 
affecting one of the envelopes (most likely for $\mathscr{G}^{+}_f$).
However, it will then make the residual (reference) dispersion term 
acting on the other envelope (e.g. $\mathscr{G}^{-}_f$) twice 
as large.
Anyway
~
\begin{eqnarray}
    \partial_{z'} 
      \mathscr{G}^{\pm}_f
  - 
    \imath 
    \omega 
    \alpha_f \beta_f
      \mathscr{G}^{\pm}_f
&=&
 \mp
  \imath 
  \left( 
    \omega \alpha_r \beta_r 
   -
    k_0
  \right)
  ~
        \mathscr{G}^{\pm}_f
 ~~
 \mp
  \frac{\imath \omega ~ a_c \beta_r}
       {2}
  \left\{
          \mathscr{G}^{\pm}_f
     +
      \mathscr{G}^{\mp}_f  ~^*
  \right\}
,
\\
\Longrightarrow ~~~~ ~~~~ ~~~~ ~~~~ ~~~~ ~~~~ ~~~~ ~~~~
    \partial_{z'} 
      \mathscr{G}^{\pm}_f
&=&
 \mp
  \imath 
  \left( 
    \omega \alpha_r \beta_r 
   \mp
    \omega \alpha_f \beta_f
   -
    k_0
  \right)
  ~
        \mathscr{G}^{\pm}_f
 ~~
 \mp
  \frac{\imath \omega ~ a_c \beta_r}
       {2}
  \left\{
          \mathscr{G}^{\pm}_f
     +
      \mathscr{G}^{\mp}_f  ~^*
  \right\}
\end{eqnarray}

% --------  --------  --------  -------- 
\subsection{Envelopes and the Second Order wave equation}\label{Ss-envelopeprops-second}

See \cite{Kinsler-FCPP}, which includes correction terms appropriate 
 to $G^\pm$ fields.
Note that there I can think of no reason to use the second order form, 
 since the first order form performs as well with fewer complications.

% ----------------------------------------------------------------------
\section{Vector identities}\label{S-vectorid}

{\bf (1)} Identity (old Physics 31.320 vector identity sheet, (I-5)):
\begin{eqnarray}
  \nabla \left( \vec{u} \cdot \vec{H} \right)
&=&
  \left( \vec{u} \cdot \nabla \right) \vec{H}
 +
  \left( \vec{H} \cdot \nabla \right) \vec{u}
 +
  \vec{u} \times \left( \nabla \times \vec{H} \right)
 +
  \vec{H} \times \left( \nabla \times \vec{u} \right)
\label{eqn-vectorid-I-5}
\\
\textrm{rearrange} ~~~~ ~~~~
  \vec{u} \times \left( \nabla \times \vec{H} \right)
&=&
  \nabla \left( \vec{u} \cdot \vec{H} \right)
 -
  \left( \vec{u} \cdot \nabla \right) \vec{H}
 -
  \left( \vec{H} \cdot \nabla \right) \vec{u}
 -
  \vec{H} \times \left( \nabla \times \vec{u} \right)
\\
\textrm{constant unit vector} ~~~~
&=&
  \nabla \left( \vec{u} \cdot \vec{H} \right)
 -
  \left( \vec{u} \cdot \nabla \right) \vec{H}
 -
  0
 -
  0
\\
\textrm{transverse field} ~~~~
&=&
  0
 -
  \left( \vec{u} \cdot \nabla \right) \vec{H}
 -
  0
 -
  0
\\
\textrm{Useful:} ~~~~
  \left( \vec{u} \cdot \nabla \right) \vec{H}
&=& 
  \nabla \left( \vec{u} \cdot \vec{H} \right)
 -
  \vec{u} \times \left( \nabla \times \vec{H} \right)
\end{eqnarray}

{\bf (2)} Identity (old Physics 31.320 vector identity sheet, (I-10)):
\begin{eqnarray}
  \nabla \times \left( \vec{u} \times \vec{H} \right)
&=&
  \vec{u} ~~ \left( \nabla \cdot \vec{H} \right)
 -
  \vec{H} ~~ \left( \nabla \cdot \vec{u} \right)
 +
  \left( \vec{H} \cdot \nabla \right) ~~ \vec{u}
 -
  \left( \vec{u} \cdot \nabla \right) ~~ \vec{H}
\label{eqn-vectorid-I-10}\\
\textrm{constant unit vector} ~~~~
&=&
  \vec{u} ~~ \left( \nabla \cdot \vec{H} \right)
 -
  0
 +
  0
 -
  \left( \vec{u} \cdot \nabla \right) ~~ \vec{H}
\\
\textrm{no monopoles} ~~~~
&=&
  0
 -
  0
 +
  0
 -
  \left( \vec{u} \cdot \nabla  \right) ~~ \vec{H}
\\
\textrm{Useful:} ~~~~
  \left( \vec{u} \cdot \nabla \right) \vec{H}
&=& 
 -
  \nabla \times \left( \vec{u} \times \vec{H} \right) 
.
\end{eqnarray}

These two identities, specialised to the case involving our unit vector
$\vec{u}$ in the propagation direction, mean that we can equate
~
\begin{eqnarray}
  \vec{u} \times \left( \nabla \times \vec{H} \right)
 -
  \nabla \left( \vec{u} \cdot \vec{H} \right)
&=&
  \nabla \times \left( \vec{u} \times \vec{H} \right)
\label{eqn-vectorid-combined}
\end{eqnarray}

Note also
~
\begin{eqnarray}
  \vec{a} \times \vec{b} \times \vec{c} 
&=& 
  \left( \vec{a} \cdot \vec{c} \right) \vec{b} - 
  \left( \vec{a} \cdot \vec{b} \right) \vec{c}
\label{eqn-vectorid-doublecross}
\end{eqnarray}

so in our case where $\vec{a}$ and $\vec{b}$ are both unit vectors 
~
\begin{eqnarray}
  \vec{u} \times \left( \vec{u} \times \vec{X} \right)
&=& 
  \left( \vec{u} \cdot \vec{X} \right) \vec{u} - 
  \left( \vec{u} \cdot \vec{u} \right) \vec{X}
\label{eqn-vectorid-doublecross-u}
\end{eqnarray}

Also, 
 I use (old Physics 31.320 vector identity sheet, (I-7)):
\begin{eqnarray}
  \nabla \cdot \left( \vec{A} \times \vec{B} \right)
&=&
  \vec{A} \cdot \left( \nabla  \times \vec{B} \right)
 -
  \vec{B} \cdot \left( \nabla \times \vec{A} \right)
\label{eqn-vectorid-I-7}
\end{eqnarray}

% ----------------------------------------------------------------------
%\newpage
\section{Units}

\begin{eqnarray}
  \left[ \mu \right] 
&=&
  \textrm{N/A}^2 =  \textrm{m} 
                  . \textrm{kg} 
                  . \textrm{s}^{-2} 
                  . \textrm{A}^{-2}
\\
  \left[ B \right] 
&=&
  \textrm{V} 
                  . \textrm{s} 
                  . \textrm{m}^{-1} 
\\
  \left[ \mu H^2 \right] 
&=&
  \left[ \mu \left(B/\mu\right)^2 \right] 
  =
    \left[ B^2/\mu \right] 
                 = 
                   \textrm{V}^2 
                  . \textrm{s}^2
                  . \textrm{m}^{-2} 
                  ~~ . \textrm{m}^{-1} 
                     . \textrm{kg}^{-1} 
                     . \textrm{s}^{2} 
                     . \textrm{A}^{2}
\\
&=&
  ~~~~
  \left[
    \textrm{V} . \textrm{A} . \textrm{s}^2 . \textrm{m}^{-2} 
  \right]^2
  ~~ . \textrm{m}^{-1} . \textrm{kg}^{-1} 
\\
\nonumber
\\
  \left[ \epsilon \right] 
&=&
  \textrm{F/m} =  \textrm{A}^{2} 
                . \textrm{s}^{4} 
                . \textrm{m}^{-3}
                . \textrm{kg}^{-1}
\\
  \left[ E \right] 
&=&
  \textrm{V} 
                  . \textrm{m}^{-1} 
\\
  \left[ \epsilon E^2 \right] 
&=&
  \textrm{A}^{2} 
                . \textrm{s}^{4} 
                . \textrm{m}^{-3}
                . \textrm{kg}^{-1}
                ~~ . \textrm{V}^2 
                   . \textrm{m}^{-2} 
\\
&=&
  ~~~~
  \left[
    \textrm{V} . \textrm{A} . \textrm{s}^2 . \textrm{m}^{-2} 
  \right]^2
  ~~ . \textrm{m}^{-1} . \textrm{kg}^{-1} 
\end{eqnarray}

% ----------------------------------------------------------------------
% ----------------------------------------------------------------------

\end{document}